\documentclass[11pt]{article}

\PassOptionsToPackage{dvipsnames}{xcolor}

\usepackage[margin=1in]{geometry}
\usepackage{pstricks,pst-node,pst-text}
\usepackage{amsthm}
\usepackage{amssymb}
\usepackage{amsmath}
\usepackage{amsfonts}
\usepackage{enumerate}
\usepackage[english]{babel}
\usepackage{graphicx}
\usepackage[cp1250]{inputenc}
\usepackage[sort,comma]{natbib}
\usepackage{float}
\usepackage{chngcntr}
\usepackage{dsfont}
\usepackage{geometry}
\usepackage[title]{appendix}
\usepackage{multicol}
\usepackage{enumitem}
\usepackage{multirow}
\usepackage{booktabs}
\usepackage{setspace}
\usepackage{makecell}

\numberwithin{equation}{section}
\numberwithin{figure}{section}
\numberwithin{table}{section}

\newcommand{\beqo}{\begin{eqnarray*}}
\newcommand{\eeqo}{\end{eqnarray*}\noindent}
\newcommand{\beq}{\begin{eqnarray}}
\newcommand{\eeq}{\end{eqnarray}\noindent}

\newtheorem{thm}{Theorem}

\newtheorem{rem}{Remark}
\newtheorem{cor}{Corollary}
\newtheorem{df}{Definition}
\newtheorem{prop}{Proposition}

\def\be{{\mathbb{E}}}
\def\E{{\mathbb{E}}}

\def\bp{{\mathbb{P}}}

\def\bq{{\mathbb{Q}}}

\def\br{{\mathbb{R}}}

\def\R{{\mathbb R}}  

\def\bx{\mathbf{x}}

\def\bX{\mathbf{X}}

\newcommand{\bl}[1]{\textcolor{black}{{#1}}}

\begin{document}

\author{{\L}ukasz Delong\footnote{University of Warsaw, Department of Statistics and Econometrics,
l.delong@uw.edu.pl} \and
Mario W\"{u}thrich\footnote{ETH Z\"{u}rich, Department of Mathematics, mario.wuethrich@math.ethz.ch}}

\date{}

\title{Universal Inference for Testing Calibration \bl{and Specification} of Mean Estimates within the Exponential Dispersion Family}

\maketitle

\begin{abstract}\noindent
\bl{Calibration of mean estimates is a crucial property in many applications, particularly in actuarial decision-making. Tests of mean-calibration are often performed under the assumption that the estimated means correctly preserve the ordering of the corresponding population means. In this paper, we first consider the classical Likelihood Ratio Test (LRT) within the Exponential Dispersion Family (EDF), and we show that applying likelihood-based methods to test mean-calibration within the EDF requires a strong ordering assumption on the mean estimates. This assumption makes the null hypothesis of the mean-calibration test equivalent to the correct specification of the mean estimates. Next, we investigate the framework of universal inference for testing mean-calibration and mean-specification, and the relationship between the tests. We develop a sub-sampled split LRT within the EDF that provides finite-sample guarantees with universally valid critical values. We investigate type I error, power, and e-power of the sub-sampled split LRT and compare its performance with that of the classical LRT. We also propose a novel test statistic based on a sub-sampled split $L_q$-likelihood ratio test (L$q$RT) to enhance the performance of the statistical test. Numerical results demonstrate that universal inference based on the sub-sampled split LRT and sub-sampled split L$q$RT provides an attractive alternative to the classical LRT within the EDF, achieving high power for detecting miscalibration and misspecification in medium- and large-sized samples.}

\medskip


\end{abstract}

\section{Introduction}\label{sec_intro}

Our focus is on prediction problems of responses $Y$. Let the response $Y$ be equipped with features (covariates, explanatory variables) $\mathbf{X}$ taking values in a feature space ${\cal X}\subseteq \br^\ell$. We are interested in (measurable) regression functions $\mu:\mathcal{X}\to \br$, $\mathbf{X}\mapsto \mu(\mathbf{X})$, serving as predictions for $Y$, under given features $\mathbf{X}$.

\begin{df}
A regression function ${\mu}:\mathcal{X}\to \br$ is mean-calibrated for the pair $(Y, \mathbf{X})$, or simply calibrated, if
\beq\label{calibration_def}
\mathbb{E}\left[Y \mid {\mu}\left(\mathbf{X}\right)\right] = {\mu}\left(\mathbf{X}\right), \qquad \text{almost surely}.
\eeq
\end{df}
\noindent The notion of calibration was introduced by \cite{murphy_decomp} in the literature of meteorology, and it has been studied in detail in the recent statistical literature; see, e.g., \cite{pohle_decomp}, \cite{dimitriadis_diagram}, \cite{gneiting_evaluation}. In different streams of the literature, (mean-)calibration \eqref{calibration_def} is also known as auto-calibration, well-calibration or moderate calibration. There is a consensus that mean-calibration is a minimal requirement that predictive models should satisfy to be sensible prediction tools.

\bl{The property of mean-calibration \eqref{calibration_def} is weaker than the requirement of ${\mu}$ being correctly mean-specified}. 
\begin{df}
\bl{A regression function ${\mu}:\mathcal{X}\to \br$ is mean-specified for the pair $(Y, \mathbf{X})$ if}
\beq\label{specification_def}
\mathbb{E}\left[Y \mid \mathbf{X}\right] = {\mu}\left(\mathbf{X}\right), \qquad \text{almost surely}.
\eeq
\end{df}

\bl{Property \eqref{calibration_def} implies that the prediction ${\mu}(\mathbf{X})$ of $Y$ is on average correct within the group of instances with an identical prediction. Clearly, this is a desirable property of predictive models, especially in actuarial applications, as discussed below. Property \eqref{specification_def} implies that whenever we predict $Y$ for a given feature combination $\mathbf{X}$, the prediction is on average correct. It checks whether our model correctly estimates the mean value for each individual feature combination. Thus, this latter property has a finer granularity, in fact, this is the finest granularity w.r.t.~the available feature information $\bX$, whereas \eqref{calibration_def} uses a one-dimensional summary of this information. In contrast to moderate calibration \eqref{calibration_def}, the property defined in \eqref{specification_def} is also sometimes referred to as strong calibration; see \cite{calster}.} 

\bl{Calibration of mean estimates is particularly important in non-life insurance pricing;  \cite{lindholm_pricing} discuss key properties. Traditionally, actuaries have focused on in-sample unbiasedness of predictions, both at the portfolio level and within predefined cohorts. These properties imply that premiums should neither under- nor overestimate claims on average. In actuarial practice, global in-sample unbiasedness is often referred to as the balance property, which means that premiums can be viewed as an allocation of the total observed claims cost across the portfolio. If predictions are in-sample unbiased within cohorts, then each pricing cohort is on average self-financing, and no systematic cross-subsidization occurs within the pricing system. The importance of unbiasedness of predictions for non-life insurance pricing has been emphasized throughout the actuarial literature; see \cite{denuit_truffin}, \cite{denuit_curves}, \cite{WZiegel}, \cite{wuthrich_auto_test}, \cite{lindholm_pricing}, and the references therein.} 

\bl{The mean-calibration property \eqref{calibration_def} can be viewed as a refinement of this cohort-level unbiasedness. In contrast to the traditional approach, in which cohorts are defined ex-ante by the actuary, mean-calibration considers groups formed endogenously by the individuals sharing the same predictive value; when the regression function ${\mu}$ is highly granular, such groups may contain only a small number of observations. Consequently, mean-calibration is a local property.}

\bl{In the context of \eqref{calibration_def} and \eqref{specification_def}, it is useful to highlight the distinction between these two properties in insurance pricing. Property \eqref{specification_def} implies that the regression function for premium calculation is both correctly specified and correctly estimated, meaning that no further improvement in predictive performance is possible w.r.t.~the information set generated by $\bX$, i.e., all risks are correctly valuated and discriminated w.r.t.~$\bX$. In contrast, the weaker property \eqref{calibration_def} states that, when grouping together all policyholders with the same premium, the true average mean value for the claims of this group coincides with the predicted mean. This is a statement at a coarser granularity and it does not necessarily imply that the regression function for premium calculation is fully adequate, because it may lack sufficient discriminatory power to distinguish the risks. For example, the global mean $\mu_0=\E[Y]$ satisfies \eqref{calibration_def}, but it does not discriminate at all. Therefore, the ultimate goal is to estimate the true regression function \eqref{specification_def}, which provides the maximal predictive power w.r.t.~the available information $\bX$.}

\bl{Calibration and specification of mean estimates are also crucial in non-life insurance claims reserving. In recent years, individual claims reserving models have gained increasing popularity, often relying on sophisticated predictive techniques; see \cite{lopez}, \cite{delong_rbns}, \cite{gabrielli}, \cite{chaoubi}, \cite{richman}, and the references therein. Obtaining accurate reserve estimates requires predictions to be locally unbiased. In other words, at least mean-calibration of mean estimates is needed to ensure that expected future claim payments are neither systematically under- nor overestimated at a granular level. These claims estimates subsequently enter pricing algorithms, so any bias in the reserves may propagate to insurance premiums, which should of course be avoided; see \cite{antonio}. Therefore, mean-calibration and mean-specification also play a crucial role in obtaining adequate claims reserves.}

\bl{Validation of mean-specification \eqref{specification_def} is a classical statistical problem, which is generally difficult to solve without distributional assumptions. The validation of mean-calibration \eqref{calibration_def} is a comparably recent problem, and it is also challenging due to its local nature. The following approaches are considered in the literature for testing mean-calibration.} \cite{dimitriadis_diagram}, \cite{gneiting_evaluation} and \cite{Dimitriadis} suggest CORP reliability diagrams in a Bernoulli setting, which can be extended to a setting with real-valued responses. A deviation of a CORP reliability diagram from its diagonal line is a sign of miscalibration of the mean estimates, this gives a visual calibration test; see also \cite{W2026}. This approach is distribution-free. \cite{delong_iso} develop Monte Carlo/bootstrap techniques to estimate consistency bands for CORP reliability diagrams within the exponential dispersion family (EDF), and \cite{delong_band} derive analytical formulas for calibration bands of mean estimates within the EDF. \cite{pohle_decomp}, \cite{dimitriadis_diagram} and \cite{gneiting_evaluation} propose to decompose the validation loss (under strictly consistent loss functions) resulting in \cite{murphy_decomp}'s score decomposition with the three components of miscalibration, discrimination (resolution) and uncertainty. The resulting miscalibration statistics summarizes the deviation of a CORP reliability diagram from its diagonal line in a single number. Using the Monte Carlo/bootstrap techniques of \cite{delong_iso}, this allows one to test for the null hypothesis of mean-calibration using the miscalibration statistics of Murphy's score decomposition. \bl{All these approaches rely on isotonic regression and the assumption that the estimated means correctly preserve the true ordering of the corresponding population means.} \cite{denuit_curves} take a different approach and derive a statistical test with an asymptotic distribution being based on the difference between Lorenz and concentration curves, and this is further discussed in simplified settings in \cite{wuthrich_auto_test}. These approaches do not depend on the distribution of the response, but they use their asymptotic variances.

\bl{The Likelihood Ratio Test (LRT) is the most popular statistical test in parametric and non-parametric hypothesis testing. Yet, it requires an assumption concerning the full distribution of the observations. To the best of our knowledge, the LRT has not been investigated so far in the context of mean-calibration. Our goal is to show how the Likelihood Ratio Statistics (LRS) can be used for verifying mean-calibration, as defined in \eqref{calibration_def}, and mean-specification, as defined in \eqref{specification_def}, within the EDF. As usually is done in the literature on mean-calibration, we assume that the estimated means correctly order the corresponding population means. However, the application of likelihood-based methods to test mean-calibration within the EDF requires a stronger ordering assumption on the mean estimates than the one adopted in the mean-calibration literature. This assumption makes the null hypothesis of the mean-calibration test equivalent to the correct specification of the mean estimate}. We show that the LRS for testing mean-calibration and mean-specification in classical inference coincides with the miscalibration statistics of Murphy's score decomposition within the EDF, up to a non-linear transformation. This seems very promising since both the LRT and Murphy's score decomposition have strong technical foundations in mathematical statistics. However, the most challenging part of the classical LRS and the miscalibration statistics in practical applications is to derive the critical region of the test; as noted in \cite{dimitriadis_diagram} and \cite{gneiting_evaluation}. One can use Monte Carlo simulation or bootstrap, as in \cite{delong_iso}, but the need for simulations clearly limits a widespread application of the test in practice. An important point here is that we do not expect that the classical LRS for testing mean-calibration/mean-specification has a standard $\chi^2$-distribution. \bl{A main contribution of this paper is to show how likelihood-based methods can be applied to test mean-calibration and mean-specification using a simplified decision rule, without estimating the quantile (critical value) of the test statistic under the null hypothesis.}

For cases in which the regularity assumptions and large sample conditions required for the classical LRT are unlikely to hold, and the asymptotic distribution of the test statistic is not tractable, \cite{wasserman_universal} have recently introduced {\it universal inference} and the {\it split LRT}. The idea of the split LRT is to split the data into two sets: a {\it training sample} used to train a model under the alternative hypothesis and a {\it validation sample} used to validate the null hypothesis against the alternative. Interestingly, thanks to this data splitting scheme, a universal critical value for the split LRT can be derived in rather general settings. As a result, simulation of the test statistic under the null hypothesis is not required. In this paper, \bl{we develop and promote universal inference for testing mean-calibration and mean-specification within the EDF}. Even though many approaches for testing mean-calibration and mean-specifcation have been introduced, see above discussion, we believe that none of them offers a straightforward and sufficiently practical application. At the same time, universal inference is gaining popularity in the statistical community; \cite{wasserman_universal} give numerous applications of universal inference in irregular parametric and non-parametric settings in which the split LRT is the first ever statistical test with finite sample guarantees, e.g., for testing the number of components in mixture models. \cite{strieder_universal} observe empirically that the split LRT may be conservative. \cite{dunn_universal} study the performance of the split LRT in Gaussian models and they conclude that the power of the split LRT is reasonable compared to the classical LRT, even though the type I error is below the pre-assumed significance level. \cite{strieder_universal} and \cite{dunn_universal} conclude that the choice of the splitting ratio equal to $1/2$ into training and validation samples is a reasonable choice in all dimensions and should lead to the highest power of the split LRT (under certain assumptions). \cite{shi_universal} recently investigated the split LRT in Gaussian mixture models. Finally, \cite{henzi_hl} use the split LRT to validate calibration of estimated probabilities in a binomial distribution setting in the sense of \eqref{calibration_def}. All this literature recommends to use sub-sampled versions of the split LRT, unfortunately most conclusions are model-specific.

\bl{From a statistical point of view, this paper contributes to the framework of universal inference by generalizing the binomial results of \cite{henzi_hl} to the entire EDF}. Following \cite{wasserman_universal} and \cite{dunn_universal}: \textit{"Our methods may not be optimal, though we do not yet fully understand how close to optimal they are beyond special cases. Many basic questions remain unanswered about the universal LRT, since its power even in very simple settings remains unknown."}  \bl{Our goal is to take a further step in this direction and contribute to the field of universal inference by investigating the performance of the sub-sampled split LRT for testing mean-calibration and mean-specification within the EDF, under the assumption that the mean estimates correctly order the corresponding population means (in an appropriate sense). We study the relationship between the mean-calibration and mean-specification tests under different ordering assumptions. As we are most interested in assessing miscalibration, we observe that, although likelihood-based methods require a strong ordering assumption on the mean estimates to test mean-calibration, the sub-sampled split LRT within the universal inference framework retains power to detect mean-calibration under the weaker ordering assumption commonly adopted in the mean-calibration literature.}

In this paper, in a theoretical framework and in a numerical example, we consider the split LRT and the sub-sampled split LRT within the EDF, their type I error, power and e-power, compare the sub-sampled split LRT to the classical LRT, and we study the effect of sampling of training and validation sets on the split LRT. Moreover, we propose a new testing procedure to verify mean-calibration and mean-specification within universal inference. The novel test statistics is based on the sub-sampled split L$q$RT and is motivated by the $L_q$-Likelihood Ratio Test (L$q$RT), see \cite{lq_estimator}, \cite{lq_test} -- the terminology $L_q$ comes from $q$-logarithm. All our results essentially draw on the newly developed framework of e-values and e-testing; an excellent reference is the book of \cite{wang_book}. From our numerical analysis, we conclude that the sub-sampled split LRT and the so-called sub-sampled split mean power L$q$RT have a high and reasonable power in medium/large samples in the case of medium/large miscalibration/mispecification errors, and the sub-sampled split mean power L$q$RT may perform better in some scenarios than the sub-sampled split LRT in terms of the power of rejecting the false null hypothesis against the true alternative for mean estimates. \bl{We also study the performance of the sub-sampled split LRT in scenarios where the test's underlying assumptions are violated. We demonstrate that the sub-sampled split LRT is robust to small/medium mispecification of the assumptions underlying the test in medium/large samples in the case of medium/large miscalibration/mispecification errors.}

\bl{From an actuarial perspective, we focus on both synthetic and real insurance data sets in our numerical experiments. The contribution of this paper is to extend the toolbox available to actuaries for validating predictive models and to introduce, for the first time in the actuarial literature, a straightforward and practical test for assessing calibration and specification of mean estimates.}

This paper is structured as follows. In Section \ref{sec_model}, we introduce our statistical set-up. In Section \ref{sec_classic}, we discuss classical approaches to test mean-calibration and mean-specification. In Section \ref{sec_universal}, we focus on universal inference and present the main theoretical results of this paper. Finally, in Section \ref{sec_examples} we study the split LRT and the split L$q$RT with numerical examples. \bl{The appendix contains: A -- possible methods for verifying the ordering assumptions imposed on the mean estimates, B -- a discussion of the case in which the dispersion parameter of the EDF is estimated, C -- additional analyses related to the split L$q$RT, and D -- an application of the results on a real data set.}

\section{Statistical set-up}\label{sec_model}

We work within the EDF. The EDF is given by densities of the form
\beq\label{edf_density}
p(y)=\exp\left(\frac{y\theta-\kappa(\theta)}{\phi/v}+a(y,v/\phi)\right),
\eeq
where $\theta$ denotes the {\it canonical parameter}, $\phi>0$ is the {\it dispersion parameter}, $v>0$ is a {\it case weight} and $a(\cdot,v/\phi)$ is a normalizing function w.r.t.~the selected $\sigma$-finite measure on $\R$ so that $p$ integrates to one. The canonical parameter $\theta$ is supported in the effective domain $\boldsymbol{\Theta}$, and the {\it cumulant function} $\kappa:\boldsymbol{\Theta}\to \br$ is smooth and $\kappa''(\theta)>0$ on the interior of the effective domain denoted by ${\rm int}\big(\boldsymbol{\Theta}\big)$.

\bl{The specific distributional choice within the EDF is characterized by the functional form of the cumulant function $\kappa$. The EDF includes, e.g., the normal, binomial, Poisson, gamma and inverse Gaussian distributions. For a specified distribution from the EDF, i.e., for a given $\kappa$, a member of the EDF with tuple $(\theta,\phi,v)$ is denoted by $EDF_{v,\phi}(\theta)$. We omit the cumulant function from the notation since we work under the fixed selected distribution. The canonical parameter $\theta$ of the distribution $EDF_{v,\phi}(\theta)$ is our main object of interest under the given cumulant function $\kappa$. The random variable $Y \sim EDF_{v,\phi}(\theta)$ has expected value $\mu:=\mathbb{E}[Y]=\kappa'(\theta)$. Based on the strict convexity of $\kappa$ on ${\rm int}\big(\boldsymbol{\Theta}\big)$, there is a one-to-one correspondence between the mean parametrization $\mu$ and the canonical parameter $\theta$ of the selected member of the EDF. More precisely, we can either work with the canonical parameter representation $\theta\in{\rm int}\big(\boldsymbol{\Theta}\big)$ or with its mean parameter representation $\mu\in\boldsymbol{\Gamma}$ where}
\beqo
\boldsymbol{\Gamma}=\kappa'\big({\rm int}\big(\boldsymbol{\Theta}\big)\big)=\big\{\kappa'(\theta); \ \theta\in{\rm int}\big(\boldsymbol{\Theta}\big)\big\}.
\eeqo
We freely switch between these two parametrizations; for more background on the EDF, see
\cite{wuthrich_st_learning_book}.

We describe our statistical set-up. Each observation in the population is characterized by a vector of features $\mathbf{X}\in\mathcal{X}$, a case weight $V\in(0,\infty)$ and a response $Y\in\br$. We assume that $(Y,\mathbf{X}, V)\sim F_{(Y,\mathbf{X}, V)}$ with distribution $F$ supported on $\br\times\mathcal{X}\times (0,\infty)$. The conditional distribution of the response is assumed to satisfy $Y|\mathbf{X}=\bx, V=v \sim EDF_{v,\phi^*}(\theta^*(\bx))$, for a given member of the EDF, and there exists a regression function $f^*:\mathcal{X}\to {\rm int}\big(\boldsymbol{\Theta}\big)$ such that, respectively,
\beq\label{sufficiency_f_star}
\theta^*(\bx)=f^*(\bx)\quad \text{ and }\quad \mu^*(\bx)= \kappa'\big(\theta^*(\bx)\big)=\kappa'\big(f^*(\bx)\big),\qquad \bx\in\mathcal{X}. 
\eeq
The upper star $^*$ labels the true data generating model. Thus, the true conditional expected value of $Y|\mathbf{X}=\bx, V=v \sim EDF_{v,\phi^*}(\theta^*(\bx))$ is equal to $\mu^*(\bx)$, for all $\bx\in\mathcal{X}$. In the context of regression modeling, it is more common to assume that there exists a regression function $g^*:\mathcal{X}\to \boldsymbol{\Gamma}$ such that $\mu^*(\bx)=g^*(\bx)$, thus, $\theta^*(\bx)=f^*(\bx)=(\kappa')^{-1}\big(\mu^*(\bx)\big)$, for all $\bx\in\mathcal{X}$; the function $(\kappa')^{-1}$ is the canonical link of the selected EDF member.

\bl{We start from an i.i.d.~{\it learning sample} $\mathcal{L}=(y^\ell_j,\bx^\ell_j, v_j^\ell)_{j=1}^m$ generated from the same law as $(Y,\mathbf{X}, V)\sim F_{(Y,\mathbf{X}, V)}$. Working under the true member of the EDF, we estimate the true regression function $f^*:\mathcal{X}\to {\rm int}\big(\boldsymbol{\Theta}\big)$ by a regression function $\widehat{f}:\mathcal{X}\to {\rm int}\big(\boldsymbol{\Theta}\big)$ from the learning sample $\mathcal{L}$. We note that when estimating $f$ we have access to the entire set of risk factors denoted by $\mathbf{X}$. Our goal is to verify whether the resulting mean estimates are mean-calibrated \eqref{calibration_def} and mean-specified \eqref{specification_def}. For this, we consider an independent {\it test sample} $\mathcal{D}=(y_i,\bx_i, v_i)_{i=1}^n$, generated by i.i.d.~observations $(Y_i,\mathbf{X}_i, V_i)\sim F_{(Y,\mathbf{X}, V)}$. Given the estimated regression function $\widehat{f}$ on the learning sample $\mathcal{L}$, we estimate the canonical parameters and the mean values, respectively, on the test sample $\mathcal{D}$ by}
\begin{equation}\label{mean estimates 19}
\widehat{\theta}(\bx_i)=\widehat{f}(\bx_i),\quad \widehat{\mu}(\bx_i) = \kappa'\big(\widehat{f}(\bx_i)\big), \quad i=1,\ldots,n.
\end{equation}

\bl{Note that we assume that the true canonical parameters and their estimates lie in the interior of the effective domain. Under the assumption that the cumulant function $\kappa$ is steep, a distribution in the EDF is degenerate if and only if its canonical parameter lies on the boundary $\partial\boldsymbol{\Theta}$, see \citet[Theorem 9.2]{Barndorff} and \citet[Section 2.2.4]{wuthrich_st_learning_book}. Hence, it is natural to consider $f^*:\mathcal{X}\to {\rm int}\big(\boldsymbol{\Theta}\big)$ for the true canonical parameters and $\widehat{f}:\mathcal{X}\to {\rm int}\big(\boldsymbol{\Theta}\big)$ for their estimates. Moreover, under steepness of $\kappa$, the set $\boldsymbol{\Gamma}$ covers all valid mean values of the distribution. In practice, steepness of $\kappa$ is not restrictive at all, as all commonly used EDF members have steep cumulant functions $\kappa$.}

\bl{In the sequel, the learning sample $\mathcal{L}$ is fixed, and all subsequent considerations are conditional on $\mathcal{L}$, but for ease of notation we drop this conditioning from the notation. Consequently, the estimated regression function $\widehat{f}$ is treated as given (and fixed). We could also drop the hat in the notation of $\widehat{f}$, but we prefer to keep $\widehat{f}$ to emphasize that this function needs to be estimated in practice from a learning sample $\mathcal{L}$ in the first place. Yet, in our study, the learning sample $\mathcal{L}$ is not random, as is typically the case in mathematical statistics. However, the test sample $\mathcal{D}$ is treated as random. Consequently, the mean estimates $\big(\widehat{\mu}(\mathbf{X}_i)\big)_{i=1}^n$ in the test sample are random implied by the features $(\mathbf{X}_i)_{i=1}^n$, but fixed in terms of the hat notation.}

We make the following assumptions under which the LRT and the sub-sampled split LRT are constructed below.
\begin{itemize}
\item[(A)] \bl{The member of the EDF is fixed and known, i.e., the correct cumulant function $\kappa$ is given.}
\end{itemize}
This is a standard assumption in regression modelling, where we typically assume the distribution family and fit the regression function given the distributional assumption, e.g., with maximum likelihood estimation (MLE).
\begin{itemize}
\item[(B)] \bl{The dispersion parameter $\phi^*$ of the member of the EDF is known.}
\end{itemize}
\bl{From a theoretical perspective, assumption (B) corresponds to considering a one-parameter EDF with known dispersion. From a practical point of view, we assume that $\phi^*$ can be estimated accurately (in large samples). Below, we will also allow for misspecification in the dispersion parameter and assess the impact of assumption (B) through numerical experiments in Section \ref{sec_examples}. In Appendix \ref{sec_app_2} we remove this assumption and show that our results continue to hold when a consistent estimator of the dispersion parameter is used.}

\bl{As discussed in the introduction, tests of mean-calibration are usually performed using isotonic regression under the assumption that the estimated means correctly preserve the ordering of the corresponding population means; see, e.g., \cite{dimitriadis_diagram}, \cite{gneiting_evaluation}, \cite{Dimitriadis}, \cite{delong_iso}, \cite{delong_band}, \cite{henzi_hl}. We formalize this assumption.}
\begin{itemize}
\item[(C)] \bl{There exists a non-decreasing mapping $\mathcal{T}:\br\rightarrow\br$ such that}
\beq\label{ranking_function_weak_2}
\mathbb{E}\big[Y \mid \widehat{\mu}(\mathbf{X})\big]=\mathcal{T}(\widehat{\mu}(\mathbf{X})), \quad \text{almost surely.}
\eeq
\end{itemize}
\bl{This assumption is related to the estimation of $\mu^*$. Although our theoretical results treat $\widehat{\mu}$ as fixed, the quality of its estimation plays an important role in our analysis. Assumption (C) can be interpreted as requiring that the learning sample $\mathcal{L}$ allows us to estimate a regression function $\widehat{\mu}$ that preserves the correct ordering of the true conditional means, given the mean estimates, possibly with ties. Hence, (C) intuitively implies that the mean estimates obtained using $\widehat{\mu}$ are assumed to be reasonable up to a monotone transformation. Interpreted in an actuarial context, if we have two groups of policyholders with predicted risks $\widehat{\mu}_1$ and $\widehat{\mu}_2$ such that $\widehat{\mu}_1 > \widehat{\mu}_2$, then the first group is, on average, at least as risky as the second group. The impact of assumption (C) is investigated in our numerical experiments in Section \ref{sec_examples}. A possible verification method of (C) is discussed in Appendix \ref{sec_app_1}}.

\bl{However, in case of testing mean-calibration with likelihood-based methods, we have to strenghten assumption (C).}
\begin{itemize}
\item[(D)] \bl{There exists a strictly increasing mapping $\mathcal{S}:\br\rightarrow\br$ such that}
\beq\label{ranking_function}
\mathbb{E}\big[Y \mid \mathbf{X}\big] = \mathcal{S}(\widehat{\mu}(\mathbf{X})), \quad \text{almost  surely.}
\eeq
\end{itemize}
\bl{In other words, we assume strict comonotonicity between $\widehat{\mu}(\mathbf{X})$ and $\mu^*(\mathbf{X})=\be[Y|\mathbf{X}]$. This assumption is again related to the estimation method for $\mu^*$. It can be interpreted as requiring that the learning sample $\mathcal{L}$ enables us to estimate a regression function $\widehat{\mu}$ that preserves the ordering of the true conditional means, given the risk factors. In an actuarial context, if we have two policyholders with predicted risks $\widehat{\mu}_1$ and $\widehat{\mu}_2$ such that $\widehat{\mu}_1 > \widehat{\mu}_2$, then the first policyholder is indeed riskier than the second. We observe that (D) implies (C) since}
\beq\label{eq_cal_spec}
\mathbb{E}\big[Y \mid \widehat{\mu}\left(\mathbf{X}\right)\big]=\mathbb{E}\big[Y \mid \mu^*\left(\mathbf{X}\right)\big]=\be\big[Y \mid \mathbf{X}\big]=\mu^*\left(\mathbf{X}\right)=\mathcal{S}\big(\widehat{\mu}\left(\mathbf{X}\right)\big),\quad \text{almost surely},
\eeq
\bl{where we have used (D) for the first and last equalities, and properties of the EDF distribution for the remaining equalities.} 

\bl{It is worth noting that assumption (D) is difficult to verify, since it involves the unknown true mean values for individuals. Moreover, due to random noise and the inherent inaccuracies of estimation methods on finite samples, strict comonotonicity of $(\widehat{\mu}(\mathbf{X}),\mu^*(\mathbf{X}))$ is unlikely to hold in practice, as the estimated regression function generally exhibits local deviations from the true regression function. A theoretical justification for (D) is provided in Appendix \ref{sec_app_1}, where we argue that (D) is reasonable when our mean estimates are uniformly consistent for the true means. Consequently, we may expect (D) to hold only approximately on sufficiently large samples. Most importantly for our research, we investigate the impact of (D) and the effects of deviations from strict comonotonicity, both theoretically in Section \ref{sec_universal_split} and through numerical experiments in Section \ref{sec_examples} and a real data application in Appendix \ref{sec_app_4}. A possible method for verifying (D) is discussed in Appendix \ref{sec_app_1}.}

\bl{Assumption (D) has important implications for our statistical test of mean-calibration. Under (D), see \eqref{eq_cal_spec}, we immediately derive that}
\beqo
\mathbb{E}\big[Y \mid \widehat{\mu}\left(\mathbf{X}\right)\big] = \mathbb{E}\big[Y \mid \mathbf{X}\big], \quad \text{almost surely}.
\eeqo
\bl{Consequently, within our statistical set-up based on the EDF and under assumption (D), the property of calibration of mean estimates \eqref{calibration_def} is equivalent to the correct specification of the mean estimates \eqref{specification_def}. We elaborate in the sequel that if we want to apply the LRT and the sub-sampled split LRT to test mean-calibration, we are forced to test mean-specification, and they are equivalent under (D).}

\bl{Assumption (D) plays a different role than assumption (C). Assumption (D) allows us to present the null hypothesis of mean-calibration as the null hypothesis of mean-specification and to apply likelihood-based methods within the EDF. Assumption (C), stated directly or as a conclusion from (D), allows us to use isotonic regression to construct reasonable mean estimates under the alternative hypothesis that the mean estimates are not calibrated/not correctly specified, while still preserving the correct ordering of the mean responses in the appropriate sense.} 

\begin{rem}\normalfont
Within the EDF, there holds
$\mathbb{E}[Y]=\mathbb{E}[Y|V=v]$, this is to say that the case weight $v>0$ does not impact the expected value of $Y$ within the EDF, only the higher moments depend on $v$. For this reason, within the EDF, conditioning on the weights $V$ has no impact on the above statements; for a more general discussion of exposure weighting, see \cite{W2026}.
\end{rem}

We now formally introduce our statistical hypotheses. \bl{We consider a measurable space $(\Omega,\mathcal{F})$ and let $\mathcal{M}$ denote the set of probability measures under consideration on $(\Omega,\mathcal{F})$. This set $\mathcal{M}$ contains the distributions of the i.i.d.~observations $(Y_i,\mathbf{X}_i,V_i)_{i=1}^n$, where, for each $i=1,\ldots,n$, the conditional distribution of the response is given by}
\beqo
Y_i \mid \mathbf{X}_i=\bx_i,V_i=v_i \sim EDF_{v_i,\phi^*}\big(\Psi(\bx_i)\big).
\eeqo
\bl{Here, $EDF_{v_i,\phi^*}\big(\Psi(\bx_i)\big)$ denotes the member of the EDF with the assumed cumulant function $\kappa$, dispersion parameter $\phi^*$, case weight $v_i$, and canonical parameter $\theta(\bx_i)=\Psi(\bx_i)$, where $\Psi:\mathcal{X}\to{\rm int}\big(\boldsymbol{\Theta}\big)$ is a measurable function determining the canonical parameter from the vector of risk factors $\bx_i$.} Let $\widehat{\mu}: \mathcal{X}\to \boldsymbol{\Gamma}$ denote the estimated regression function. First, using the definition of mean-calibration \eqref{calibration_def}, we would like to verify the null hypothesis
\beq\label{h_0_mu}
\mathcal{H}_0&=&\Big\{\mathbb{P}\in\mathcal{M}; \quad \be_\mathbb{P}\big[Y_i \mid\widehat{\mu}\big(\mathbf{X}_i\big)\big]=\widehat{\mu}\big(\mathbf{X}_i\big), 
~\bl{\bp\text{-almost surely}}, \ \text{for all} \ i=1,\ldots,n\Big\},
\qquad
\eeq
against the alternative
\beq\label{h_1_mu}
\mathcal{H}_1&=&\Big\{\mathbb{Q}\in\mathcal{M}; \quad \be_\mathbb{Q}\big[Y_i \mid \widehat{\mu}\big(\mathbf{X}_i\big)\big]\neq\widehat{\mu}\big(\mathbf{X}_i\big), \nonumber\\ 
&& \quad \bl{\text{with positive probability} \ \bq}, \ \text{for at least one} \ i=1,\ldots,n\Big\}.
\eeq
Adapting to the notation of \citet[Chapter 3.1]{wang_book}, we introduce $\mathcal{P}$ and $\mathcal{Q}$ to denote the sets of probability measures from $\mathcal{M}$ which satisfy the null and the alternative hypothesis, respectively. Hence, the null (alternative) hypothesis is interpreted both as the statistical statement $\mathcal{H}_0$ ($\mathcal{H}_1$) and as the set of probability measures $\mathcal{P}$ ($\mathcal{Q}$). An element of $\mathcal{P}$ is denoted by $\bp$, and an element of $\mathcal{Q}$ by $\bq$.

We use the following notation in the sequel. We focus on the test sample $\mathcal{D}$ given by the i.i.d.~observations $(Y_i,\mathbf{X}_i, \widehat{\Theta}_i, V_i)_{i=1}^n$, where $(\widehat{\Theta}_i)_{i=1}^n$ denote the estimates of the canonical parameters using the transformations $\widehat{\Theta}_i = (\kappa ')^{-1}\big(\widehat{\mu}(\mathbf{X}_i))$ under the given regression function $\widehat{\mu}$. Under (D), we can now reformulate the null hypothesis of mean-calibration as
\beq\label{h_0_edf}
\mathcal{H}_0&=&\Big\{\mathbb{P}\in\mathcal{M}; \quad Y_i \mid \mathbf{X}_i, V_i \ \overset{\mathbb{P}}{\sim} \ EDF_{V_i,\phi^*}(\widehat{\Theta}_i), 
~\bl{\bp\text{-almost surely}}, \ \text{for all} \ i=1,\ldots,n\Big\},
\qquad
\eeq
and the alternative hypothesis by
\beq\label{h_1_edf}
\mathcal{H}_1&=&\Big\{\mathbb{Q}\in\mathcal{M}; \quad Y_i \mid \mathbf{X}_i, V_i \ \overset{\mathbb{Q}}{\not\sim} \ EDF_{V_i,\phi^*}(\widehat{\Theta}_i), 
\nonumber\\
&& 
\quad \bl{\text{with positive probability} \ \bq}, \ \text{for at least one} \ i=1,\ldots,n\Big\}.
\eeq
\bl{Thus, this results in the null hypothesis of mean-specification \eqref{specification_def}. Alternatively, we can start directly with the null hypothesis of mean-specification \eqref{specification_def}.}

\bl{Summing up, in this paper we focus on the null hypothesis \eqref{h_0_edf} against the alternative hypothesis \eqref{h_1_edf}, which can be interpreted as a statistical test of mean-calibration or mean-specification within a known EDF class under a known dispersion parameter $\phi^*$. If we are interested in testing mean-calibration, we assume (A), (B), (D), and (D) implies (C). If we are interested in testing mean-specification, we assume (A), (B), (C).}

\section{Classical inference for mean estimates}\label{sec_classic}

We begin by considering classical statistical approaches to testing the null hypothesis \eqref{h_0_edf} against the alternative hypothesis \eqref{h_1_edf}. These methods serve as benchmarks for universal inference presented in Section \ref{sec_universal}, below.

\subsection{Isotonic regression and recalibration of mean estimates}\label{sec_iso_intro}

\bl{We start by briefly recalling isotonic regression. Let us consider a response $Y$ and a one-dimensional real-valued feature $Z$. Assume that the true mean functional $z\mapsto\be[Y|Z=z]$ is non-decreasing. Based on data points
$(y_i,z_i, v_i)_{i=1}^n$ with positive case weights $v_i>0$, we aim at fitting a non-parametric regression function to $z\mapsto\be[Y|Z=z]$.} This motivates the following isotonic estimation 
\beq\label{iso_optimization}
  \widehat{\boldsymbol{\gamma}} &=&\underset{\boldsymbol{\gamma}=(\gamma_1,\ldots, \gamma_n)^\top \in \R^n}{\arg\min}~
                     \sum_{i=1}^n v_i \left(y_i - \gamma_i\right)^2, \\ 
                 && \text{subject to \, $z_j \le z_k \Longrightarrow \gamma_j \le \gamma_k$, \, for all $1\le j,k \le n$.}
                    \nonumber
\eeq
The solution $(\widehat{\gamma}_i)_{i=1}^n$ to \eqref{iso_optimization} preserves the ranks of $(z_i)_{i=1}^n$. This isotonic estimation is solved with the pool adjacent violators (PAV) algorithm; see \cite{Leeuw}. Typically, the solution $  \widehat{\boldsymbol{\gamma}}\in \R^n$ is interpolated by a step function.

\bl{Note that in \eqref{iso_optimization}, we can replace the square loss function by any EDF-deviance loss function yielding the same solution $\widehat{\boldsymbol{\gamma}}$, that is, the isotonic regression solution is invariant under different deviance loss choices; see \cite{banerjee} and \cite{banerjee_2}.}

\bl{Under (C), it is justified to use an isotonic regression to obtain a non-parametric estimate of the conditional mean $\be[Y|\widehat{\mu}(\mathbf{X}_i)]$, based on the ranking generated by $(\widehat{\mu}(\mathbf{X}_i))_{i=1}^n$.} This estimation step is called {\it recalibration step}; see, e.g., \cite{WZiegel}. For the purpose of our applications below, we shall apply the recalibration step to all observations in the test set $\mathcal{D}$. 

\noindent\hrulefill

{\sc Isotonic recalibration of mean estimates}

\noindent\hrulefill

\begin{itemize}
\item Consider the sample $\big(Y_i,\widehat{\mu}(\mathbf{X}_i), V_i\big)_{i\in\mathcal{D}}$ with case weights $V_i>0$.
\item Compute the isotonic regression $m\mapsto \widehat{\mu}^{\rm iso}_\mathcal{D}(m)$ based on the sample $\big(Y_i,\widehat{\mu}(\mathbf{X}_i), V_i\big)_{i\in\mathcal{D}}$ with ranking $\big(\widehat{\mu}(\mathbf{X}_i)\big)_{i\in\mathcal{D}}$ (we interpolate by a step function).
\item Define the recalibrated mean estimates
\beqo
\widehat{\mu}^{\rm rc}\big(\mathbf{X}_i)=\widehat{\mu}^{\rm iso}_\mathcal{D}\Big(\widehat{\mu}\big(\mathbf{X}_i\big)\Big),\quad i\in\mathcal{D},
\eeqo
as estimates of $\be[Y_i|\widehat{\mu}(\mathbf{X}_i)]$.
\end{itemize}

\cite{dimitriadis_diagram} and \cite{gneiting_evaluation} have popularized the so-called CORP reliability diagram (Consistent, Optimally binned, Reproducible, Pool adjacent violators algorithm) to assess mean-calibration. The CORP reliability diagram shows the two-dimensional plot of $(\widehat{\mu}\big(\mathbf{X}_i\big), \widehat{\mu}^{\rm rc}\big(\mathbf{X}_i))_{i=1}^n$, which should roughly lie on the diagonal under mean-calibration. In the next section, we show that the deviations from the diagonal observed in the CORP reliability diagram are the building blocks of the classical LRT for mean-calibration within the EDF.

\subsection{Classical LRT for mean-calibration and mean-specification within the EDF}\label{sec_classic_lrt}

\bl{Within the statistical set-up of Section \ref{sec_model}, we can write the likelihood of the observations $(y_i,\bx_i,v_i)_{i=1}^n$ of the test sample as}
\begin{equation}\label{definition likelihood data}
\mathcal{L}_{(y_i,\bx_i,v_i)_{i=1}^n}\big((\Psi(\bx_i))_{i=1}^n\big)=\prod _{i=1}^n\exp\left(\frac{y_i\Psi(\bx_i)-\kappa(\Psi(\bx_i))}{\phi^*/v_i}+a(y_i,v_i/\phi^*)\right)w(\mathbf{X}_i=\bx_i,V_i=v_i),
\end{equation}
\bl{where use the definition of the probability measures of $\mathcal{M}$ on $(\Omega, \mathcal{F})$, and the joint distribution of $(\mathbf{X},V)$ is denoted by $w$. We have to construct the MLEs under the null hypotheses of mean-calibration \eqref{h_0_mu} and mean-specification \eqref{h_0_edf}, as well as under the corresponding alternative hypotheses for testing.} 

\bl{Under the null hypothesis of mean-specification \eqref{h_0_edf}, the MLE of $\big(\Psi(\mathbf{X}_i)\big)_{i=1}^n$ is obtained by setting $\widehat{\Psi}(\mathbf{X}_i)=\widehat{\Theta}(\mathbf{X}_i)=(\kappa')^{-1}\big(\widehat{\mu}(\mathbf{X}_i)\big)$. This conclusion is immediate when testing either mean-specification or mean-calibration under (D), since these two hypotheses are equivalent under (D). However, if mean-calibration is tested without assumption (D), but only under (C), this conclusion about the MLE of $\big(\Psi(\mathbf{X}_i)\big)_{i=1}^n$ under the null hypothesis of mean-calibration \eqref{h_0_mu} no longer holds, see also the points below. This yielded a motivation for introducing assumption (D). In our framework, (D) is needed to uniquely define the MLE under the null hypothesis of mean-calibration in terms of the estimated regression function $\widehat{\mu}$. This is also essential for the application of universal inference in the next section. Thus, assumption (D) cannot be weakened in our framework. The MLE under the alternative hypothesis also deserves attention. If we test mean-calibration or mean-specification both under (D), then $\be^\bq[Y|\mathbf{X}]=\be^\bq[Y|\widehat{\mu}(\mathbf{X})]=\mathcal{S}(\widehat{\mu}(\mathbf{X}))$ for some strictly increasing function $\mathcal{S}$. By the results from Section \ref{sec_iso_intro}, see \cite{banerjee} and \cite{banerjee_2}, if we assume (D), under the alternative hypothesis of mean-calibration \eqref{h_1_mu} or mean-specification \eqref{h_1_edf}, the MLE of $\big(\Psi(\mathbf{X}_i)\big)_{i=1}^n$ is obtained through the recalibration procedure described in Section \ref{sec_iso_intro}, namely, by fitting an isotonic regression to $\big(Y_i,\widehat{\mu}(\mathbf{X}_i),V_i\big)_{i=1}^n$. However, if mean-specification is tested under (C) alone, the recalibration procedure from Section \ref{sec_iso_intro} yields an estimate of $\big(\be[Y|\widehat{\mu}(\mathbf{X}_i)]\big)_{i=1}^n$, rather than the MLE of $\big(\be[Y|\mathbf{X}_i]\big)_{i=1}^n$ under the alternative hypothesis of mean-specification \eqref{h_1_edf}, unless assumption (D) is imposed. Fortunately, as discussed below, this does not cause any problems for the universal inference procedure.}

\bl{Let us assume that (A), (B), (D) hold for both cases of testing mean-calibration and mean-specification. The LRT to validate the null hypothesis \eqref{h_0_edf} against the alternative \eqref{h_1_edf} for the observations in the test set $\mathcal{D}$ then takes the following form:}

\noindent\hrulefill

{\sc Classical LRT for mean-calibration/mean-specification within the EDF}

\noindent\hrulefill

\begin{itemize}
\item Consider the sample $\big(Y_i,\kappa'(\widehat{\Theta}_i), V_i\big)_{i\in\mathcal{D}}$ with case weights $V_i>0$.
\item Estimate the isotonic regression $m\mapsto \widehat{\mu}^{\rm iso}_{\mathcal{D}}(m)$ based on $(Y_i,\kappa'(\widehat{\Theta}_i), V_i)_{i\in \mathcal{D}}$ using the ranks $(\kappa'(\widehat{\Theta}_i))_{i\in \mathcal{D}}$.
\item Estimate the canonical parameter $\Xi_i$ for all observations $i\in\mathcal{D}$ by
\beqo
\widehat{\Xi}_i = (\kappa')^{-1}\left(\widehat{\mu}^{\rm iso}_{\mathcal{D}}\big(\kappa'(\widehat{\Theta}_i)\big)\right),\quad i\in\mathcal{D}.
\eeqo
\item Define the Likelihood Ratio Statistics (LRS) for testing $\mathcal{H}_0$ against $\mathcal{H}_1$
\beq\label{def_lrt}
E^{LRT}= \frac{\mathcal{L}_{\mathcal{D}}\big(\widehat{\Xi}\big)}{\mathcal{L}_{\mathcal{D}}\big(\widehat{\Theta}\big)}
=\prod_{i\in\mathcal{D}} E_{\widehat{\Xi}_i}\big(Y_i,\widehat{\Theta}_i, V_i\big)= \prod_{i\in\mathcal{D}}\exp\left(\frac{Y_i\,(\widehat{\Xi}_i-\widehat{\Theta}_i)-(\kappa(\widehat{\Xi}_i)-\kappa(\widehat{\Theta}_i))}{\phi^*/V_i}\right),
\eeq
where $(\widehat{\Xi}_i)_{i=1}^n$ and $(\widehat{\Theta}_i)_{i=1}^n$ are the estimates of the canonical parameters constructed under $\mathcal{H}_1$ and $\mathcal{H}_0$, respectively, and $\mathcal{L}_\mathcal{D}(\cdot)$ denotes the likelihoods of the observations in $\mathcal{D}$ under the two estimates of the canonical parameters.
\item Reject the null hypothesis $\mathcal{H}_0$ at significance level $\alpha \in (0,1)$ if
\beq\label{def_lrt_rejection}
E^{LRT}>F^{-1}_{E^{LRT}|\mathcal{H}_0}(1-\alpha).
\eeq
\end{itemize}

\noindent\hrulefill

\medskip

\bl{In order to define the LRS \eqref{def_lrt} and have non-degenerate distributions, we need one more assumption:}
\begin{itemize}
\item \bl{(E) The recalibrated mean estimates $\big(\widehat{\mu}^{\rm iso}_{\mathcal{D}}\big(\kappa'(\widehat{\Theta}_i)\big)\big)_{i=1}^n$ are in the domain of the mean parameter space $\boldsymbol{\Gamma}=\kappa'\big({\rm int}\big(\boldsymbol{\Theta}\big)\big)$, or equivalently, the recalibrated canonical parameters $\big(\widehat{\Xi}_i\big)_{i=1}^n$ are in the interior of the effective domain ${\rm int}\big(\boldsymbol{\Theta}\big)$.}
\end{itemize}
\bl{In practice, under steepness of $\kappa$, assumption (E) may force us to modify the isotonic estimates at the observation boundaries. E.g., in the Poisson case in low frequency problems, we often get recalibrated mean estimates equal to zero for very low original mean estimates, and a simple modification is to merge the two smallest classes to preserve calibration.}

The most challenging aspect of the classical LRT is to determine the quantiles of the test statistic $E^{LRT}$ under the null hypothesis, which are needed to determine the critical value of the test in \eqref{def_lrt_rejection}. We highlight two closely related papers by \cite{banerjee} and \cite{banerjee_2}. This author assumes that $Y|Z=z$ belongs to a one-parameter EDF with known dispersion parameter, as we do under (A)-(B), and that the true mean function $z\mapsto\be[Y | Z=z]$ is non-decreasing. Then, they consider a statistical test of the null hypothesis $\be[Y | Z=z_0]=\mu_0$ against the alternative hypothesis $\be[Y | Z=z_0]\neq\mu_0$. It is shown that under the null hypothesis, the LRS has an asymptotic distribution that differs from the $\chi^2$-distribution, and instead it can be characterized as a functional of Brownian motion. However, this characterization is difficult to use in practice. Therefore, a more feasible approach to testing the null hypothesis is desirable, both in their setting and in ours.

\begin{rem}\normalfont
\begin{itemize}
\item \bl{Testing mean-calibration without (D), only under (C), the MLE of the canonical parameters $\big(\Psi(\mathbf{X}_i)\big)_{i=1}^n$ is not uniquely determined under the null hypothesis \eqref{h_0_mu}, as there are many possible estimates of $\big(\be[Y|\mathbf{X}_i]\big)_{i=1}^n$ that produce $\be[Y | \widehat{\mu}(\mathbf{X})] = \widehat{\mu}(\mathbf{X})$.}
\item \bl{Testing mean-specification only under (C), the MLE of the canonical parameters $\big(\Psi(\mathbf{X}_i)\big)_{i=1}^n$ is not uniquely determined under the alternative hypothesis \eqref{h_1_edf}, as there are many possible estimates of $\big(\be[Y|\mathbf{X}_i]\big)_{i=1}^n$ that produce $\be[Y | \widehat{\mu}(\mathbf{X})] = \mathcal{T}\big(\widehat{\mu}(\mathbf{X})\big)$ for a non-decreasing $\mathcal{T}$.}
\item \bl{If we test mean-specification without (C)-(D), then the MLE of $\big(\Psi(\mathbf{X}_i)\big)_{i=1}^n$ under the alternative hypothesis \eqref{h_1_edf} leads to a saturated model with $\widehat{\Psi}(\mathbf{X}_i)=(\kappa')^{-1}\big(Y_i\big)$.}
\item \bl{Since we do not assume any asymptotic distribution for the test statistics in \eqref{def_lrt}, we instead have to rely on simulations to obtain their critical values. We can use the statistics to test mean-specification under assumptions (A), (B) and (C), however, the resulting test would not correspond to the classical LRS.}
\end{itemize}
\end{rem}

\subsection{Murphy's score decomposition and miscalibration statistics}\label{sec_classic_murphy}

It is well-established in statistical learning that the predictive accuracy of models should be evaluated with strictly consistent loss functions; see \cite{GneitingRaftery}. Let $L$ denote \bl{the unit deviance loss function of the selected EDF} (which is a strictly consistent loss function for mean estimation); see \cite{wuthrich_st_learning_book}. The predictive accuracy of the mean estimates $(\widehat{\mu}(\mathbf{X}_i))_{i=1}^n$ on the test set $\mathcal{D}$ is evaluated by the score (smaller means better)
\begin{equation}\label{Murphy score 1}
S(\mathbf{Y}, \widehat{\boldsymbol{\mu}}, \mathbf{V}) = \frac{1}{\sum_{i=1}^n V_i} \,\sum_{i=1}^n V_i \, L(Y_i, \widehat{\mu}(\mathbf{X}_i)),
\end{equation}
with responses $\mathbf{Y}=(Y_1,\ldots, Y_n)^\top$, mean estimates $\widehat{\boldsymbol{\mu}}=\big(\widehat{\mu}(\mathbf{X}_1),\ldots,\widehat{\mu}(\mathbf{X}_n)\big)^\top$ and case weights
$\mathbf{V}=(V_1,\ldots, V_n)^\top$ from the test set $\mathcal{D}$.
Following \cite{murphy_decomp}, \cite{pohle_decomp}, \cite{dimitriadis_diagram} and \cite{gneiting_evaluation}, we decompose the score into the uncertainty (UNC), discrimination (DSC) and miscalibration (MCB) statistics
\begin{equation}\label{Murphy score 2}
S(\mathbf{Y}, \widehat{\boldsymbol{\mu}}, \mathbf{V})
={\rm UNC}(\mathbf{Y},  \mathbf{V})
-{\rm DSC}(\mathbf{Y}, \widehat{\boldsymbol{\mu}}, \mathbf{V})+{\rm MCB}(\mathbf{Y}, \widehat{\boldsymbol{\mu}}, \mathbf{V}),
\end{equation}
where we set
\begin{eqnarray*}
  {\rm UNC}(\mathbf{Y},  \mathbf{V})
&=& S(\mathbf{Y}, \bar{\boldsymbol{\mu}}, \mathbf{V}),  
\\
{\rm DSC}(\mathbf{Y}, \widehat{\boldsymbol{\mu}}, \mathbf{V})
&=& S(\mathbf{Y}, \bar{\boldsymbol{\mu}}, \mathbf{V})
-S(\mathbf{Y}, \widehat{\boldsymbol{\mu}}^{\rm rc}, \mathbf{V})~\ge ~0,\\
{\rm MCB}(\mathbf{Y}, \widehat{\boldsymbol{\mu}}, \mathbf{V})
&=& S(\mathbf{Y}, \widehat{\boldsymbol{\mu}}, \mathbf{V})
-S(\mathbf{Y}, \widehat{\boldsymbol{\mu}}^{\rm rc}, \mathbf{V})~\ge ~0,
\end{eqnarray*}
with $\bar{\boldsymbol{\mu}}=(\bar{\mu}, \ldots, \bar{\mu})^\top \in\mathbb{R}^n$ having weighted mean entries $\bar{\mu}=\sum_{i=1}^n V_iY_i/\sum_{i=1}^n V_i$ not considering any features, and $\widehat{\boldsymbol{\mu}}^{\rm rc}=(\widehat{\mu}^{\rm rc}(\mathbf{X}_1),\ldots,\widehat{\mu}^{\rm rc}(\mathbf{X}_n))^\top$ is the recalibrated version of $\widehat{\boldsymbol{\mu}}=(\widehat{\mu}(\mathbf{X}_1),\ldots,\widehat{\mu}(\mathbf{X}_n))^\top$. Moreover, ${\rm MCB}=0$ if and only if $\widehat{\boldsymbol{\mu}}^{\rm rc}\equiv\widehat{\boldsymbol{\mu}}$, and ${\rm DSC}=0$ if and only if $\widehat{\boldsymbol{\mu}}^{\rm rc}\equiv\bar{\boldsymbol{\mu}}$;  this follows from the strict consistency of $L$. We observe that the miscalibration statistics MCB expresses deviations of the CORP reliability diagram from the diagonal in terms of the score under consideration. Hence, there is a close link to the techniques presented in Section \ref{sec_iso_intro}.

In our statistical set-up within the EDF, the miscalibration statistics for the mean estimates $(\widehat{\mu}(\mathbf{X}_i))_{i=1}^n$ takes the form
\beq\label{def_mcb}
\bl{{\rm MCB}(\mathbf{Y}, \widehat{\boldsymbol{\mu}}, \mathbf{V}) = \frac{2}{\sum_{i=1}^n V_i}\,\sum_{i=1}^n\left(Y_i\,(\widehat{\Xi}_i-\widehat{\Theta}_i)-(\kappa(\widehat{\Xi}_i)-\kappa(\widehat{\Theta}_i))\right)V_i,}
\eeq 
where $(\widehat{\Theta}_i)_{i=1}^n$ are the predictions of the canonical parameters derived from $(\widehat{\mu}(\mathbf{X}_i))_{i=1}^n$, and $(\widehat{\Xi}_i)_{i=1}^n$ denote the canonical parameters derived from the recalibrated mean estimates, see Sections \ref{sec_iso_intro}-\ref{sec_classic_lrt}. We conclude that the miscalibration statistics from Murphy's score decomposition is related to the LRS \eqref{def_lrt} by the simple transformation
\beq\label{relation_lrt_mcb}
\bl{{\rm MCB}(\mathbf{Y}, \widehat{\boldsymbol{\mu}}, \mathbf{V}) = \frac{\phi^*}{\sum_{i=1}^n V_i}2\log\left(\frac{\mathcal{L}_{\mathcal{D}}\big(\widehat{\Xi}\big)}{\mathcal{L}_{\mathcal{D}}\big(\widehat{\Theta}\big)}\right).}
\eeq
\bl{Henceforth, we can equivalently use the LRS and the MCB to test calibration of the mean estimates within the EDF, i.e., under the corresponding deviance loss choice and appropriate assumptions. This provides further support for using the LRT to test mean-calibration within the EDF. }

\section{Universal inference for mean estimates}\label{sec_universal}

In this section, we develop our new approach to assess calibration and specification of mean estimates based on universal inference and the {\it split LRT}. This approach uses e-values and e-testing, and it is crucially based on the work of \cite{wang_book}.

\medskip

\subsection{Split Likelihood Ratio Test}
\label{sec_universal_intro}

We first introduce basic tools of universal inference; for a comprehensive mathematical introduction, we refer to \cite{wang_book}. In this section we work with classical parametric statistical models. Consider i.i.d.~observations $\mathbf{Y}=(Y_1,\ldots ,Y_n)$ from a distribution $\bp_{\theta^*}$ that belongs to a family of distributions $ \{\bp_\theta;\, \theta\in\boldsymbol{\Theta}\}$.
A general aim is to test
\beqo
\mathcal{H}_0:\theta^*\in\boldsymbol{\Theta}_0 \qquad \text{ against} \qquad \mathcal{H}_1:\theta^*\in
\boldsymbol{\Theta}_1=\boldsymbol{\Theta}\setminus\boldsymbol{\Theta}_0,
\eeqo
for a non-empty null hypothesis set $ \boldsymbol{\Theta}_0\subsetneq\boldsymbol{\Theta}$.
Consider the LRS,  and assume that under the null hypothesis $\mathcal{H}_0$ the LRS satisfies the asymptotic behavior
\beqo
2\log\left(\frac{\mathcal{L}_{\mathbf{Y}}\big(\widehat{\theta}_1\big)}{\mathcal{L}_{\mathbf{Y}}\big(\widehat{\theta}_0\big)}\right)\sim \chi^2\big({\rm dim}(\boldsymbol{\Theta}_1)-{\rm dim}(\boldsymbol{\Theta}_0)\big),\qquad \text{ as $n\to\infty$,}
\eeqo
where $\widehat{\theta}_1$ and $\widehat{\theta}_0$ are the MLEs of the parameters under $\mathcal{H}_1$ and $\mathcal{H}_0$, respectively, and $\mathcal{L}_\mathbf{Y}(\cdot)$ denotes the data likelihoods of $\mathbf{Y}$ under the given parameter. Such a LRS limit relies on large sample asymptotics requiring regularity conditions. These conditions are likely not fulfilled in many practical applications. \cite{wasserman_universal} have introduced universal inference which yields tests and confidence sets for arbitrary statistical models with finite-sample guarantees and without restrictive regularity conditions. In particular, \cite{wasserman_universal} propose the use of the so-called split LRS.

\noindent\hrulefill

{\sc Split Likelihood Ratio Test}

\noindent\hrulefill

\begin{itemize}
    \item Partition the test sample $\mathbf{Y}=(Y_1,\ldots, Y_n)$ at random into two disjoint, independent and non-empty sub-samples $\mathbf{Y}_0$ and $\mathbf{Y}_1$.
    \item Let $\widehat{\theta}_1$ denote an estimator of the model parameter $\theta^*\in\boldsymbol{\Theta}_1$ under ${\cal H}_1$ based on sub-sample $\mathbf{Y}_1$ (e.g., the MLE constructed under $\mathcal{H}_1$).
    \item Let $\widehat{\theta}_0$ denote the MLE constructed under $\mathcal{H}_0$ based on $\mathbf{Y}_0$.
    \item Calculate the split LRS for testing $\mathcal{H}_0$ against $\mathcal{H}_1$
    \beq\label{def_split_lrt_0}
    E^{sLRT} = \frac{\mathcal{L}_{\mathbf{Y}_0}\big(\widehat{\theta}_1\big)}{\mathcal{L}_{\mathbf{Y}_0}\big(\widehat{\theta}_0\big)}.
    \eeq
    \item Reject the null hypothesis $\mathcal{H}_0$ at a pre-assumed significance level $\alpha \in (0,1)$ if 
    \beq\label{def_split_lrt_rejection_0}
    E^{sLRT} \geq 1/\alpha.
    \eeq
\end{itemize}

\noindent\hrulefill

\medskip

The key difference between the classical LRS \eqref{def_lrt} and the split LRS \eqref{def_split_lrt_0} is that the former uses an in-sample likelihood ratio whereas the latter uses an out-of-sample likelihood ratio. We emphasize that when implementing the split LRT, a part of the sample $\mathbf{Y}$ is used for estimating the parameter of a model under the alternative $\mathcal{H}_1$ (training a model) and the remaining part is used for validating the null hypothesis $\mathcal{H}_0$ about the parameter of a model (testing a model). Thus, the test sample $\mathbf{Y}$ is split into a {\it training set} $\mathbf{Y}_1$ and a {\it validation set} $\mathbf{Y}_0$ for hypothesis testing.

The main advantage of the universal approach developed by \cite{wasserman_universal} is that it considers a universal critical value at the selected significance level $\alpha$ for the statistics $E^{sLRT}$ given by $1/\alpha$; we also refer to \citet[Section 5.2]{wang_book}. This is justified by the following theorem.

\begin{thm}\label{thm_universal}
The split LRT \eqref{def_split_lrt_rejection_0} controls the type I error of rejecting $\mathcal{H}_0$ in favor of $\mathcal{H}_1$ at significance level $\alpha \in (0,1)$, i.e., 
\[
\sup_{\theta^*\in\boldsymbol{\Theta}_0}\bp_{\theta^*}\left(E^{sLRT} \geq 1/\alpha\right) \leq \alpha.
\]
\end{thm}
\noindent
Generally, one can expect that the type I error is much lower than $\alpha$ if the universal critical value $1/\alpha$ is used for rejecting the null hypothesis. 

\medskip

The proof of Theorem \ref{thm_universal} is straightforward. It is based on Markov's inequality and the property that the expected value of the split LRS \eqref{def_split_lrt_0} is less or equal to one under the null hypothesis ${\cal H}_0$; this comes from the fact that
$\mathcal{L}_{\mathbf{Y}_0}({\theta}^*)/\mathcal{L}_{\mathbf{Y}_0}(\widehat{\theta}_0)\le 1$ for all $\theta^* \in \boldsymbol{\Theta}_0$, since $\widehat{\theta}_0$ is the MLE over $\boldsymbol{\Theta}_0$, for observations $\mathbf{Y}_0$. In other words, the split LRS \eqref{def_split_lrt_0} is an {\it e-variable} under the null hypothesis ${\cal H}_0$. 

For an e-variable, we recall Definition 1.2 of \cite{wang_book}. In the definition below and in the sequel, we now interpret the null hypothesis $\mathcal{H}_0$ and the alternative $\mathcal{H}_1$ as sets of probability measures $\mathcal{P}$ and $\mathcal{Q}$, see also Section \ref{sec_model}. 

\begin{df}
Let $\mathcal{P}$ denote the set of probability measures representing the null hypothesis $\mathcal{H}_0$. 
\begin{itemize}
\item An e-variable $E$ for $\mathcal{P}$ is an $[0,\infty]$-valued random variable satisfying $\be_\bp[E]\leq 1$ for all $\bp\in\mathcal{P}$.
\item An e-variable $E$ is exact if $\be_\bp[E]=1$ for all $\bp\in\mathcal{P}$.
\end{itemize}
\end{df}

The split LRS involves statistical randomness (irreducible risk), due to the randomness in the observations $\mathbf{Y}$, and it involves algorithmic randomness, due to the partition of the data $\mathbf{Y}$ into the two sub-samples $\mathbf{Y}_0$ and $\mathbf{Y}_1$. In the classical LRS, only the statistical randomness is present. This motivates one to try to reduce the algorithmic randomness in the split LRS by (independently) repeating the partition of $\mathbf{Y}$ and averaging the resulting split LRSs. This suggests to consider a repeated independent sampling of $\mathbf{Y}_0$ and $\mathbf{Y}_1$. Choose $B\geq 1$. We define the {\it sub-sampled split LRS}
\beq\label{def_sub_split_lrt_0}
\overline{E}^{sLRT}_B = \frac{1}{B}\sum_{b=1}^BE_b^{sLRT},
\eeq
where $E_b^{sLRT}$ denotes the split LRS of the $b$-th (independent) random partition of $\mathbf{Y}$. The test statistics $\overline{E}^{sLRT}_B$ is an e-variable, if $E_b^{sLRT}$ are e-variables, and one immediately concludes with the following statement from Theorem \ref{thm_universal}.

\begin{cor}
Let $\mathcal{P}$ denote the set of probability measures representing the null hypothesis $\mathcal{H}_0$. Reject the null hypothesis $\mathcal{H}_0$ at a pre-assumed significance level $\alpha \in (0,1)$ if $\overline{E}^{sLRT}_B \geq 1/\alpha$. This test controls the type I error of rejecting $\mathcal{H}_0$ in favor of $\mathcal{H}_1$ at  significance level $\alpha \in (0,1)$, i.e.,
\[
\sup_{\bp\in\mathcal{P}}\bp\left(\overline{E}^{sLRT}_B \geq 1/\alpha\right) \leq \alpha.
\] 
\end{cor}

In practical applications, we are interested in the power of the statistical tests. Traditionally, we could estimate the probability that the null hypothesis is rejected under the true alternative, as we would do in classical inference. However, in the context of e-variables, one rather relies on \emph{e-power} as the suitable measure in universal inference; see \citet[Chapter 3]{wang_book}.

\begin{df}\label{df_power}
Let $\mathcal{Q}$ denote the set of probability measures representing the alternative $\mathcal{H}_1$.
\begin{itemize}
\item The power of a test based on an e-variable $E$ against a simple alternative $\bq\in\mathcal{Q}$ is $\be_\bq[E]$. An e-variable $E$ is powered against $\mathcal{Q}$ if $\be_\bq[E]>1$ for all $\bq\in\mathcal{Q}$. 
\item The e-power of an e-variable $E$ against a simple alternative $\bq\in\mathcal{Q}$ is $\be_\bq[\log(E)]$, assuming the expectation exists. An e-variable $E$ has a positive e-power against $\mathcal{Q}$ if $\be_\bq[\log(E)]>0$ for all $\bq\in\mathcal{Q}$. 
\end{itemize}
\end{df}

Jensen's inequality implies that $\be_\bq[\log(E)]>0$ yields $\be_\bq[E]>1$. The intuition behind the above definitions is that the test statistics should be large under the true alternative to correctly reject a false null hypothesis. As discussed in \citet[Section 3.8]{wang_book}, the larger the e-power of a test statistic, the better the test is considered to be. Thus, the e-power of a test statistic can be used as a comparative measure of test performance. The link between the e-power and the power of our statistical test of mean-calibration/mean-specification is discussed in the next section; we also refer to \citet[Proposition 5.3]{wang_book} for a general discussion. 

\subsection{Sub-sampled split LRT for \bl{mean-calibration and mean-specification} within the EDF}\label{sec_universal_split}

The previous section on the sub-sampled split LRS was rather generic. We translate this to evaluate mean-calibration and mean-specification within the EDF. \bl{The results of \cite{wasserman_universal} motivate the following statistical test of the null hypothesis \eqref{h_0_edf}, $\mathcal{H}_0$, against the alternative hypothesis \eqref{h_1_edf}, $\mathcal{H}_1$. We also recall that our statistical test can be interpreted either as a test of mean-calibration or of mean-specification. As discussed in Section \ref{sec_model}, we assume condition (D) for mean-calbration and (C) for mean-specification.}

\noindent\hrulefill

{\sc Sub-sampled split LRT for mean-calibration/mean-specification within the EDF}

\noindent\hrulefill

\begin{itemize}
\item Consider the sample $(Y_i,\kappa'(\widehat{\Theta}_i), V_i)_{i\in \mathcal{D}}$ with case weights $V_i>0$.
\item Choose a fixed split ratio $s\in(0,1)$.
\item Select a sub-sample $\mathcal{D}_0$ by randomly choosing $[ns]$ tuples from $(Y_i,\widehat{\Theta}_i, V_i)_{i\in\mathcal{D}}$ without replacement, and set $\mathcal{D}_1 = \{1,\ldots,n\}\setminus \mathcal{D}_0$.
\item Fit an isotonic regression $m\mapsto \widehat{\mu}^{\rm iso}_{\mathcal{D}_1}(m)$ on $(Y_i,\kappa'(\widehat{\Theta}_i),V_i)_{i\in \mathcal{D}_1}$ with ranks $(\kappa'(\widehat{\Theta}_i))_{i\in \mathcal{D}_1}$. 
\item Estimate $\Xi_i$ for all instances $i\in\mathcal{D}_0$ by
\begin{equation}\label{isotonic regression under H1}
\widehat{\Xi}_i = (\kappa')^{-1}\Big(\widehat{\mu}^{\rm iso}_{\mathcal{D}_{1}}\big(\kappa'(\widehat{\Theta}_i)\big)\Big).
\end{equation}
\item Define the split LRS for testing $\mathcal{H}_0$ against $\mathcal{H}_1$
\begin{equation}\label{def_split_lrt}
E^{sLRT}=\frac{\mathcal{L}_{\mathcal{D}_0}\big(\widehat{\Xi}\big)}{\mathcal{L}_{\mathcal{D}_0}\big(\widehat{\Theta}\big)}=\prod_{i\in\mathcal{D}_0} E_{\widehat{\Xi}_i}\big(Y_i,\widehat{\Theta}_i, V_i\big)
= \prod_{i\in\mathcal{D}_0}\exp\left(\frac{Y_i\,(\widehat{\Xi}_i-\widehat{\Theta}_i)-(\kappa(\widehat{\Xi}_i)-\kappa(\widehat{\Theta}_i))}{\phi^*/V_i}\right),
\end{equation}
and, by repeating the previous items, the sub-sampled split LRS for testing $\mathcal{H}_0$ against $\mathcal{H}_1$
\beq\label{def_sub_split_lrt}
\overline{E}^{sLRT}_{B}=\frac{1}{B}\sum_{b=1}^BE^{sLRT}_b,
\eeq
where $E^{sLRT}_b$ are conditional i.i.d.~copies of $E^{sLRT}$,  with the i.i.d.~applying to the partitioning of the set ${\cal D}$ into ${\cal D}_0$ and ${\cal D}_1$ (i.e., this is repeated $B$ times).
\item Reject the null hypothesis $\mathcal{H}_0$ at a pre-assumed significance level $\alpha \in (0,1)$ if 
    \beq\label{def_split_lrt_rejection}
    \overline{E}^{sLRT}_B \geq 1/\alpha.
    \eeq
\end{itemize}
\noindent\hrulefill

\medskip

\noindent
For $B=1$, the sub-sampled split LRS turns into the split LRS.

\begin{thm}\label{splitLRT_EDF}
\bl{Let (A)-(B) and (E) hold. Consider the null hypothesis $\mathcal{H}_0$ against the alternative $\mathcal{H}_1$. Assume (D) for the statistical test of mean-calibration or (C) for the statistical test of mean-specification}
\begin{itemize}
\item The split LRS \eqref{def_split_lrt} and the sub-sampled split LRS \eqref{def_sub_split_lrt} are exact e-variables for the set $\mathcal{P}$ representing the null hypothesis $\mathcal{H}_0$.
\item The split and sub-sampled split LRTs \eqref{def_split_lrt_rejection} control the type I error of rejecting $\mathcal{H}_0$ in favor of $\mathcal{H}_1$ at significance level $\alpha \in (0,1)$.
\end{itemize}
\end{thm}
\noindent \textbf{Proof:} Select $\bp\in\mathcal{P}$ from the null hypothesis $\mathcal{H}_0$ set given by \eqref{h_0_edf}. 
We prove $\be_\bp[E^{sLRT}]=1$. We compute
\beqo
\be_\bp\left[E^{sLRT}\right]=\be_\bp\left[\be_\bp\left[E^{sLRT}\;\middle|\;\mathcal{D}_1,\left(\mathbf{X}_i,\widehat{\Theta}_i, V_i\right)_{i\in\mathcal{D}_0}\right]\right].
\eeqo
Given $\{\mathcal{D}_1,(\mathbf{X}_i,\widehat{\Theta}_i, V_i)_{i\in\mathcal{D}_0}\}$, the variables $\big(\widehat{\Xi}_i,\widehat{\Theta}_i, V_i\big)_{i\in\mathcal{D}_0}$ are fixed, and the responses $(Y_i)_{i\in\mathcal{D}_0}$ are conditionally independent. \bl{The moment generating function of the conditional distribution of $Y_i|\mathcal{D}_1,\mathbf{X}_i,\widehat{\Theta}_i,V_i$ under $\mathcal{H}_0$ is that of the assumed member of the EDF with canonical parameter $\widehat{\Theta}_i$ and dispersion $\phi^*$, conditional on fixed $\mathcal{D}_1,\mathbf{X}_i,\widehat{\Theta}_i,V_i$. It is well-defined at $\frac{\widehat{\Xi}_i-\widehat{\Theta}_i}{\phi^*}V_i$, conditional on fixed $\mathcal{D}_1,\mathbf{X}_i,\widehat{\Theta}_i,V_i$}. We deduce that
\begin{equation}\label{the statement is as}
\be_\bp\left[E^{sLRT}\;\middle|\;\mathcal{D}_1,\left(\mathbf{X}_i,\widehat{\Theta}_i, V_i\right)_{i\in\mathcal{D}_0}\right]=1,\quad \bp\text{-almost surely}.
\end{equation}
Consequently, the first claim follows. The second assertion follows from the theory of e-variables and Theorem \ref{thm_universal}.
\qed

\medskip

There are the following remarks to  the above theorem.

\begin{rem}\normalfont \label{remark 5}
\begin{itemize}
\item \bl{We can identify a second point at which assumption (D) is required for testing mean-calibration. Recall that in Section \ref{sec_classic_lrt}, we identified one reason why (D) is needed -- it allows us to uniquely determine the MLE of the canonical parameters under the null hypothesis of mean-calibration. We now observe, from the proof of Theorem \ref{splitLRT_EDF}, that in order to use the moment generating function of the member of the EDF under consideration, we must specify the distribution of $Y|(\mathbf{X},V)$ under the null hypothesis (under the measure $\bp$). This is possible under $\be_\bp[Y|\mathbf{X}]=\widehat{\mu}(\mathbf{X})$, i.e., under mean-specification, but not under $\be_\bp[Y|\widehat{\mu}(\mathbf{X})]=\widehat{\mu}(\mathbf{X})$, i.e., under mean-calibration alone. Therefore, to prove Theorem \ref{splitLRT_EDF} when testing mean-calibration, we need assumption (D). However, when testing mean-specification, assumption (C) is not formally needed for the proof.}
\item \bl{Regarding the estimators used in the split LRS under the null and alternative hypotheses, recall from the general exposition of universal inference in Section \ref{sec_universal_intro} that we require the MLE of the canonical parameters $(\Psi(\mathbf{X}_i))_{i=1}^n$ under $\mathcal{H}_0$, based on $\mathcal{D}_0$, and an arbitrary estimator of the canonical parameters $(\Psi(\mathbf{X}_i))_{i=1}^n$ under $\mathcal{H}_1$, based on $\mathcal{D}_1$. The MLE under $\mathcal{H}_0$ was discussed in Section \ref{sec_classic_lrt}. It is straightforward to obtain this MLE when testing mean-calibration under (D) or mean-specification, even without assumption (C). This raises the question of why (C) is needed when testing mean-specification. The answer is that when testing either mean-calibration or mean-specification, the estimator of $(\Psi(\mathbf{X}_i))_{i=1}^n$ under $\mathcal{H}_1$ obtained by isotonic regression can be only formally justified under (C); see Sections \ref{sec_iso_intro} and \ref{sec_classic_lrt}. This estimator coincides with the MLE when testing mean-calibration under (D), but it is not the MLE when testing mean-specification under (C) alone. Nevertheless, in both cases, assumptions (C) and (D), respectively, provide a justification for using isotonic recalibration of the mean estimates as a suitable approach for constructing an estimator of the mean values under $\mathcal{H}_1$. At the same time, we highlight that although the sub-sampled split LRS defined in \eqref{def_split_lrt} can formally be used to test mean-specification under assumption (C) alone within the framework of universal inference, we expect that the power of the test in rejecting the null hypothesis of mean-specification might be low if assumption (D) does not hold, unless miscalibration is sufficiently large to be detected by the isotonic regression and consequently leading to rejection of the correct specification of the mean estimates; see also the point below}.
\item \bl{Ideally, we would like to apply our test statistics \eqref{def_split_lrt} under assumption (D). Under (D), mean-calibration is equivalent to mean-specification, and miscalibration is equivalent to misspecification. However, as discussed above, assumption (D) is unlikely to hold in practice, and we can only hope that it holds approximately; see Section \ref{sec_examples_c_d} for an investigation of the impact of violation of (D) on the statistical test and Appendix \ref{sec_app_1} for a discussion of (D). In practice, it is therefore safer to adopt assumption (C). If we work under (C), the null hypothesis $\mathcal{H}_0$ formally concerns mean-specification, with mean-specification constituting a strict subset of mean-calibration. By construction, our test statistics \eqref{def_split_lrt}, which is based on isotonic regression fitted under assumption (C), rejects the null hypothesis of mean-specification only when the mean estimates are not calibrated and not correctly specified. Conversely, it does not reject the null when the mean estimates are correctly specified, but it also does not reject the null when the mean estimates are misspecified yet remain calibrated. Since our main interest is in detecting miscalibration, we conclude that our sub-sampled split LRT, designed to test the null hypothesis of mean-specification at a pre-assumed significance level, should still have power to detect departures from mean-calibration under the mean ordering assumption (C), which is commonly adopted in the mean-calibration literature and is weaker than the mean ordering assumption (D). This is particularly the case for larger sample sizes and greater degrees of miscalibration; see Appendix \ref{sec_app_4} for a real data example. Yet, the type I error of such a test based on the test statistics \eqref{def_split_lrt}, whose null hypothesis encompasses both correctly specified mean estimates and misspecified but calibrated mean estimates, cannot be controlled in our framework under assumption (C).}
\item \bl{If (C) does not hold, i.e., if there exists a (not necessary monotone) mapping $\mathcal{U}$ such that
    \beqo
    \mathbb{E}\big[Y \mid \widehat{\mu}(\mathbf{X})\big]=\mathcal{U}(\widehat{\mu}(\mathbf{X})), \quad \text{almost surely,}
    \eeqo
    then by applying isotonic regression under $\mathcal{H}_1$ we end up with the $L^2$-projection of $\mathcal{U}$ onto the set of monotone functions, in the limit as $n\rightarrow\infty$, see \cite{Lim}. This still may be a reasonable estimate of the mean values under the alternative hypothesis if we test mean-specification and assumption (C) holds approximately;  see also the previous remarks}. 
\item \bl{The situation is different in the binomial case investigated by \cite{henzi_hl}. The authors do not formally consider likelihood functions and the split LRT. They introduce the test statistics \eqref{def_split_lrt} and prove it is an e-variable under the hypothesis of pure mean-calibration, $\be_\bp[Y|\widehat{\mu}(\mathbf{X})]=\widehat{\mu}(\mathbf{X})$. This is possible because in the binomial case the test statistics \eqref{def_split_lrt} can be linearized with respect to $Y$, and $\be_\bp[Y|\widehat{\mu}(\mathbf{X})]$ can be calculated under the null hypothesis of mean-calibration without requiring correct mean-specification for the conditional binomial distribution.}
\item \bl{With reference to the remark above, we could consider the test statistics}
\begin{equation}\label{def_split_lrt_linear}
E^{slogLRT}=1+\sum_{i\in\mathcal{D}_0}\left(\frac{\big(Y_i-\kappa'\big(\widehat{\Theta}_i\big)\big)\,(\widehat{\Xi}_i-\widehat{\Theta}_i)}{\phi^*/V_i}\right).
\end{equation}
\bl{One can prove that the expected value of $E^{slogLRT}$ is equal to 1 under the pure hypothesis of mean-calibration, $\be_\bp[Y|\widehat{\mu}(\mathbf{X})]=\widehat{\mu}(\mathbf{X})$. However, in general, we cannot guarantee that $E^{slogLRT}$ is non-negative. Hence, $E^{slogLRT}$ is not an e-variable under $\bp\in\mathcal{P}$ under the null hypothesis of mean-calibration, and universal inference fails.}
\item \bl{From \eqref{the statement is as} in the proof of Theorem \ref{splitLRT_EDF}, we observe that one could strengthen the unconditional result to a conditional one, given $(\bX, V)$. This also applies to other steps below.}
\end{itemize}
\end{rem}

\medskip

The next result links e-power and classical power of the split LRT in our statistical setting.

\begin{prop}\label{prop_asymptotic_power}
\bl{Let (A)-(B) and (E) hold. Consider the null hypothesis $\mathcal{H}_0$ against the alternative $\mathcal{H}_1$. Assume (D) for the statistical test of mean-calibration or (C) for the statistical test of mean-specification. Let $\mathcal{D}$ denote the test sample of size $n=|{\cal D}|$, and split $\mathcal{D}$ randomly into $\mathcal{D}_0$ and $\mathcal{D}_1$ with sizes $n_0=|{\cal D}_0|$ and $n_1=|{\cal D}_1|$. Consider the split LRS given by, see also \eqref{def_split_lrt},}
\beq\label{def_multiplicative_split_lrt}
E^{sLRT}_{[n_0]}=\prod_{i\in\mathcal{D}_0} E_{\widehat{\Xi}_i}\big(Y_i, \widehat{\Theta}_i, V_i\big).
\eeq
\bl{Choose $\bq\in\mathcal{Q}$ from the set of probability measures representing the alternative hypothesis $\mathcal{H}_1$. Assume}
\beqo
\be_\bq\left[\big|\log\big(E_{\widehat{\Xi}}\big(Y, \widehat{\Theta}(\mathbf{X}), V\big)\big)\big| \;\middle|\; \mathcal{D}_1\right]<\infty, \quad \text{$\bq$-a.s.},
\eeqo
\bl{for $(Y,\mathbf{X},V)$ sampled from the population with distribution $F_{Y,\mathbf{X},V}$. Define}
\beqo
C(\mathcal{D}_1)=\be_\bq\big[\log\big(E_{\widehat{\Xi}}\big(Y, \widehat{\Theta}(\mathbf{X}), V\big)\big) \mid \mathcal{D}_1\big].
\eeqo
\bl{If $C(\mathcal{D}_1)>0, \ \bq\text{-a.s.}$, then}
\beqo
\bq\big(E^{sLRT}_{[n_0]}>1/\alpha\big)\rightarrow 1\quad \text{as} \quad n_0\rightarrow\infty, 
\eeqo
\bl{for any $\alpha>0$ with $n_1$ fixed. If $C(\mathcal{D}_1)<0, \ \bq\text{-a.s.}$, then}
\beqo
\bq\big(E^{sLRT}_{[n_0]}>1/\alpha\big)\rightarrow 0\quad \text{as} \quad n_0\rightarrow\infty, 
\eeqo
\bl{for any $\alpha>0$ with $n_1$ fixed.}
\end{prop}
\noindent \textbf{Proof:} We analyze the probability
\beqo
\bq\left(E^{sLRT}_{[n_0]}>1/\alpha\right)=\be_\bq\left[\bq\left(E^{sLRT}_{[n_0]}>1/\alpha \;\middle|\; \mathcal{D}_1\right)\right].
\eeqo
Given $\mathcal{D}_1$, the variables $\big(E_{\widehat{\Xi}_i}(Y_i,\widehat{\Theta}_i,V_i)\big)_{i\in\mathcal{D}_0}$ are i.i.d., and we are in the classical setting of multiplicative i.i.d.~e-variables. \bl{We deduce that}
\beqo\label{maxima_rate}
\frac{1}{n_0}\log\Big(E^{sLRT}_{[n_0]}\Big) = \frac{1}{n_0}\sum_{i\in\mathcal{D}_0}\log\big(E_{\widehat{\Xi}_i}\big(Y_i, \widehat{\Theta}_i, V_i\big)\big),
\eeqo
and
\beqo
\frac{1}{n_0}\sum_{i\in\mathcal{D}_0}\log\big(E_{\widehat{\Xi}_i}\big(Y_i, \widehat{\Theta}_i, V_i\big)\big)\rightarrow C(\mathcal{D}_1),\quad \bq\text{-a.s.},
\eeqo
\bl{conditional on $\mathcal{D}_1$. Let us fix $\mathcal{D}_1$ such that $C(\mathcal{D}_1)>0$. For $\bq$-a.a.~$\mathcal{D}_0$ (conditional on fixed $\mathcal{D}_1$), there exists $N(\omega)$ such that for all $n_0\geq N(\omega)$, we have}
\beqo
\frac{1}{n_0}\sum_{i\in\mathcal{D}_0}\log\big(E_{\widehat{\Xi}_i}\big(Y_i, \widehat{\Theta}_i, V_i\big)\big)>\frac{C(\mathcal{D}_1)}{2}\quad \text{ and } \quad \log\Big(E^{sLRT}_{[n_0]}\Big)>n_0\frac{C(\mathcal{D}_1)}{2}>0.
\eeqo
\bl{Hence, $\lim_{n_0\rightarrow\infty}\mathbf{1}\{E^{sLRT}_{[n_0]}>1/\alpha\}=1$ for $\bq$-a.a.~$\mathcal{D}_0$ (conditional on fixed $\mathcal{D}_1$). We conclude that for $\bq$-a.a.~$\mathcal{D}_1$ such that $C(\mathcal{D}_1)>0$, we have $\bq\big(E^{sLRT}_{[n_0]}>1/\alpha\big|\mathcal{D}_1\big)\rightarrow 1$ as $n_0\rightarrow\infty$, for any $\alpha>0$. In the same way, for $\bq$-a.a.~$\mathcal{D}_1$ such that $C(\mathcal{D}_1)<0$, we have $\bq\big(E^{sLRT}_{[n_0]}>1/\alpha\big|\mathcal{D}_1\big)\rightarrow 0$ as $n_0\rightarrow\infty$, for any $\alpha>0$. The results then follow by the dominated convergence theorem; see also \citet[Proposition 3.12]{wang_book} for a similar result in a simplified setting.} \qed

\medskip

The conclusion from Proposition \ref{prop_asymptotic_power} is that an a.s.~positive conditional e-power of the split LRS \eqref{def_multiplicative_split_lrt} leads to an asymptotic power-1 test as desired, and an a.s.~negative conditional e-power of the split LRS \eqref{def_multiplicative_split_lrt} leads to an asymptotic power-0 test. \bl{Hence, we investigate the e-power of the split LRS in the sequel, both from a theoretical point of view and via numerical examples. Note that $\be_\bq\big[\log\big(E_{\widehat{\Xi}}\big(Y, \widehat{\Theta}(\mathbf{X}), V\big)\big) | \mathcal{D}_1\big]$ and $\be_\bq\big[\log\big(E^{sLRT}_{[n_0]}\big)|\mathcal{D}_1\big]$ have the same sign.}

\subsection{Sub-sampled split L$q$RT for \bl{mean-calibration and mean-specification} within the EDF}\label{sec_universal_modified}

The e-variable based on the LRS \eqref{def_split_lrt} is not the only possible e-variable in our statistical setting of testing the null hypothesis \eqref{h_0_edf} against the alternative hypothesis \eqref{h_1_edf}. We introduce a novel test statistics in this section. The motivation for this novel test statistics comes from robust likelihood and $L_q$-likelihood; see \cite{lq_estimator}, \cite{lq_test} and \citet[Section 5.6.3]{wang_book}.

\begin{df}
Choose $q\in(0,1]$. Define the split $L_q$-Likelihood Ratio Statistics (L$q$RS) by
\beq\label{def_split_power_lrt}
E^{sLRT,q}&=&\prod_{i\in\mathcal{D}_0}E^q_{\widehat{\Xi}_i}\big(Y_i, \widehat{\Theta}_i,V_i\big)\nonumber\\
& = &\prod_{i\in\mathcal{D}_0}\exp\left(\frac{q\,Y_i\left(\widehat{\Xi}_i-\widehat{\Theta}_i\right)-\left(\kappa\left(q\,\widehat{\Xi}_i+(1-q)\widehat{\Theta}_i\right)-\kappa\left(\widehat{\Theta}_i\right)\right)}{\phi^* /V_i}\right).
\eeq
\end{df}
\noindent \bl{Since (E) is assumed and the interior of the effective domain ${\rm int}\big(\boldsymbol{\Theta}\big)$ is convex, then $q\,\widehat{\Xi}_i+(1-q)\widehat{\Theta}_i\in {\rm int}\big(\boldsymbol{\Theta}\big)$ for all $i\in\mathcal{D}_0$ and the L$q$RS is well-defined}.

\begin{thm}\label{splitpowerLRT_EDF}
\bl{Let (A)-(B) and (E) hold. Consider the null hypothesis $\mathcal{H}_0$ against the alternative $\mathcal{H}_1$. Assume (D) for the statistical test of mean-calibration or (C) for the statistical test of mean-specification.} For any $q\in(0,1]$, the split L$q$RS \eqref{def_split_power_lrt} is an exact e-variable for $\mathcal{P}$ representing the null hypothesis $\mathcal{H}_0$.
\end{thm}
\noindent The proof is analogous to the one of Theorem \ref{splitLRT_EDF}.
Choosing $q=1$ in \eqref{def_split_power_lrt}, we rediscover the split LRS \eqref{def_split_lrt}. The key question is which choice of $q\in(0,1]$ yields the largest power of the split L$q$RT? The next two theorems shed more light on the power of the split L$q$RT. 

Within our statistical set-up of Section \ref{sec_model} and under the alternative hypothesis ${\cal H}_1$, for every measure $\mathbb{Q} \in \mathcal{Q}$ from the set representing the alternatives, there exists at least one instance $i \in \{1,\ldots, n\}$ such that
\beqo
\bl{Y_i \mid \mathbf{X}_i, V_i}&\bl{\overset{\mathbb{Q}}{\sim}}& 
\bl{EDF_{V_i,\phi^*}(\Pi_i),\quad \bq\text{-almost surely},} \\
\bl{\Pi_i}&\bl{\neq}& \bl{\widehat{\Theta}_i,\quad \text{with positive $\bq$-probability}}.
\eeqo
\bl{From the definition of the probability measures $\mathcal{M}$ on $(\Omega, \mathcal{F})$, we know that $\Pi_i=\Psi(\mathbf{X}_i)$.} 

\begin{thm}\label{thm_power_1}
\bl{Let (A)-(B) and (E) hold. Consider the null hypothesis $\mathcal{H}_0$ against the alternative $\mathcal{H}_1$. Assume (D) for the statistical test of mean-calibration or (C) for the statistical test of mean-specification.} Choose an element $\bq_\Pi =\bq \in \mathcal{Q}$ from the set of measures representing the alternative $\mathcal{H}_1$, yielding the canonical parameters $\Pi=(\Pi_1,\ldots, \Pi_n)$. For $i\in\mathcal{D}_0$, consider the function
\beq\label{function_f}
f_i(q) = \be_{\bq_\Pi}\left[E^q_{\widehat{\Xi}_i}\big(Y_i, \widehat{\Theta}_i,V_i\big) \;\middle|\; \mathcal{D}_1, (\mathbf{X}_i,\widehat{\Theta}_i, V_i)_{i\in\mathcal{D}_0}\right], \quad q\in(0,1].
\eeq

\noindent Assume $\widehat{\Xi}_i>\widehat{\Theta}_i$. This gives us the following property:
if $\Pi_i>\widehat{\Theta}_i$, then $f_i(q)>1$ for all $q\in(0,1]$.
\end{thm}

\noindent \bl{The statement is understood in the following formal way. We introduce the event}
\beqo
A=\Big\{\omega\in\Omega: \widehat{\Xi}_i(\omega)>\widehat{\Theta}_i(\omega) \ \text{and} \ \Pi_i(\omega)>\widehat{\Theta}_i(\omega)\Big\}.
\eeqo
\bl{For $\bq_\Pi$-almost all $\omega\in A$ such that $\widehat{\Xi}_i(\omega)>\widehat{\Theta}_i(\omega)$ and $\Pi_i(\omega)>\widehat{\Theta}_i(\omega)$, we have $f_i(q)(\omega)>1$ for all $q\in(0,1]$.}

\medskip

\noindent \textbf{Proof:} Assume $\{\mathcal{D}_1,(\mathbf{X}_i,\widehat{\Theta}_i, V_i)_{i\in\mathcal{D}_0}\}$ is given. The variables $\big(\widehat{\Xi}_i,\widehat{\Theta}_i\big)$ are fixed, and we can use the moment generating function for the conditional distribution of $Y_i|\mathcal{D}_1,\mathbf{X}_i,\widehat{\Theta}_i, V_i$, with canonical parameters $\Pi_i$ under $\bq_\Pi \in \mathcal{Q}$ (within the EDF). We derive from this moment generating function
\beqo
\frac{\phi^*}{V_i}\,\log(f_i(q))&=&\left(\kappa\left(\Pi_i+q\left(\widehat{\Xi}_i-\widehat{\Theta}_i\right)\right)-\kappa\left(\Pi_i\right)-\left(\kappa\left(q\,\widehat{\Xi}_i+(1-q)\,\widehat{\Theta}_i\right)-\kappa\left(\widehat{\Theta}_i\right)\right)\right)\\
&=&\left(\kappa\left(\Pi_i+q\left(\widehat{\Xi}_i-\widehat{\Theta}_i\right)\right)-\kappa\left(\Pi_i\right)-
\left(\kappa\left(\widehat{\Theta}_i +q\left(\widehat{\Xi}_i-\widehat{\Theta}_i\right)\right)-\kappa\left(\widehat{\Theta}_i\right)\right)\right)>0,
\eeqo
the last inequality follows from strict convexity of the cumulant function $\kappa$. This leads to increasing differences and proves the claim.
\qed

\medskip

Function \eqref{function_f} allows us to identify the alternative hypotheses (the probability measures) in \eqref{h_1_edf} against which the e-variable \eqref{def_split_power_lrt} is powered; see Definition \ref{df_power}. We investigate a conditional version of \eqref{def_split_power_lrt} for one instance $i$. Under the assumption $\widehat{\Xi}_i>\widehat{\Theta}_i$, we should require that $\Pi_i>\widehat{\Theta}_i$ in order to obtain $f_i(q)>1$ for all $q\in(0,1]$. By complete symmetry, under the assumption $\widehat{\Xi}_i<\widehat{\Theta}_i$, we should require that $\Pi_i<\widehat{\Theta}_i$ to force that $f_i(q)>1$ for all $q\in(0,1]$. Thus, the conclusions from Theorem \ref{thm_power_1} are that the isotonic estimate $\widehat{\Xi}_i$ and the ground truth $\Pi_i$ should be on the same side of the estimate $\widehat{\Theta}_i$ (specified in the null hypothesis) for the statistical test to have power against the true alternative $\Pi_i$. This property has been identified by \cite{henzi_hl} within the class of binomial models, and we have just extended it to the entire EDF.

From a practical point of view, if we under-estimate (over-estimate) the true value $\Pi_i$ by the original estimate $\widehat{\Theta}_i$, i.e., if $\Pi_i>\widehat{\Theta}_i$ ($\Pi_i<\widehat{\Theta}_i$), then the split L$q$RT for any $q\in(0,1]$ --  including the split LRT --  has power in rejecting the false $\widehat{\Theta}_i$ from the null hypothesis against the true alternative $\Pi_i$, if the isotonic regression estimate $\widehat{\Xi}_i$ uncovers this under-estimation (over-estimation) of $\Pi_i$ with $\widehat{\Theta}_i$, i.e., if $\widehat{\Xi}_i>\widehat{\Theta}_i$ ($\widehat{\Xi}_i<\widehat{\Theta}_i$). This property may be hard to guarantee in small samples because it also relies on the accuracy of the isotonic recalibration step, but it is likely to hold in large samples where the isotonic regression attains good accuracy. If we look for an e-variable which is powered against the true alternative, then the split LRT and the split L$q$RT with any $q\in(0,1]$ have the same property in this regard.

\begin{thm}\label{thm_power_2}
\bl{Let (A)-(B) and (E) hold. Consider the null hypothesis $\mathcal{H}_0$ against the alternative $\mathcal{H}_1$. Assume (D) for the statistical test of mean-calibration or (C) for the statistical test of mean-specification.} Choose an element $\bq_\Pi =\bq \in \mathcal{Q}$ from the set of measures representing the alternative $\mathcal{H}_1$, yielding the canonical parameters $\Pi=(\Pi_1,\ldots, \Pi_n)$. For $i\in\mathcal{D}_0$, consider the function
\beq\label{function_F}
F_i(q) = \be_{\bq_\Pi}\left[\log\big(E^q_{\widehat{\Xi}_i}\big(Y_i, \widehat{\Theta}_i,V_i\big)\big) \;\middle|\; \mathcal{D}_1, (\mathbf{X}_i,\widehat{\Theta}_i, V_i)_{i\in\mathcal{D}_0}\right], \quad q\in(0,1].
\eeq

\noindent Assume $\widehat{\Xi}_i>\widehat{\Theta}_i$ and $\Pi_i>\widehat{\Theta}_i$. We have the following properties:
\begin{itemize}
\item[(a)] If $\Pi_i>\widehat{\Xi}_i$, then $\arg\sup_{q\in(0,1]}F_i(q) = 1$.
\item[(b)] If $\Pi_i\in(\widehat{\Theta}_i,\widehat{\Xi}_i)$, then $\arg\sup_{q\in(0,1]}F_i(q) \in(0,1)$.
\item[(c)] There is a unique $\Pi_i^*\in(\widehat{\Theta}_i,\widehat{\Xi}_i)$ such that
\beqo
\kappa'\left(\Pi_i^*\right)=\frac{\kappa\left(\widehat{\Xi}_i\right)-\kappa\left(\widehat{\Theta}_i\right)}{\widehat{\Xi}_i-\widehat{\Theta}_i}.
\eeqo  
\item[(d)] If $\Pi_i>\Pi_i^*$, then we have $F_i(q)>0$ for all $q\in(0,1]$.
\item[(e)] If $\Pi_i\in(\widehat{\Theta}_i,\Pi_i^*)$, then there exists $q^*\in(0,1)$ such that $F_i(q)>0$ for all $q\in(0,q^*)$ and $F_i(q)<0$ for all $q\in(q^*,1]$.
\end{itemize}
\end{thm}

\medskip

\noindent \textbf{Proof:}
Assume $\{\mathcal{D}_1,(\mathbf{X}_i,\widehat{\Theta}_i, V_i)_{i\in\mathcal{D}_0}\}$ is given. Similar to the proof of Theorem \ref{thm_power_1}, one derives
\beq\label{formula_epower}
\frac{V_i}{\phi^*}\,F_i(q)=q \, \kappa'\left(\Pi_i\right)\left(\widehat{\Xi}_i-\widehat{\Theta}_i\right)-\left(\kappa\left(q\,\widehat{\Xi}_i+(1-q)\widehat{\Theta}_i\right)-\kappa\left(\widehat{\Theta}_i\right)\right).
\eeq
We deduce that $F_i(0)=0$. Moreover, the first derivative w.r.t.~$q$ is equal to
\beqo
\frac{V_i}{\phi^*}\,F_i'(q)&=&\kappa'\left(\Pi_i\right)\left(\widehat{\Xi}_i-\widehat{\Theta}_i\right)-\kappa'\left(q\,\widehat{\Xi}_i+(1-q)\widehat{\Theta}_i\right)\left(\widehat{\Xi}_i-\widehat{\Theta}_i\right)\\
&=&\left(\widehat{\Xi}_i-\widehat{\Theta}_i\right)\left(\kappa'\left(\Pi_i\right)-\kappa'\left(\widehat{\Theta}_i+q\left(\widehat{\Xi}_i-\widehat{\Theta}_i\right)\right)\right),
\eeqo
and from the second derivative we conclude concavity
\begin{equation*}
\frac{V_i}{\phi^*}\,F_i''(q)=-\left(\widehat{\Xi}_i-\widehat{\Theta}_i\right)^2 \kappa''\left(\widehat{\Theta}_i+q\left(\widehat{\Xi}_i-\widehat{\Theta}_i\right)\right) <0.
\end{equation*}
Moreover, the first derivative in zero satisfies
\beqo
\frac{V_i}{\phi^*}\, F_i'(0)=\left(\widehat{\Xi}_i-\widehat{\Theta}_i\right)\left(\kappa'\big(\Pi_i\big)-\kappa'\big(\widehat{\Theta}_i\big)\right).
\eeqo
We immediately conclude that $F_i'(0)>0$ and $F_i''(q)<0$ for all $q\in(0,1]$ under the assumptions of the theorem. Hence, function $F_i$ is increasing in the neighbourhood of zero. It is increasing for all $q\in(0,1]$ if $F_i'(q)>0$ for all $q\in(0,1]$. Using the concavity of $F_i$ on $(0,1]$, the latter is equivalent to requiring
\beqo
\frac{V_i}{\phi^*}\, F_i'(1)=\left(\widehat{\Xi}_i-\widehat{\Theta}_i\right)\left(\kappa'\big(\Pi_i\big)-\kappa'\big(\widehat{\Xi}_i\big)\right)>0,
\eeqo
which is satisfied only if $\Pi_i>\widehat{\Xi}_i$. In this case, $\arg\sup_{q\in(0,1]}F_i(q) = 1$. This proves item (a).

Otherwise, if $\Pi_i<\widehat{\Xi}_i$, then $\arg\sup_{q\in[0,1]}F_i(q) \in (0,1)$, and the concave function $F_i$ is increasing for $q\in(0,q^{\rm opt})$ and decreasing for $q\in(q^{\rm opt},1]$; this proves item (b). We investigate $F_i(1)$ and its sign. We have
\beqo
\frac{V_i}{\phi^*}\,F_i(1)=\kappa'\big(\Pi_i\big)\left(\widehat{\Xi}_i-\widehat{\Theta}_i\right)-\left(\kappa\big(\widehat{\Xi}_i\big)-\kappa\big(\widehat{\Theta}_i\big)\right).
\eeqo
By the mean value theorem, there exists a unique $\Pi^*_i \in(\widehat{\Theta}_i,\widehat{\Xi}_i)$ such that
\beqo
\kappa'\big(\Pi^*_i\big)\left(\widehat{\Xi}_i-\widehat{\Theta}_i\right)-\left(\kappa\big(\widehat{\Xi}_i\big)-\kappa\big(\widehat{\Theta}_i\big)\right)=0;
\eeqo
recall that $\kappa'$ is strictly increasing; this proves item (c). If $\Pi_i>\Pi_i^*$, then $F_i(1)>0$ and $F_i(q)>0$ for all $q\in(0,1]$; this proves item (d). Otherwise, $F_i(1)<0$ and the concave function $F_i$ crosses zero from above at some point $q^*\in(0,1)$. This concludes the proof.
\qed

\medskip

By symmetry, there is a similar result for $\widehat{\Xi}_i<\widehat{\Theta}_i$ and $\Pi_i<\widehat{\Theta}_i$. Function \eqref{function_F} allows us to identify the alternative hypotheses (the probability measures) in \eqref{h_1_edf} for which our e-variables \eqref{def_split_power_lrt} have a positive e-power; see Definition \ref{df_power}. Consider a conditional version of \eqref{def_split_power_lrt} for one observation. As expected, the conditions of Theorem \ref{thm_power_2} are stronger than the ones of Theorem \ref{thm_power_1}. From Theorem \ref{thm_power_1}, we know that we should require the isotonic estimate $\widehat{\Xi}_i$ and the ground truth $\Pi_i$ to be on the same side of the estimate $\widehat{\Theta}_i$, specified through the null hypothesis for a given data set. Under this assumption, the split LRT and the split L$q$RT for any $q\in(0,1]$ are powered against the true alternative. Yet, the tests have different e-powers against the true alternative. If $\widehat{\Xi}_i$ and $\Pi_i$ are both larger (smaller) than $\widehat{\Theta}_i$, and in addition if $\Pi_i>\widehat{\Xi}_i$ ($\Pi_i<\widehat{\Xi}_i$) for a given data split -- i.e., if we under-estimate (over-estimate) the true value $\Pi_i$ with the isotonic regression $\widehat{\Xi}_i$ but we correctly identify with $\widehat{\Xi}_i$ that the original estimate $\widehat{\Theta}_i$ from the null hypothesis is under-estimated (over-estimated) -- then the split L$q$RS with $q=1$ (the split LRT) has the largest e-power. However, if $\widehat{\Xi}_i$ and $\Pi_i$ are both larger (smaller) than $\widehat{\Theta}_i$, but if $\Pi_i<\widehat{\Xi}_i$ ($\Pi_i>\widehat{\Xi}_i$) for a given data split -- i.e., if we over-estimate (under-estimate) the true value $\Pi_i$ with the isotonic regression $\widehat{\Xi}_i$ but we still correctly identify with $\widehat{\Xi}_i$ that the original estimate $\widehat{\Theta}_i$ from the null hypothesis is under-estimated (over-estimated) -- then the split L$q$RT with $q^{\rm opt}\in(0,1)$ has a larger e-power than the split L$q$RT with $q=1$ (the split LRT). From this we conclude that in small samples, we should try to use the split L$q$RT for an optimal $q\in(0,1]$ to mitigate the inaccuracy of the isotonic regression on small samples. In large samples, we can safely choose $q=1$ in the split L$q$RT (i.e., the split LRT) as the isotonic regression should sufficiently accurately estimate the true mean values based on the original mean estimates specified in the null hypothesis.

\medskip

\begin{figure}[htb!]
\begin{center}
\includegraphics[width=0.72\textwidth]{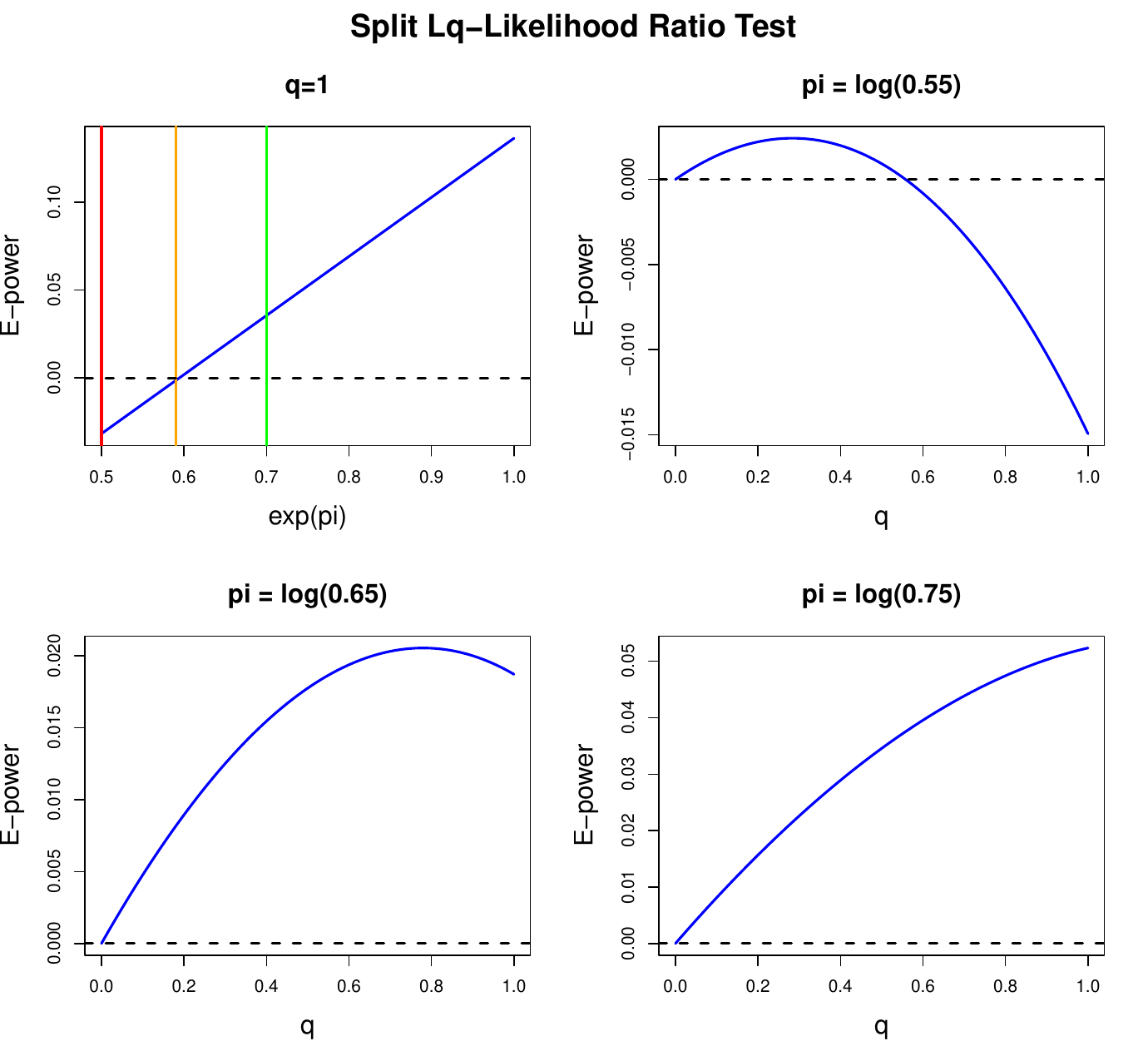}
\end{center}
\vspace{-.7cm}
\caption{E-power of the split L$q$RT for Poisson responses as the function determined by \eqref{function_F} for $(\widehat{\theta}, \widehat{\xi})=(\log(0.5),\log(0.7))$ being the original estimate and its recalibrated version.}
\label{fig_function_F}
\end{figure}

\noindent \textbf{Example:} Theorem \ref{thm_power_2} can be nicely illustrated with the example of Figure \ref{fig_function_F}. We study the e-power of the L$q$RT computed from a Poisson distributed response. Let the original estimate of the canonical parameter be equal to $\widehat{\theta}=\log(0.5)$, and its recalibrated version equal to $\widehat{\xi}=\log(0.7)$. We consider different true values of the canonical parameter $\pi$ satisfying $\pi>\widehat{\theta}=\log(0.5)$. If the true value of the canonical parameter satisfies $\pi>\widehat{\xi}=\log(0.7)$, e.g., $\pi=\log(0.75)$, then the split L$q$RT with $q=1$ (the split LRT) has the largest e-power; see Figure \ref{fig_function_F} (bottom-right). However, if the true value of the canonical parameter satisfies $\widehat{\theta}=\log(0.5)<\pi<\widehat{\xi}=\log(0.7)$, then the split L$q$RT with some $q^{\rm opt}\in(0,1)$ has the largest e-power, in particular, it has a larger e-power than the split L$q$RT with $q=1$ (the split LRT). If the true value of the canonical parameter satisfies $\log(0.59)<\pi<\widehat{\xi}=\log(0.7)$, e.g., $\pi=\log(0.65)$, then the split L$q$RT with $q=1$ has a positive e-power (but not optimal); see Figure \ref{fig_function_F} (bottom-left). If the true value of the canonical parameter satisfies $\widehat{\theta}=\log(0.5)<\pi<\log(0.59)$, e.g., $\pi=\log(0.55)$, then the split L$q$RT with $q=1$ has negative e-power; see Figure \ref{fig_function_F} (top-right). Hence, if we over-estimate the true mean value with an isotonic regression fitted to the original mean estimate, given that the true mean value is under-estimated with the original mean estimate specified in the null hypothesis, then the split L$q$RT should be more powerful than the split LRT in rejecting the false null hypothesis \eqref{h_0_edf} against the true alternative \eqref{h_1_edf}. If we severely over-estimate the true mean value with an isotonic regression fitted to the original mean estimate, given the true mean value is only slightly under-estimated with the original mean estimate specified in the null hypothesis, then the split LRT has a negative e-power, which may imply an asymptotic power-0 test, see Proposition \ref{prop_asymptotic_power}.\qed

\medskip

From Theorems \ref{thm_power_1}-\ref{thm_power_2} we conclude that we should use the split L$q$RT with the optimal power parameter
\beq\label{optimal_L$q$RT}
q^*=\underset{q \in (0,1]}{\arg\max}\sum_{i\in\mathcal{D}_0}\be_{\bq_\Pi}\left[\log\big(E^q_{\widehat{\Xi}_i}\big(Y_i, \widehat{\Theta}_i,V_i\big)\big) \;\middle|\; \mathcal{D}_1, (\mathbf{X}_i,\widehat{\Theta}_i, V_i)_{i\in\mathcal{D}_0}\right],
\eeq
under the true alternative $\bq_\Pi$. Yet, Theorems \ref{thm_power_1}-\ref{thm_power_2} are of theoretical nature and an optimal choice \eqref{optimal_L$q$RT} cannot generally be found. The expected value $\be_{\bq_\Pi}[\cdot]$ is not known and we can only try to estimate it. However, this estimate, and the choice of $q$, would be accurate only in large samples, and in large samples the split LRT is good enough; the analyses are presented in Appendix \ref{sec_app_3}.

For practical applications of the conclusions from Theorems \ref{thm_power_1}-\ref{thm_power_2}, we now proceed as follows: in order to enhance the e-power and, consequently, the power of the split LRT, one could try to average or maximize over $q\in (0,1]$ in the split L$q$RS defined by \eqref{def_split_power_lrt}. Select a finite subset $\mathcal{I}\subset (0,1]$ including $\{1\}$, and define \textit{the split mean power L$q$RS} w.r.t.~$\mathcal{I}$
\beq\label{def_split_mean power_lrt}
E^{sLRT,\bar{q}}~=~\frac{1}{|\mathcal{I}|}\sum_{q\in\mathcal{I}}\,E^{sLRT,q},
\eeq
and \textit{the split maximal power L$q$RS} w.r.t.~$\mathcal{I}$
\beq\label{def_split_max power_lrt}
E^{sLRT,\tilde{q}}~=~\max_{q\in\mathcal{I}}\,E^{sLRT,q}.
\eeq
We also introduce sub-sampled versions of the above test statistics by defining for i.i.d.~repetitions
\beq
\overline{E}^{sLRT,\bar{q}}_{B}=\frac{1}{B}\sum_{b=1}^BE^{sLRT,\bar{q}}_b, \label{def_sub_split_mean power_lrt}\\
\overline{E}^{sLRT, \tilde{q}}_{B}=\frac{1}{B}\sum_{b=1}^BE^{sLRT,\tilde{q}}_b, \label{def_sub_split_max power_lrt}
\eeq
The variables \eqref{def_split_mean power_lrt} and \eqref{def_sub_split_mean power_lrt} are e-variables and the standard theory of universal inference applies. However, \eqref{def_split_max power_lrt} and \eqref{def_sub_split_max power_lrt} are no longer e-variables, which means that it is not immediately clear whether one can control type I errors for these two test statistics.

\begin{thm}\label{split error I}
\bl{Let (A)-(B) and (E) hold. Consider the null hypothesis $\mathcal{H}_0$ against the alternative $\mathcal{H}_1$. Assume (D) for the statistical test of mean-calibration or (C) for the statistical test of mean-specification.} Reject $\mathcal{H}_0$ at a pre-assumed significance level $\alpha \in (0,1)$ if a test statistics exceeds the value of $1/\alpha$. The split mean power L$q$RS \eqref{def_split_mean power_lrt}, the split maximal power L$q$RS \eqref{def_split_max power_lrt} and the sub-sampled split mean power L$q$RS \eqref{def_sub_split_mean power_lrt} control the type I error of rejecting $\mathcal{H}_0$ in favor of ${\cal H}_1$ at significance level $\alpha \in (0,1)$.
\end{thm}

\noindent \textbf{Proof:} The results for \eqref{def_split_mean power_lrt} and \eqref{def_sub_split_mean power_lrt} are obvious since the test statistics are e-variables. We prove the result for \eqref{def_split_max power_lrt}. Select $\bp\in\mathcal{P}$, where $\mathcal{P}$ is the set of probability measures representing the null hypothesis $\mathcal{H}_0$ defined by \eqref{h_0_edf}. The split L$q$RS is given by, 
see \eqref{def_split_power_lrt},
\begin{eqnarray*}
\lefteqn{E^{sLRT,q}=\prod_{i\in\mathcal{D}_0}\exp\left(\frac{q\,Y_i\left(\widehat{\Xi}_i-\widehat{\Theta}_i\right)-\left(\kappa\left(q\,\widehat{\Xi}_i+(1-q)\widehat{\Theta}_i\right)-\kappa\left(\widehat{\Theta}_i\right)\right)}{\phi^* /V_i}\right)}
\\
&=&\exp\left(q\sum_{i\in\mathcal{D}_0}\frac{V_iY_i}{\phi^*}\left(\widehat{\Xi}_i-\widehat{\Theta}_i\right)-\sum_{i\in\mathcal{D}_0}\frac{V_i}{\phi^*}\left(\kappa\left(q\,\widehat{\Xi}_i+(1-q)\widehat{\Theta}_i\right)-\kappa\left(\widehat{\Theta}_i\right)\right)\right)
=\exp\Big(q Z-c_q\Big),
\end{eqnarray*}
where we have defined
\begin{eqnarray*}
Z = \sum_{i\in\mathcal{D}_0}\frac{V_iY_i}{\phi^*}\left(\widehat{\Xi}_i-\widehat{\Theta}_i\right)&\text{~and~}&
c_q= \sum_{i\in\mathcal{D}_0}\frac{V_i}{\phi^*}\left(\kappa\left(q\,\widehat{\Xi}_i+(1-q)\widehat{\Theta}_i\right)-\kappa\left(\widehat{\Theta}_i\right)\right).
\end{eqnarray*}
Given $\{\mathcal{D}_1,\big(\mathbf{X}_i,\widehat{\Theta}_i, V_i\big)_{i\in\mathcal{D}_0}\}$, first, (i) the variables $\big(\widehat{\Xi}_i,\widehat{\Theta}_i, V_i\big)_{i\in\mathcal{D}_0}$ are fixed, and second, (ii) the responses $(Y_i)_{i\in\mathcal{D}_0}$ are conditionally independent and EDF distributed with canonical parameters $\big(\widehat{\Theta}_i\big)_{i\in\mathcal{D}_0}$ under $\bp$. The first item (i) implies that given $\{\mathcal{D}_1,\big(\mathbf{X}_i,\widehat{\Theta}_i, V_i\big)_{i\in\mathcal{D}_0}\}$, the only random term in the split L$q$RS 
$E^{sLRT,q}$ is the random variable $Z$. An immediate consequence is that the variables $\big(E^{sLRT,q}\big)_{q\in\mathcal{I}}$ are comonotonic,
given $\{\mathcal{D}_1,\big(\mathbf{X}_i,\widehat{\Theta}_i, V_i\big)_{i\in\mathcal{D}_0}\}$. This allows us to apply \citet[Lemma 15.4]{wang_book} to show that
\beqo
\bp\Big(E^{sLRT,\tilde{q}}>1/\alpha\Big)&=&\bp\Big(\max_{q\in\mathcal{\mathcal{I}}}E^{sLRT,q}>1/\alpha\Big)\\
&=&\be_{\bp}\left[\bp\left(\max_{q\in\mathcal{I}}E^{sLRT,q}>1/\alpha \;\middle|\;\mathcal{D}_1,\big(\mathbf{X}_i,\widehat{\Theta}_i, V_i\big)_{i\in\mathcal{D}_0}\right)\right]\\
&=& \be_{\bp}\left[\max_{q\in\mathcal{I}}\bp\left(E^{sLRT,q}>1/\alpha \;\middle|\; \mathcal{D}_1,\big(\mathbf{X}_i,\widehat{\Theta}_i, V_i\big)_{i\in\mathcal{D}_0}\right)\right]\leq \alpha,
\eeqo
where the last step follows by Markov's inequality and by computing the conditional expected value of $E^{sLRT,q}$ equal to 1 for each $q\in\mathcal{I}$, see Theorem \ref{splitpowerLRT_EDF}.
This completes the proof.
\qed

\bigskip

\noindent Unfortunately, a similar result does not obviously hold for the sub-sampled split maximal power L$q$RT \eqref{def_sub_split_max power_lrt} and it is an open question if $\bp\big(\overline{E}_B^{sLRT,\tilde{q}} \geq 1/\alpha\big) \leq \alpha$ for $B>1$ within the EDF.

\section{Numerical analysis}\label{sec_examples}

We present a numerical analysis that studies the performance of classical and universal inference in \bl{testing mean-calibration and mean-specification}.

\subsection{Data set}
\label{sec_examples_data}

We describe how the test set $\mathcal{D}$ is sampled for our numerical experiments. For the test sample ${\cal D}$, we consider the sample sizes $n \in \{1000, 2000, 5000, 10000, 20000, 50000\}$. If the test sample ${\cal D}$ is 10\% of the entire data (often 90\% of the data is used for the learning sample ${\cal L}$ to fit the regression function $\widehat{g}$ for the mean values, see Section \ref{sec_model}), then $n=1000, 2000, 5000$ correspond to small insurance portfolios, $n=10000, 20000$ to medium size portfolios and $n=50000$ to a large portfolio.
 
First, we generate Poisson distributed responses $(Y_i)_{i=1}^n$ with expected annual frequencies $(\mu^*_i)_{i=1}^n$ being realistic choices for insurance claim counts modeling. Based on the real French motor insurance data set {\tt freMTPL2freq} from the {\sf R} package CASdatasets of \cite{dutang}, we select an expected annual frequency range of $[\mu_{\rm min}, \mu_{\rm max}]=[0.02, 0.25]$, the lower bound reflects excellent car drivers and the upper bound typically corresponds to inexperienced ones. This is a usual range of expected annual frequencies in car insurance pricing.

\begin{figure}[htb!]
\begin{center}
\includegraphics[width=0.43\textwidth]{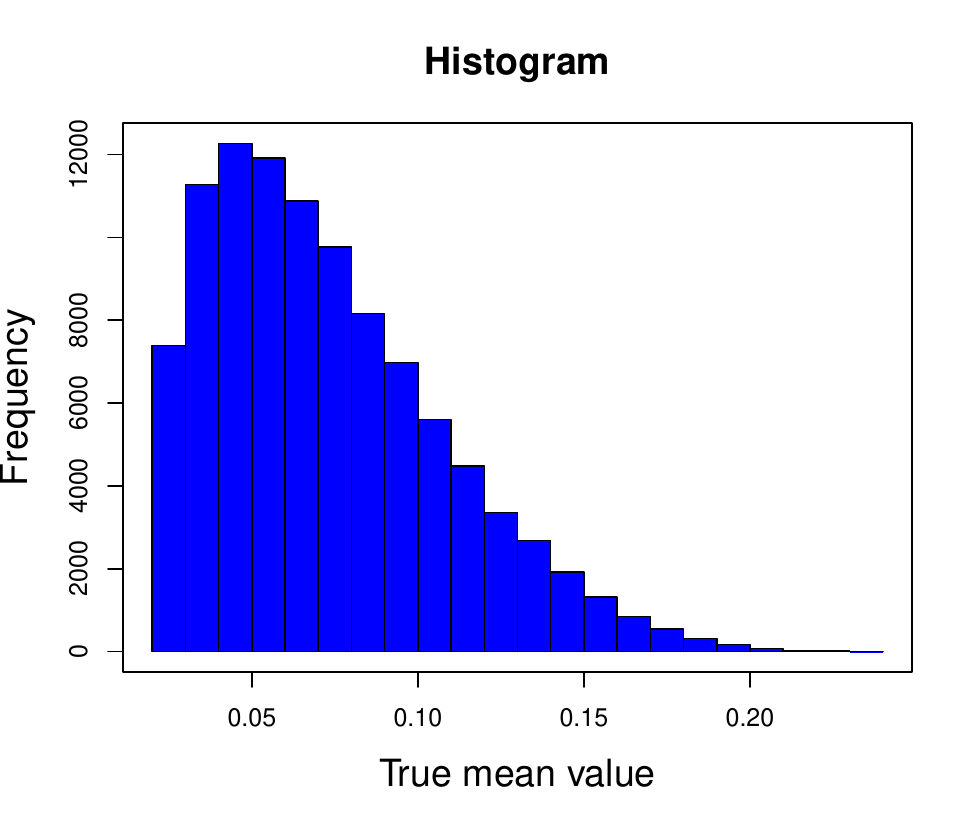}
\end{center}
\vspace{-.7cm}
\caption{Histogram of the simulated true means in the test sample of sample size $n=50000$.}
\label{fig_histogram_mean_values}
\end{figure}

We simulate the true expected annual frequencies $(\mu^*_i)_{i=1}^n$ of the test sample ${\cal D}$. Since the expected frequency profile of an insurance motor portfolio is typically right-skewed, we simulate the portfolio from a beta distribution. For this, we simulate i.i.d.~random variables $R_i\sim {\rm Beta}(a=1.5,b=5)$, and set for the true expected annual frequencies
\beq\label{true_mean_ex}
\mu^*_i=\mu_{\rm min}+R_i\left(\mu_{\rm max}-\mu_{\rm min}\right), \qquad i=1,\ldots,n. 
\eeq
Figure \ref{fig_histogram_mean_values} shows the histogram of the selected true expected annual frequencies $(\mu^*_i)_{i=1}^n$ for the test sample ${\cal D}$ of sample size $n=50000$ (one simulated scenario with the true mean values for the test sample). For simplicity, we set for all instances a unit exposure $V_i=1$. Given simulated $(\mu^*_i)_{i=1}^n$, we simulate the Poisson claims by independent $Y_i \sim {\rm Poi}(\mu^*_i)$ for all $i=1,\ldots,n$. This provides us with the test sample $(Y_i,\log(\mu^*_i))_{i=1}^n $; note that the cumulant function $\kappa$ of the Poisson EDF distribution is the exponential one, which provides us with the (canonical) log-link that maps the means $\mu^*_i$ to the true canonical parameters $\theta^*_i=\log(\mu^*_i)$. The Poisson dispersion is $\phi^*=1$.

\begin{figure}[htb!]
\begin{center}
\includegraphics[width=0.72\textwidth]{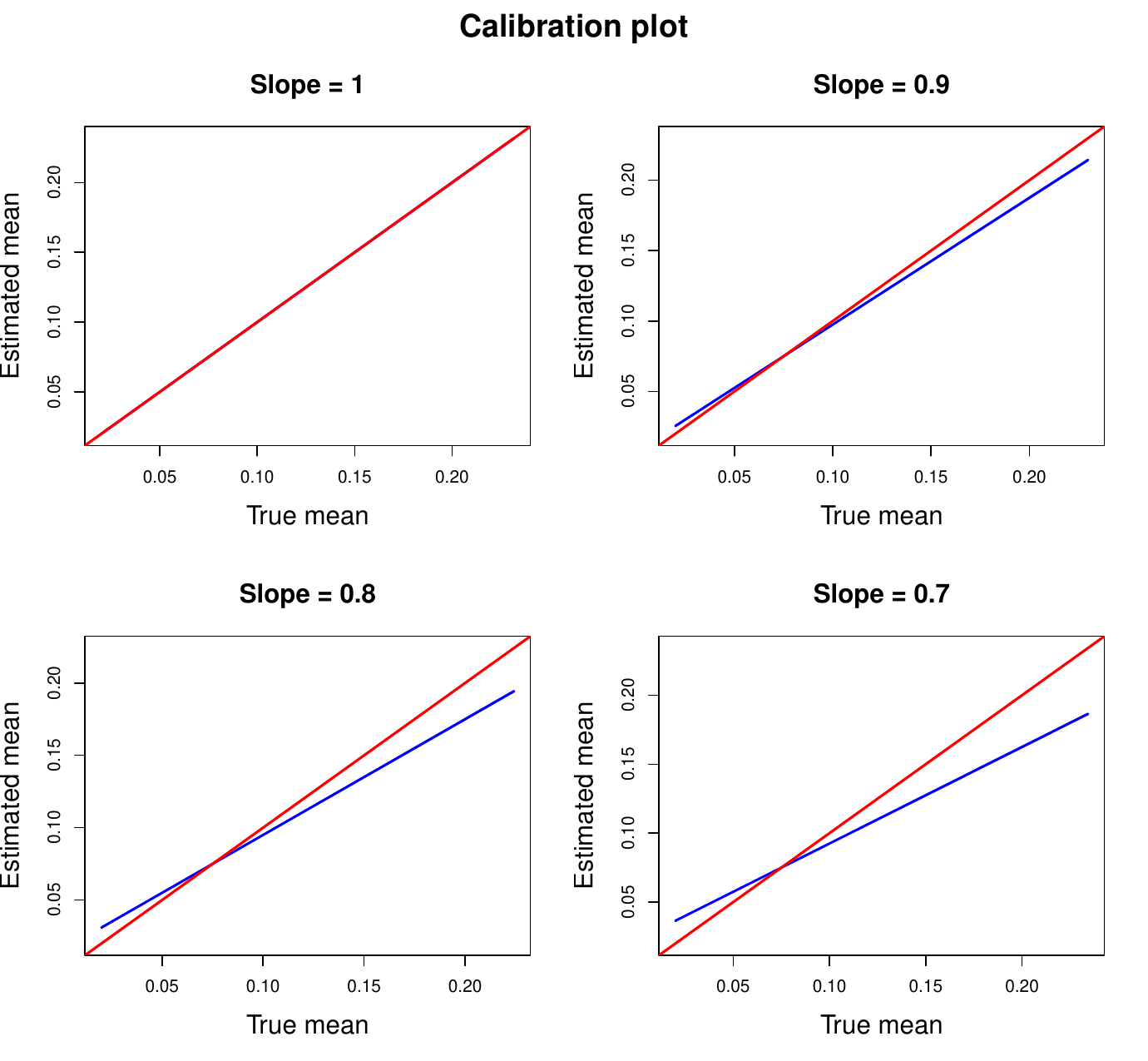}
\end{center}
\vspace{-.7cm}
\caption{Estimated means $\widehat{\mu}_i^j$ vs.~true means $\mu^{*}_i$ in the four considered calibration scenarios; the red line is the diagonal and the blue line indicates the misspecification relation.}
\label{fig_misspecification}
\end{figure}

Second, the main goal is to find these true mean parameters $(\mu^*_i)_{i=1}^n$ and canonical parameters $(\theta^*_i=\log(\mu^*_i))_{i=1}^n$, respectively. We proceed as follows. Based on a learning sample ${\cal L}$, we fit a regression model that provides us with estimated expected annual frequencies $\widehat{\mu}_i$; see Section \ref{sec_model}. Our aim is to assess a potential calibration error in these estimates. For our synthetic example, we are going to consider four scenarios of increasing miscalibration, and we try to understand whether the above calibration tests are able to identify these scenarios.
We select the slopes
\beqo
\operatorname{slope}_j =
\begin{cases} 1.0, \qquad j=1 \text{ (no miscalibration)},\\
0.9, \qquad j=2 \text{ (small miscalibration)},\\
0.8, \qquad j=3 \text{ (medium miscalibration)},\\
0.7,\qquad j=4\text{ (strong miscalibration)}.
\end{cases}
\eeqo 
Based on these slopes, we consider the four different scenarios $j=1,2,3,4$ of estimated means
\beq\label{slope_miscalibration}
\widehat{\mu}_i^j = \bar{\mu} + \operatorname{slope}_j\,(\mu^{*}_i-\bar{\mu}), \qquad i=1,\ldots,n.
\eeq
where $\bar{\mu}=0.075$ is the average mean value over the entire sample.
The value of the slope in \eqref{slope_miscalibration} represents the size of miscalibration of the mean estimates. These four scenarios of (mis)calibration are illustrated in Figure \ref{fig_misspecification}, and for the distribution of miscalibration one also needs to take into account the (right-skewed) sample distribution of Figure \ref{fig_histogram_mean_values}.

\bl{For each scenario $j=1,2,3,4$, our goal is to validate the following null hypothesis}
\beq\label{null_exp}
\mathcal{H}_0&=&\Big\{\bp\in\mathcal{M}, \ Y_i \mid \mathbf{X}_i \ \overset{\mathbb{P}}{\sim} \ {\rm Poi}\big(\widehat{\mu}_i^j\big),
~\bp\text{-almost surely}, \ \text{for all} \ i=1,\ldots,n\Big\},
\eeq
\bl{against the alternative}
\beq\label{alt_exp}
\mathcal{H}_1 &=& \Big\{\bq\in\mathcal{M}, \ Y_i \mid \mathbf{X}_i \ \overset{\mathbb{Q}}{\not\sim} \ {\rm Poi}\big(\widehat{\mu}_i^j\big),\nonumber\\ 
&& \quad \text{with positive probability} \ \bq, \ \text{for at least one} \ i=1,\ldots,n\Big\}.
\eeq
\bl{The set of probability measures $\mathcal{M}$ contains distributions for conditionally i.i.d.~responses such that $Y_i|\mathbf{X}_i=\bx_i \sim {\rm Poi}(\Psi(\bx_i))$ for some $\Psi$. For our test, for each $j=1,2,3,4$, we consider a random test sample $\mathcal{D}=(Y_i,\widehat{\mu}_i^j)_{i=1}^n$. The above steps of generating the true means $(\mu_i^*)_{i=1}^n$ and the test sample $\mathcal{D}=(Y_i,\widehat{\mu}_i^j)_{i=1}^n$ are repeated 1000 times for the following analyses. Figure \ref{fig_misspecification} shows that assumption (D) holds in our example and that the assumption of strict comonotonicity \eqref{ranking_function} is satisfied. Consequently, the null hypothesis \eqref{null_exp} against the alternative hypothesis \eqref{alt_exp} can be interpreted as a test of mean-calibration and mean-specification, both equivalent under (D).} Please also note that the vector of features $\bX_i$ does not play any role in our experiments because we directly selected the estimated mean values $\widehat{\mu}_i^j$, i.e., implicitly this includes the construction of the regression model.

\subsection{Classical inference: CORP reliability diagram, Murphy's score decomposition and LRT}

For a suitable benchmark, we use the classical methods for testing the null hypothesis \eqref{null_exp} of mean-calibration against the alternative \eqref{alt_exp}.

\begin{figure}[htb!]
\begin{center}
\includegraphics[width=0.72\textwidth]{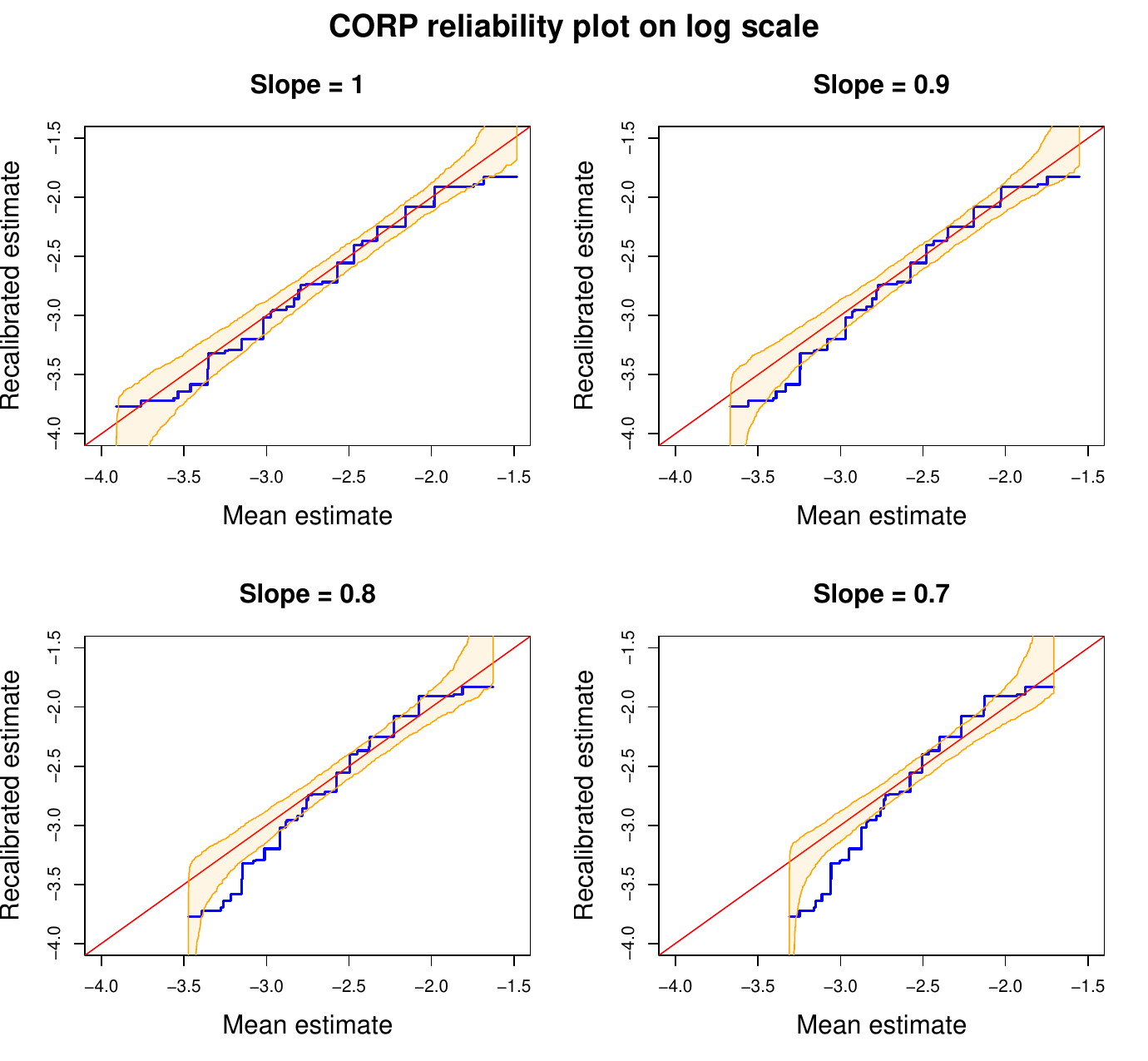}
\end{center}
\vspace{-.7cm}
\caption{CORP reliability diagram (blue) for the mean estimates together with $95\%$ point-wise consistency band (orange) for one random test sample of size $n=50000 $; the red line is diagonal.}
\label{fig_CORP}
\end{figure}

\begin{figure}[htb!]
\begin{center}
\includegraphics[width=0.72\textwidth]{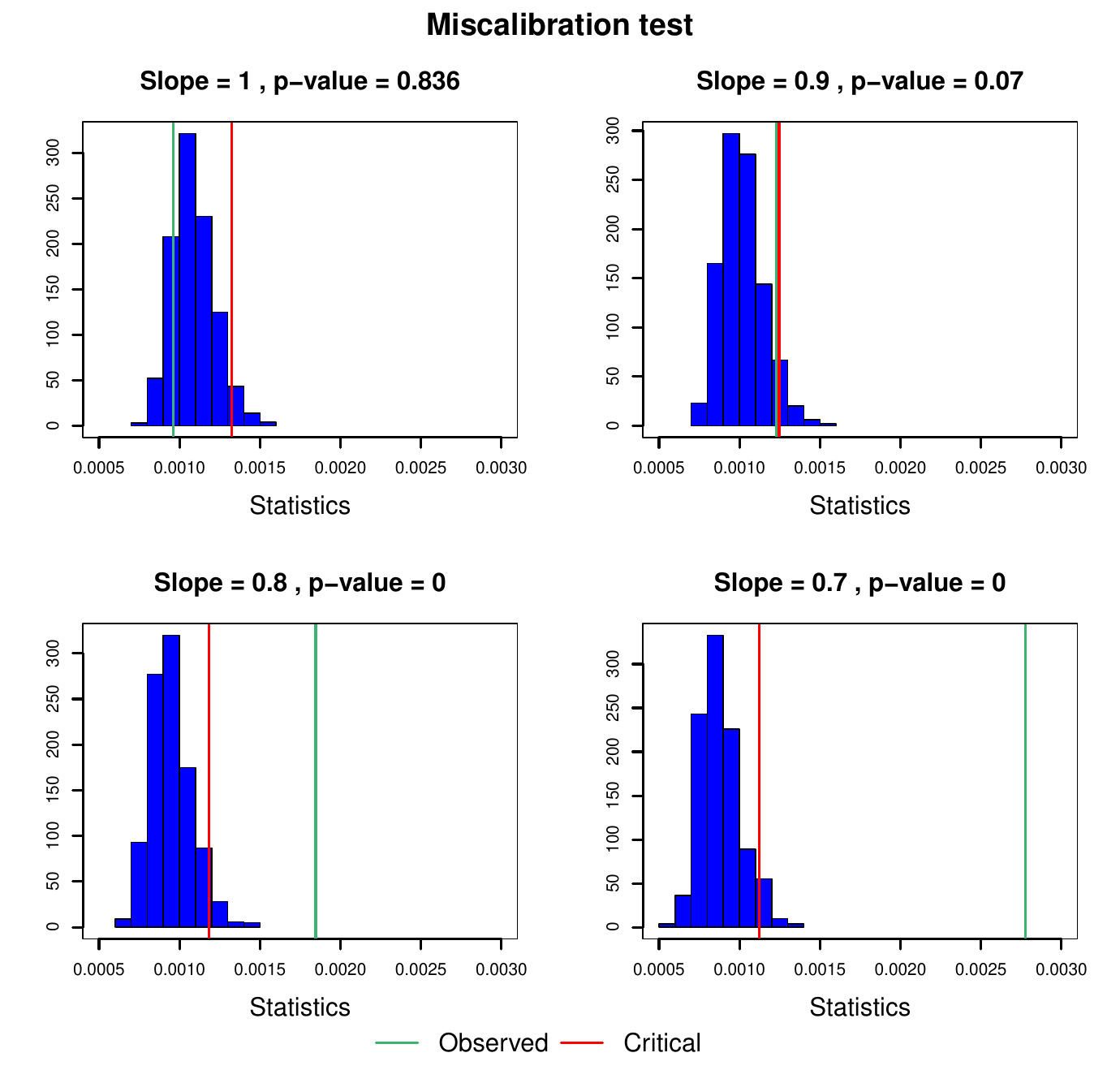}
\end{center}
\vspace{-.7cm}
\caption{Distribution of the miscalibration statistics under the null hypothesis:
green vertical line is the observed miscalibration statistics and the critical value at significance level $\alpha=0.05$ corresponds to the red vertical line; sample size $n=50000$.}
\label{fig_murphy}
\end{figure}

\subsubsection{CORP reliability diagram and Murphy's score decomposition}

We begin with the CORP reliability diagram for mean estimates using the techniques from Section \ref{sec_iso_intro}. The results for one random test sample ${\cal D}$, sample size $n=50000$, are presented in Figure \ref{fig_CORP}. Once the CORP reliability diagram is constructed for the mean estimates, we compute a $95\%$ point-wise consistency band with 1000 simulations using the parametric bootstrap approach of \cite{delong_iso}. Broadly speaking, under the null hypothesis of mean-calibration, equivalent to correct specification of the mean values under (D), we generate new (parametric bootstrap) observations $Y^\star_i\sim {\rm Poi}(\widehat{\mu}_i^j), i=1,\ldots,n,$ and re-fit a new isotonic regression to $(Y^\star_i,\widehat{\mu}_i^j, V_i=1)_{i=1}^n$ to quantify the randomness in the recalibrated mean estimates under the assumption of mean-calibration. This is illustrated in Figure \ref{fig_CORP}.

Next, we map these bootstrap results to test statistics by assessing the deviations of the CORP reliability diagram from the diagonal line under the Poisson deviance loss. In other words, we calculate the miscalibration statistics of Murphy's score decomposition \eqref{def_mcb}. The values of the miscalibration statistics observed for one random test sample of size $n=50000$ (green vertical lines) together with the bootstrap distributions of the miscalibration statistics under the null hypothesis of calibration of the mean estimates are presented in Figure \ref{fig_murphy}; for technical details see \cite{delong_iso}. From Figure \ref{fig_murphy} we conclude that the null hypothesis is (clearly) rejected at significance level $\alpha=5\%$ in the medium and large miscalibration scenarios $j=3,4$ (the 95\% quantile of the distribution of the test statistics under the null hypothesis of mean-calibration corresponds to the red vertical line), but the test does not reveal the small miscalibration generated with $\operatorname{slope}_2=0.9$, i.e., in that case the consistency band violations indicated in Figure \ref{fig_CORP} (top-right) are not sufficiently severe so that the calibration test could detect the miscalibration (reliably) on a sample size of $n=50000$. In the case of the scenario $j=1$ of mean-calibration, the test does (correctly) not reject the null hypothesis.

\subsubsection{Statistical power}

\begin{figure}[htb!]
\begin{center}
\includegraphics[width=0.43\textwidth]{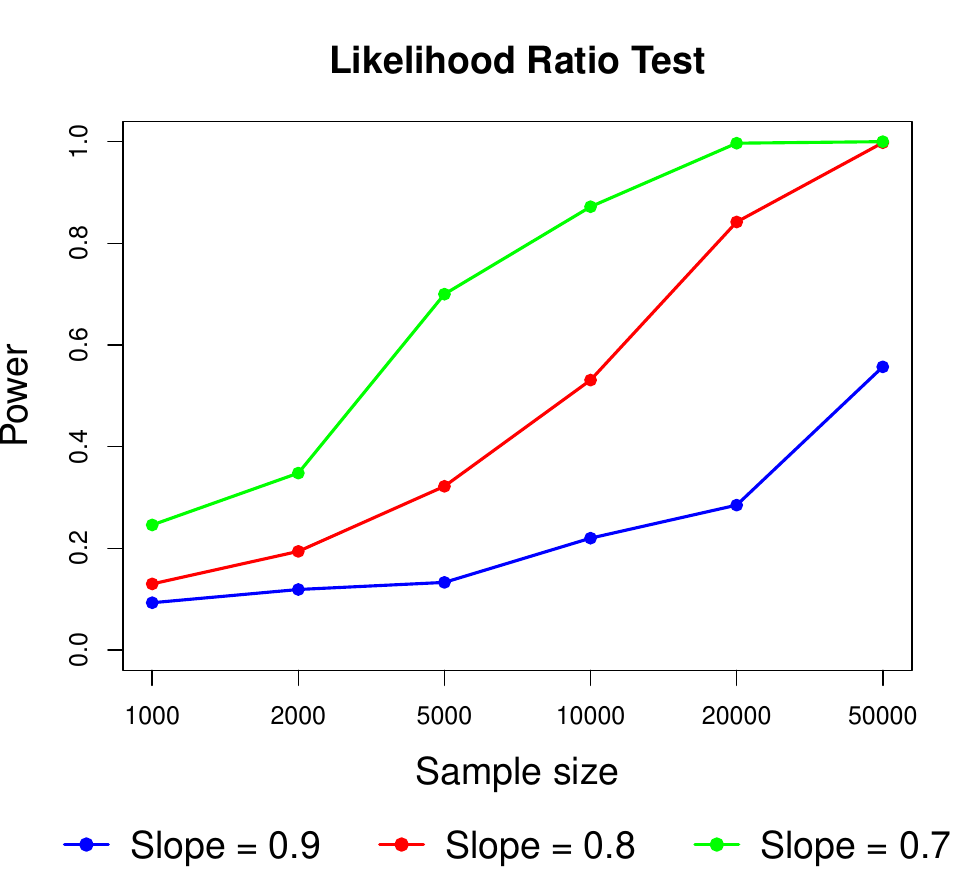}
\end{center}
\vspace{-.7cm}
\caption{Power of the classical LRT at significance level $\alpha=0.05$.}
\label{fig_lrt_power}
\end{figure}

In Section \ref{sec_classic_murphy}, equation \eqref{relation_lrt_mcb}, we have shown the equivalence between Murphy's miscalibration statistics and the classical LRS. In the sequel, we work on the LRS scale, which allows us to directly connect the classical inference to the universal inference proposals of Section \ref{sec_universal}. In particular, the results in Figure \ref{fig_murphy} reflect the classical LRT for the given test sample $\mathcal{D}=(Y_i,\widehat{\mu}_i^j)_{i=1}^n$. This classical LRT is based on one single test sample $\mathcal{D}=(Y_i,\widehat{\mu}_i^j)_{i=1}^n$, as illustrated in Figures \ref{fig_CORP} and \ref{fig_murphy}. In order to understand the power of the classical LRT, we repeat this experiment 1000 times by simulating 1000 i.i.d.~test samples $\mathcal{D}$ to evaluate the power of this classical LRT. Figure \ref{fig_lrt_power} shows the results for significance level $\alpha=5\%$, for the miscalibration scenarios $j=2,3,4$, and the different sample sizes between 1000 and 50000 -- by construction the type I error of this classical LRT is $\alpha=5\%$. The large miscalibration scenario $j=4$ can reliably be detected for sample sizes bigger than 10000, but the small miscalibration scenario $j=2$ is difficult to detect, even a comparably large sample size of 50000 only provides a power of roughly 0.6; see Figure \ref{fig_lrt_power}.

\medskip

We conclude this section by confirming that classical inference provides us with suitable tools to detect medium and large miscalibrations in medium and large samples. Yet, these tools are based on simulating the LRS under the null hypothesis, which may hinder their widespread use practice.

\subsection{Universal inference: sub-sampled split LRT}

This section applies the techniques from universal inference to test the null hypothesis \eqref{null_exp} of mean-calibration against the alternative \eqref{alt_exp}. We begin with the sub-sampled split LRS defined in \eqref{def_sub_split_lrt}, see Section \ref{sec_universal_split}. We choose a split ratio of $s=0.5$ and number of sub-samples of $B=1000$ for the sub-sampled split LRS $\overline{E}^{sLRT}_{B}$, see \eqref{def_sub_split_lrt}. The split ratio $s=0.5$ is advised by \cite{dunn_universal}; other values are investigated at the end of this section.

\subsubsection{Type I error}

\begin{figure}[htb!]
\begin{center}
\includegraphics[width=0.43\textwidth]{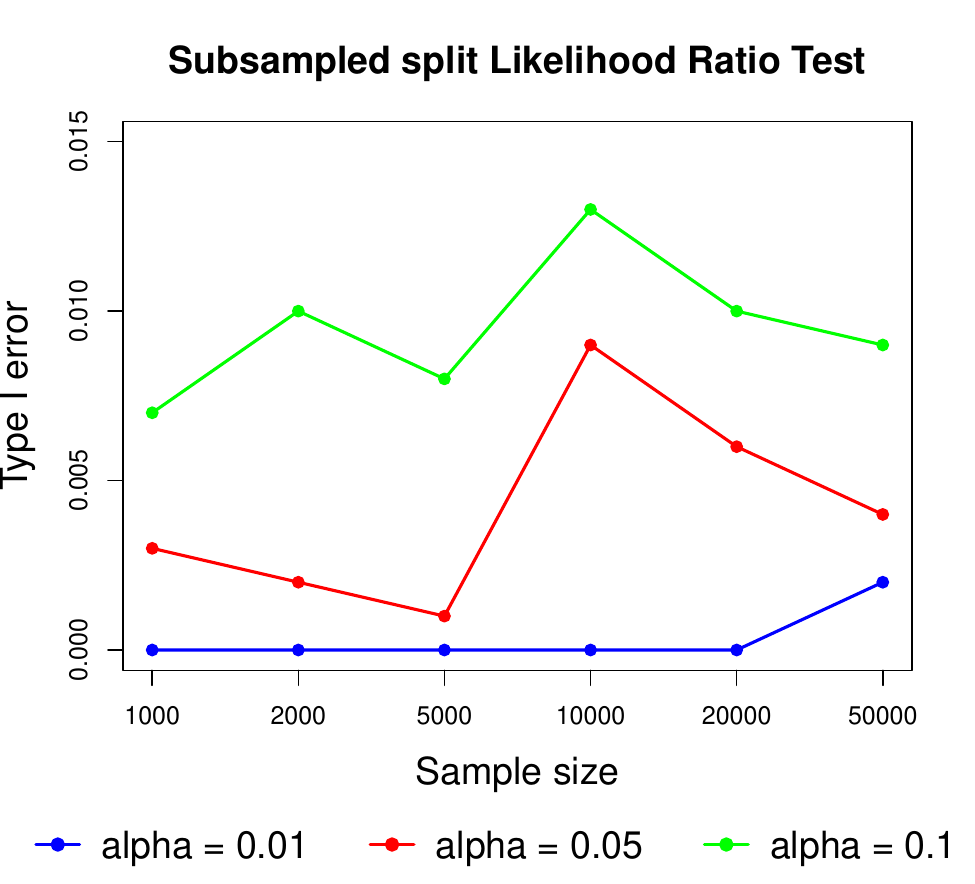}
\end{center}
\vspace{-.7cm}
\caption{Type I error of the sub-sampled split LRT with the (universal) critical value equal to $1/\alpha$ at the pre-assumed significance levels $\alpha= 0.01, 0.05, 0.1$.}
\label{fig_LRT_type_1}
\end{figure}

We start by investigating the type I error of the sub-sampled split LRT. Since we use the universal critical value of $1/\alpha=20$ at the significance level $\alpha=0.05$, we expect the type I error of the test to be much lower than $0.05$. Figure \ref{fig_LRT_type_1} confirms this expectation; the property of lower true type I errors has also been identified in other papers on universal inference. We choose the mean estimates from the scenario $j=1$ for the null hypothesis and consider the significance levels $\alpha = 0.01, 0.05, 0.1$, providing (universal) critical values $1/\alpha$ for the test. The probabilities are estimated based on 1000 i.i.d.~simulations of the test sample ${\cal D}$. \bl{We observe from Figure \ref{fig_LRT_type_1} that they are roughly 10 times lower than the assumed significance levels.}

\subsubsection{Statistical power and e-power}

\begin{figure}[htb!]
\begin{center}
\includegraphics[width=0.43\textwidth]{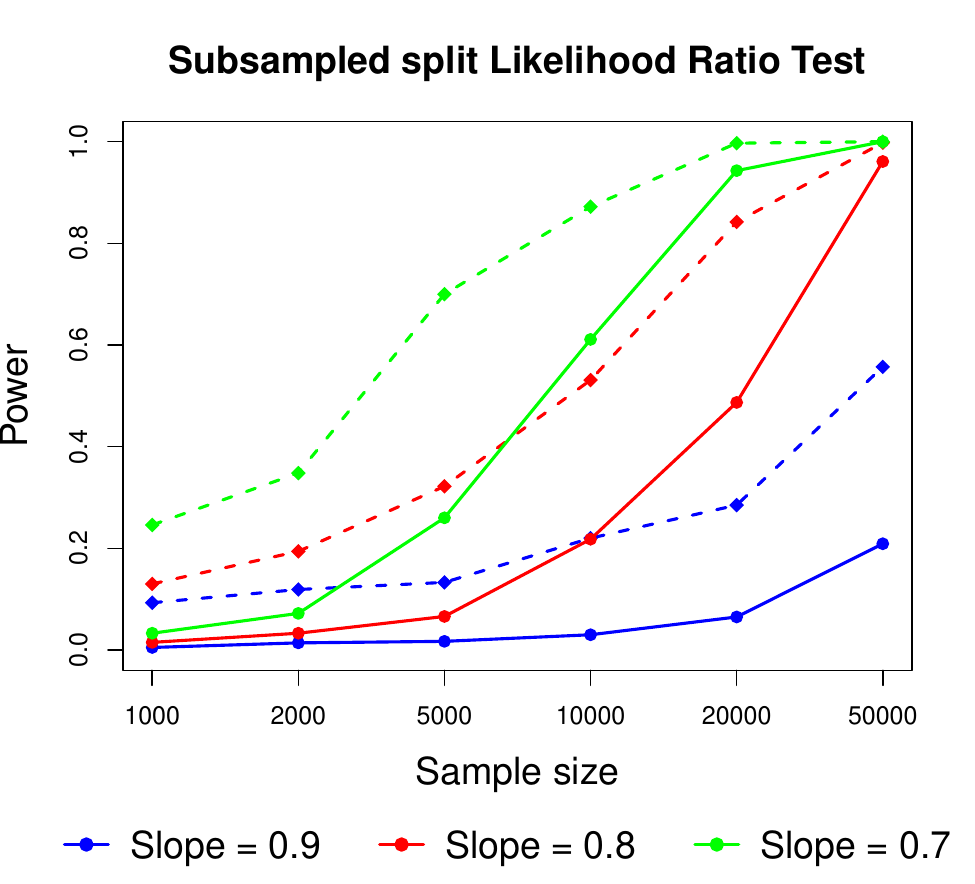}
\end{center}
\vspace{-.7cm}
\caption{Power of the sub-sampled split LRT with the (universal) critical value equal to $1/\alpha=20$; the dashed lines present the power of the classical LRT at significance level $\alpha=0.05$ (taken from Figure \ref{fig_lrt_power}).}
\label{fig_split_lrt_power}
\end{figure}

Fortunately, as our next results demonstrate, even though the type I error is below its targeted value, the power of the sub-sampled split LRT is reasonable good and large enough for many practical applications. Figure \ref{fig_split_lrt_power} shows the powers of the sub-sampled split LRTs at the pre-assumed significance level $\alpha=0.05$ with the (universal) critical value $1/\alpha=20$ for rejecting the null hypothesis. We choose the mean estimates from the scenario $j=2,3,4$. The solid lines show the results of the sub-sampled split LRTs, and the dotted lines their classical LRT counterparts taken from Figure \ref{fig_lrt_power}. The main conclusions from Figure \ref{fig_split_lrt_power} are that the medium and large miscalibrations can reliably be detected by the sub-sampled split LRT in sample sizes bigger than 20000, and the power of the test is comparably low for the miscalibration scenarios $j=3,4$ for smaller sample sizes and for the weak miscalibration $j=2$ for all sample sizes. \bl{We observe gaps in powers between the classical LRT and its sub-sampled split counterpart, particularly for weak miscalibration and for small sample sizes. The power of the sub-sampled split LRT is lower than that of the classical LRT. This is expected since the universal critical value is larger than the true critical value and universal inference uses only part of the data to validate the null hypothesis against the alternative (the remaining part is used to fit the model under the alternative hypothesis).} We deduce that for small sample sizes one should rely on the classical LRT (being evaluated with Monte Carlo and bootstrap), and for medium/large sample sizes one can rely on the sub-sampled split LRT to detect bigger miscalibration errors (which does not use any simulations). \bl{This conclusion is important for non-life insurance pricing where we deal with large sample sizes.} Weak miscalibration of mean estimates is difficult to detect in general, i.e., with both the classical and the sub-sampled split LRT even in large sample sizes.

\begin{figure}[htb!]
\begin{center}
\includegraphics[width=0.43\textwidth]{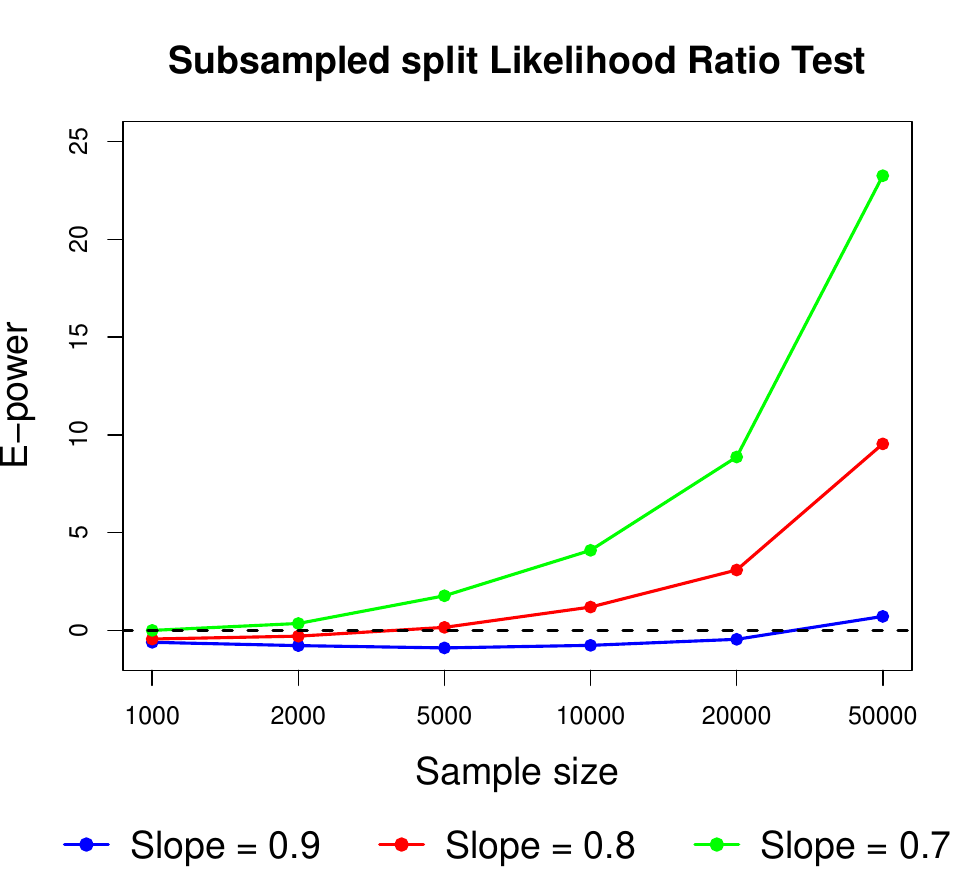}
\end{center}
\vspace{-.7cm}
\caption{E-power of the sub-sampled split LRT with the (universal) critical value $1/\alpha=20$.}
\label{fig_split_LRT_growth}
\end{figure}

The e-power of the sub-sampled split LRT is investigated in Figure \ref{fig_split_LRT_growth}. The e-power is negative for small sample sizes and small/medium miscalibration sizes and gets positive for large sample sizes for all miscalibration sizes. By Definition \ref{df_power} and Proposition \ref{prop_asymptotic_power}, we conclude that a positive e-power in large samples is a desired property of the sub-sampled split LRT in rejecting the false null hypothesis against the true alternative. The sub-sampled split LRT gains more e-power, the larger the sample size and the larger the miscalibration size. This conclusion, of course, agrees with the conclusions from Figure \ref{fig_split_lrt_power}. We summarize that the sub-sampled split LRT for different sample and miscalibration sizes behaves as expected.

\begin{figure}[htb!]
\begin{center}
\includegraphics[width=0.72\textwidth]{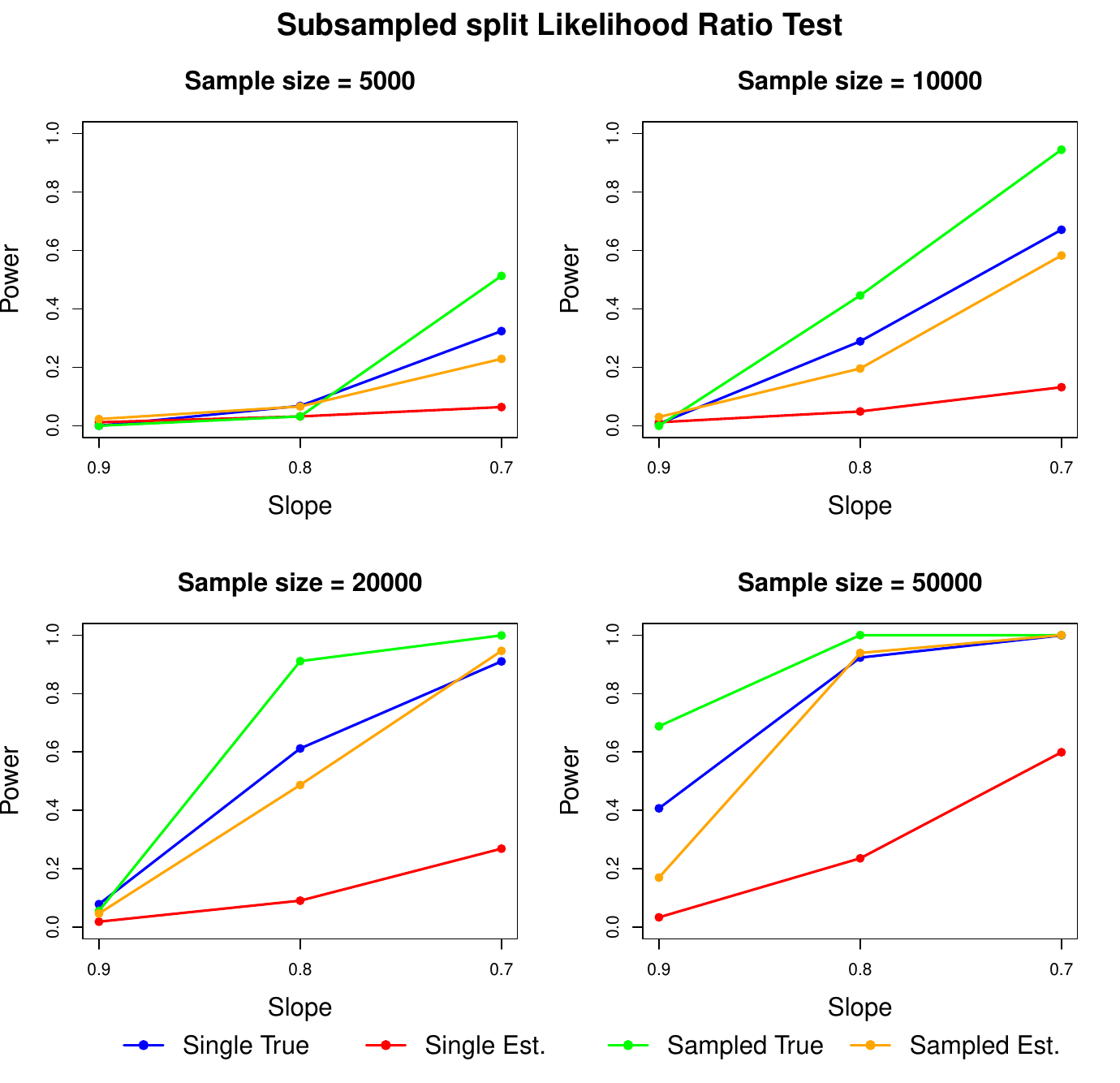}
\end{center}
\vspace{-.7cm}
\caption{Power of the (sub-sampled) split LRTs for $B\in \{1,1000\}$ with the (universal) critical value $1/\alpha=20$; we use the estimated means with isotonic regression (red/orange) and the true means (blue/green) under the alternative.}
\label{fig_LRT_comp}
\end{figure}

\subsubsection{Impact to estimating isotonic regression}

Figure \ref{fig_LRT_comp} investigates four different test statistics. In red color it considers the split LRS $E^{sLRT}$, defined in \eqref{def_split_lrt}, this corresponds to $B=1$, and in orange color the sub-sampled split LRS $\overline{E}^{sLRT}_{B}$, defined in \eqref{def_sub_split_lrt}, for $B=1000$. The difference between these two lines in Figure 
\ref{fig_LRT_comp} shows the gain in statistical power from sub-sampling $B>1$. Second, Figure \ref{fig_LRT_comp} analyzes the inaccuracy implied by the isotonic regression \eqref{isotonic regression under H1} used to estimate the mean values under the alternative hypothesis. Working with synthetic data, we can replace these isotonic regression estimates by the true means to evaluate the loss of power by this estimation step. The corresponding graphs are given in blue color ($B=1$) and in green color ($B=1000$). From these results, we conclude that the power of the split LRT is crucially depending on the accuracy of the isotonic regression step, which can be critical in small samples. This conclusion is again in support of using the sub-sampled split LRT for medium/large sample sizes with medium/large miscalibration sizes, and the classical LRT (with Monte Carlo and bootstrap) otherwise (as splitting a small/medium sample with a small miscalibration has a negative impact on the accuracy of isotonic regression and the power of the test). We may seek to improve the power of the test by refining the estimation of the mean values under the alternative hypothesis and modifying the test statistics, as discussed in Section \ref{sec_universal_modified}.

\subsubsection{Number of bootstrap sub-samples}

\begin{figure}[htb!]
\begin{center}
\includegraphics[width=0.72\textwidth]{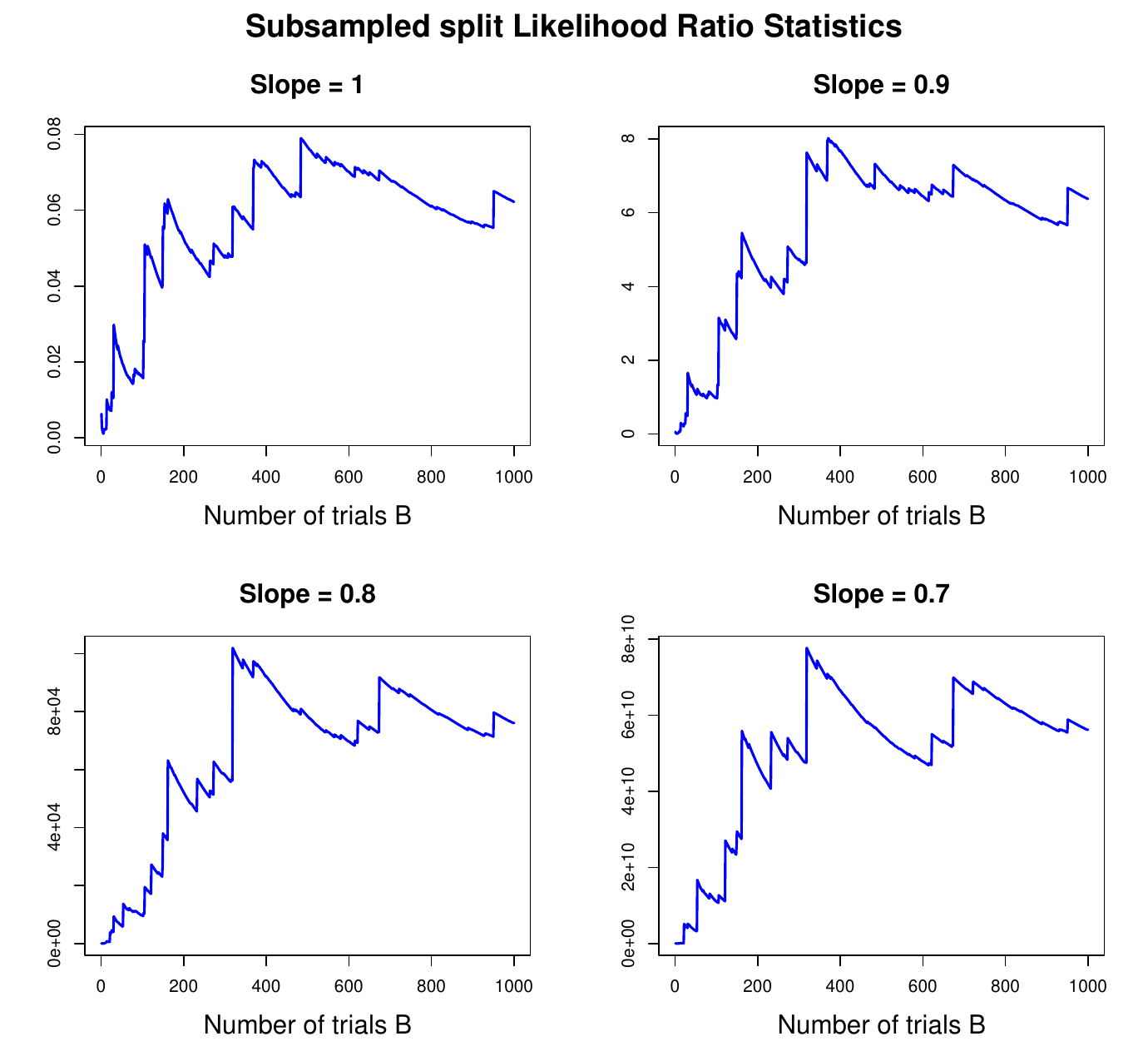}
\end{center}
\vspace{-.7cm}
\caption{Running value of the sub-sampled split LRS in repeated sampling, $B\ge 1$, for one random test sample ${\cal D}$ of sample size $n=50000$.}
\label{fig_splitLRT_ind}
\end{figure}

The above results of the sub-sampled split LRTs were based on $B=1000$ sub-samples. In Figure \ref{fig_splitLRT_ind} we investigate the running value of the sub-sampled split LRS for increasing numbers $B\ge 1$ of sub-samples in \eqref{def_sub_split_lrt}; this is for a single test sample ${\cal D}$ of sample size $n=50000$. As expected, the larger the miscalibration, the bigger the value of the sub-sampled split LRS. This figure shows that it takes roughly 500 iterations for the running average to stabilize. 

\begin{figure}[htb!]
\begin{center}
\includegraphics[width=0.9\textwidth]{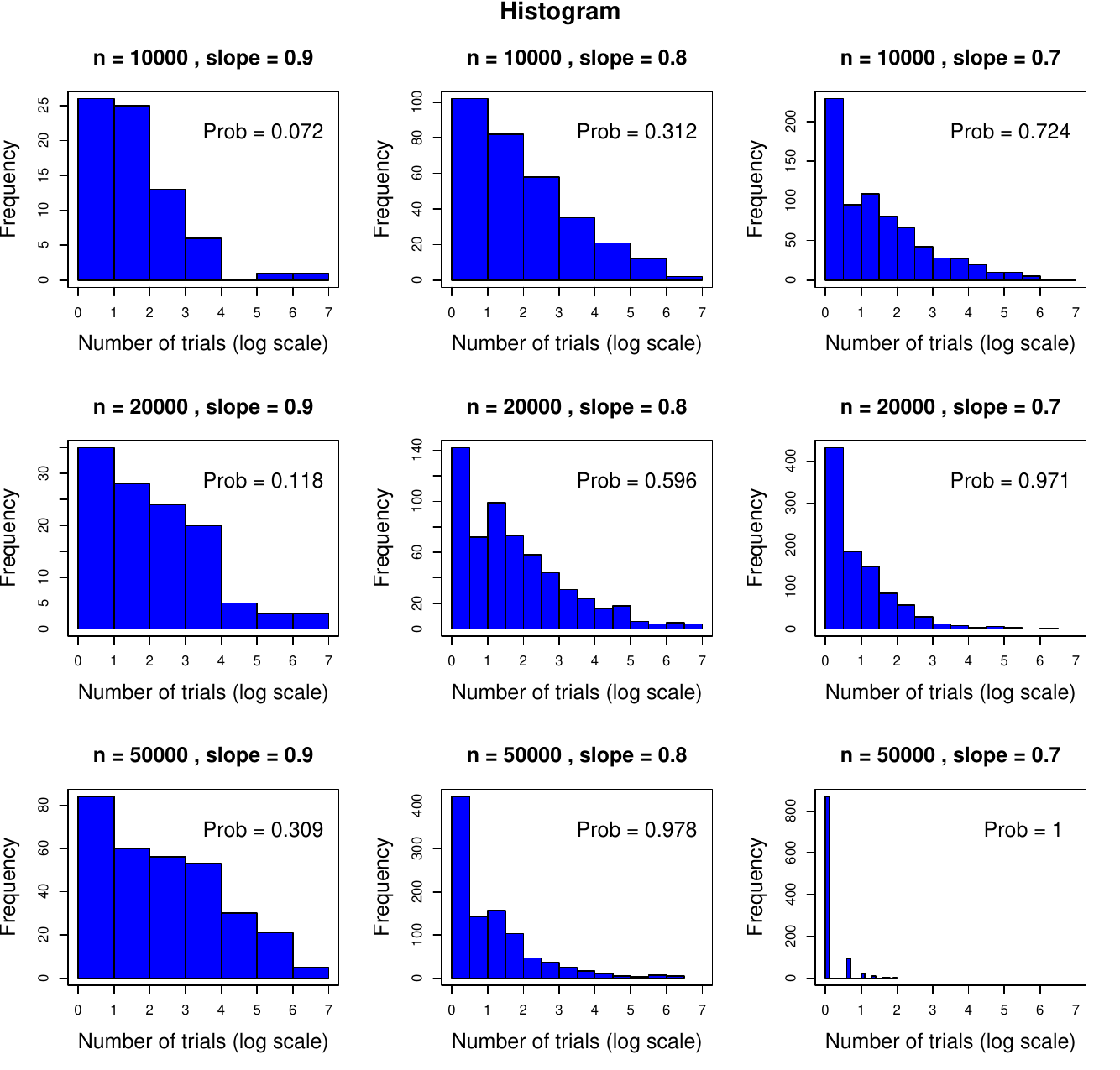}
\end{center}
\vspace{-.7cm}
\caption{Histogram of the number of sampling trials $B$ before the sub-sampled split LRS $\overline{E}^{sLRT}_{B}$ exceeds the (universal) critical value $1/\alpha=20$ for the first time; "Prob" gives power of the calibration test (empirically determined with data-dependent $B$ from at most 1000 i.i.d.~test samples ${\cal D}$).}
\label{fig_splitLRT_trials}
\end{figure}

Figure \ref{fig_splitLRT_trials} presents the histograms of the number of sampling trials $B$ before the sub-sampled split LRS $\overline{E}^{sLRT}_{B}$ exceeds for its first time the universal critical value $1/\alpha=20$. The plots show the histograms over the 1000 i.i.d.~test samples ${\cal D}$, and the upper right corner indicates the power of the calibration test. E.g., the bottom-middle plot shows the case of sample size $n=50000$ and miscalibration error $\operatorname{slope}_3=0.8$. This test has a power of 0.978, and from the histogram we see that in almost half of the 1000 i.i.d.~test samples ${\cal D}$ the critical value $1/\alpha=20$ is exceeded with $B=2$. We point out that here the power is reported for the sub-sampled split LRT with data-dependent $B$, where the sampling is terminated at the first time the running value of the sub-sampled split LRS exceeds 20; see \citet[Section 5.3]{wang_book}. From these plots we conclude that often roughly $B=200$ to $300$ sub-samples are sufficient in all our cases, and even much less for medium/large samples with medium/large miscalibrations sizes.

\subsubsection{Split ratio}

\begin{figure}[htb!]
\begin{center}
\includegraphics[width=0.72\textwidth]{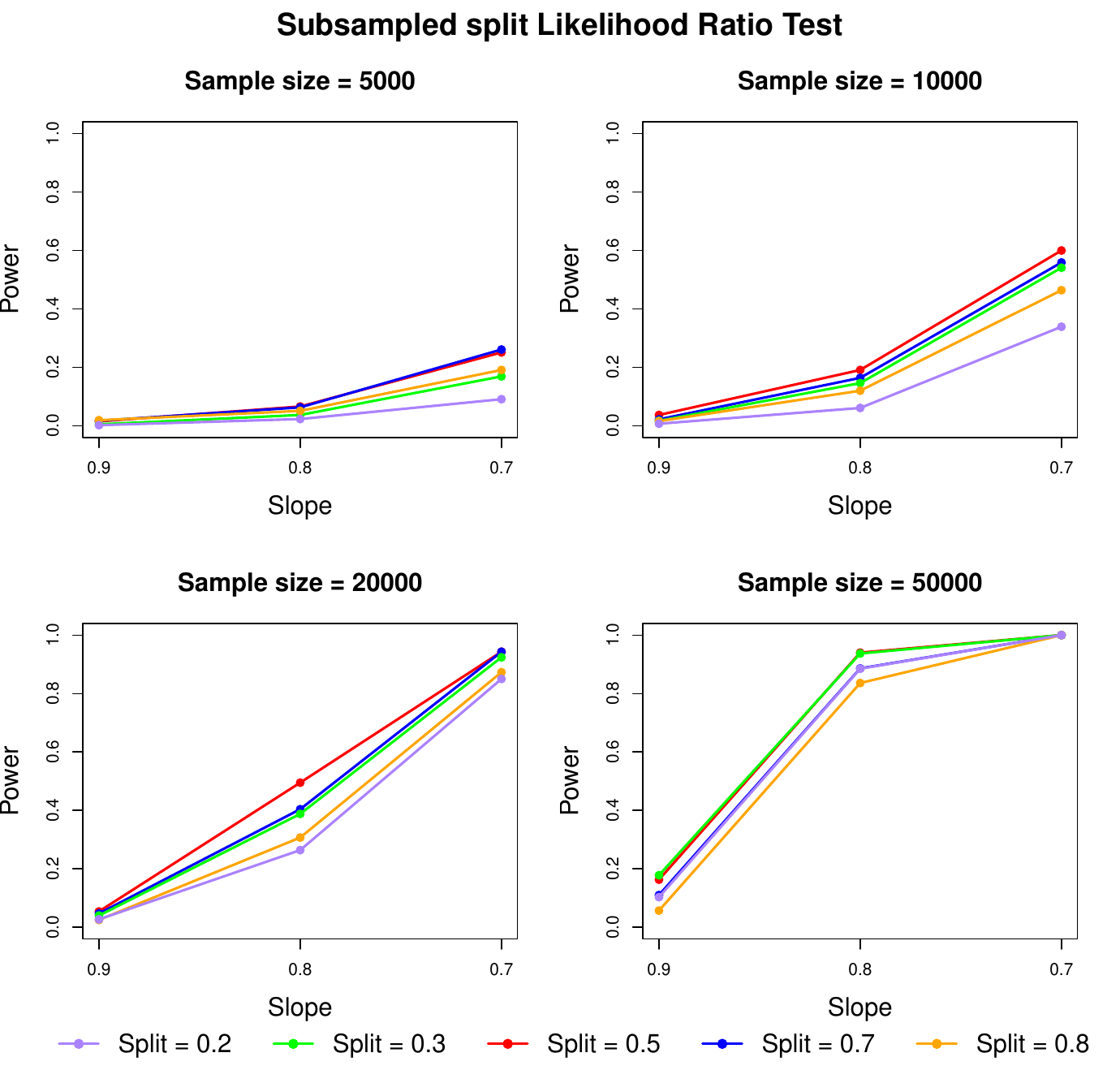}
\end{center}
\vspace{-.7cm}
\caption{Power of the sub-sampled split LRT with the (universal) critical value equal to $1/\alpha=20$ for different split ratios $s\in \{0.2, 0.3, 0.5, 0.7, 0.8\}$.}
\label{fig_LRT_best_split}
\end{figure}

From Figure \ref{fig_LRT_comp} we concluded that the loss in power of the split LRT compared to the classical LRT comes from the inaccuracy of the isotonic regression step. For the split LRT, the test sample ${\cal D}$ is partitioned at random into the training sample ${\cal D}_1$ used for estimating the isotonic regression step and the validation sample ${\cal D}_0$ used for performing the calibration test. Following \cite{dunn_universal}, we selected a split ratio of $s=0.5$. Figure \ref{fig_LRT_best_split} benchmarks this choice against $s=0.2, 0.3, 0.7, 0.8$. We run the simulations with $B=200$. The results confirm the proposal of \cite{dunn_universal}.

\subsubsection{\bl{Violation of assumption (B)}}
\label{sec_example_phi}

We investigate assumption (B) underlying our sub-sampled split LRT. We study the impact of a misspecified dispersion parameter for a fixed member of the EDF on the type I error and the power of the test. Recall that the dispersion parameter has so far been treated as correctly estimated, or fixed at 1 for the Poisson distribution in our numerical example. 

\begin{figure}[htb!]
\begin{center}
\includegraphics[width=0.72\textwidth]{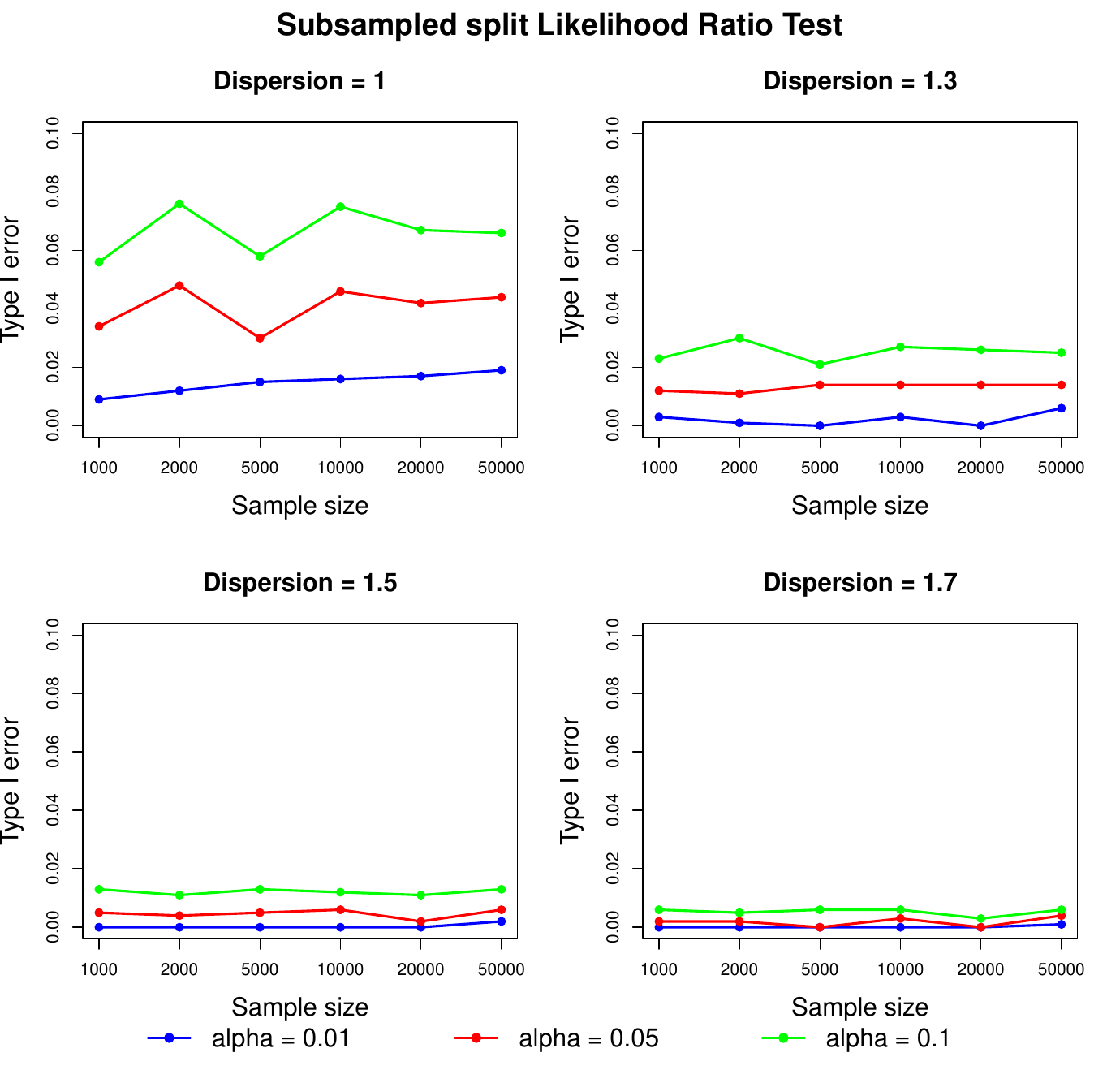}
\end{center}
\vspace{-.7cm}
\caption{Type I error of the sub-sampled split LRT with the (universal) critical value equal to $1/\alpha$ at the pre-assumed significance levels $\alpha=0.01, 0.05, 0.1$. The sub-sampled split LRS is calculated with the overdispersed Poisson deviance with dispersion $\phi\in \{1, 1.3, 1.5, 1.7\}$ and the responses follow overdispersed Poisson distributions with $\phi^*=1.5$.}
\label{fig_negbin}
\end{figure}

For our example, we focus on overdispersed Poisson distributions. In order to retain the same form of the deviance loss function that enters the LRS, while modifying only the dispersion parameter, we construct the responses for the numerical experiments in this subsection as follows. Given $(\mu_i^*)_{i=1}^n$ and $\phi^*$, we simulate independent Poisson claims $\tilde{Y}_i\sim{\rm Poi}(\mu_i^*/\phi^*)$ and set $Y_i=\phi^*\tilde{Y}_i$ for all $i=1,\ldots,n$. This gives us the test sample $(Y_i,\log(\mu_i^*))_{i=1}^n$ from an EDF with the same cumulant function and canonical parameter as the Poisson EDF, but with a dispersion parameter equal to $\phi^*$; we now have $\be[Y|\mu_i^*]=\mu_i^*$ and $Var[Y|\mu_i^*]=\phi^*\mu_i^*$. The disadvantage of this construction is that the response is no longer integer-valued, so we lose the interpretation of the response as the number of claims. The advantage, however, is that we can study the impact of using an incorrect value of $\phi$ in the sub-sampled split LRT \eqref{def_split_lrt} in the same setting as discussed above.

\begin{figure}[htb!]
\begin{center}
\includegraphics[width=0.72\textwidth]{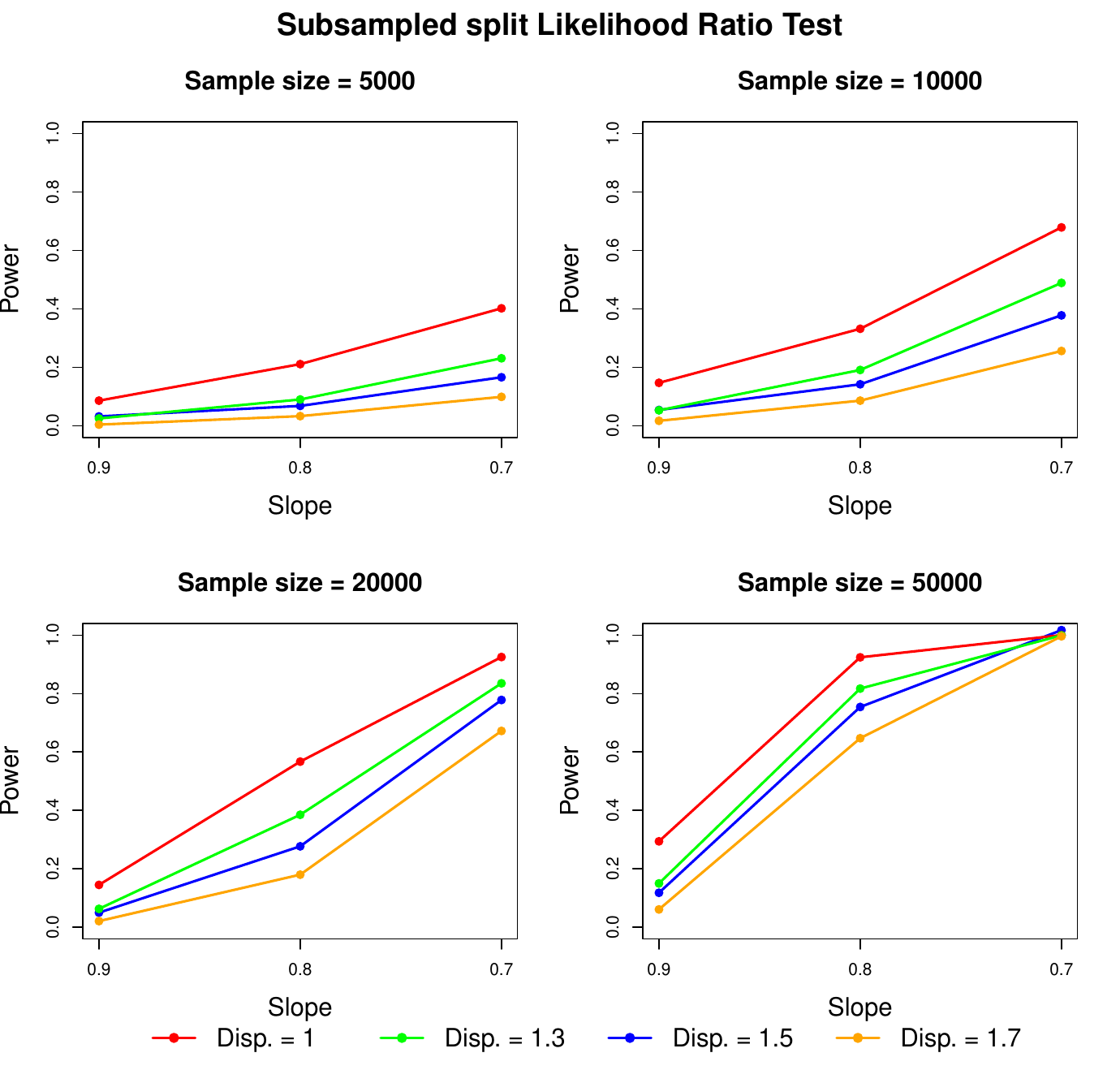}
\end{center}
\vspace{-.7cm}
\caption{Power of the sub-sampled split LRT with the (universal) critical value equal to $1/\alpha=20$. The sub-sampled split LRS is calculated with the overdispersed Poisson deviance with dispersion $\phi\in \{1, 1.3, 1.5, 1.7\}$ and the responses follow overdispersed Poisson distributions with $\phi^*=1.5$.}
\label{fig_negbin_2}
\end{figure}

\cite{wuthrich_st_learning_book} consider quasi-Poisson GLM examples for claim frequencies in Section 5. The overdispersion can reach $\phi=1.5$, and we use this value as the true dispersion parameter $\phi^*$ when simulating the responses. We consider the same four scenarios of (mis)calibration of the estimated regression function as before and apply the sub-sampled split LRT. However, we now allow the dispersion parameter $\phi$ used in the LRS to be misspecified. Specifically, we calculate the sub-sampled split LRS \eqref{def_split_lrt} with $\phi\in\{1, 1.3, 1.5, 1.7\}$ instead of the true $\phi^*=1.5$. The results are presented in Figures \ref{fig_negbin}-\ref{fig_negbin_2}. To estimate the type I error, we use the mean estimates from the scenario $j=1$ to represent the null hypothesis.

Directly from the formula for the split LRT \eqref{def_split_lrt}, we can deduce that using a smaller $\phi$ than the true value $\phi^*$ leads to an increase in the power to reject the null hypothesis when it is false. It also increases the type I error of rejecting the null hypothesis when it is true. This is confirmed by our figures. It is interesting to note that even though we use $\phi=1$ in the sub-sampled split LRS, i.e. we severely underestimate the dispersion, the type I error of the test with the (universal) critical value $1/\alpha$ is still below $\alpha$, for $\alpha=0.05, 0.1$. In general, we observe that misspecification of the dispersion parameter $\phi^*$ can affect both the type I error and the power of the sub-sampled split LRT, especially when the dispersion parameter is poorly estimated or when the Poisson deviance is used instead of the overdispersed Poisson deviance in settings where the overdispersion in the data is substantial. The impact is smaller in larger samples and in the case of larger miscalibration errors. 

We point out that we expect the dispersion parameter to be estimated accurately in large samples, making the impact of its misspecification negligible; we note that, in our sensitivity analysis, we use the values of $\phi$ quite different from the true value $\phi^*$ and independently of the size of the sample. In Appendix \ref{sec_app_2}, we show that our universal test remains asymptotically valid when the dispersion parameter is estimated consistently.

\subsubsection{\bl{Violation of assumptions (C) and (D)}}
\label{sec_examples_c_d}

Our statistical test relies on assumption (D) for testing mean-calibration and on assumption (C) for testing mean-specification. These assumptions may be difficult to verify in practice, but they are fundamental to the theoretical results presented in the paper. It is therefore important to study how the power of the sub-sampled split LRT is affected when assumption (C) or (D) is violated. 
  
In the setting introduced in Section \ref{sec_examples_data}, we violate the strict monotonic relationship between the mean estimates and the true mean values considered so far. Under the miscalibration scenarios introduced in Section \ref{sec_examples_data}, the mean estimates preserve the true ordering of the corresponding population means. We perturb these orderings by adding noise of varying magnitudes to the mean estimates shown in Figure \ref{fig_misspecification}. Recall the mean estimates have been defined as
\beqo
\widehat{\mu}_i^j = \bar{\mu} + \operatorname{slope}_j(\mu^*_i - \bar{\mu}), \qquad i = 1, \dots, n,
\eeqo
under miscalibration scenarios $j=1, 2, 3, 4$ with different values of $(\operatorname{slope}_j)_{j=1}^4$. We introduce a fixed perturbation function
\[
\Upsilon : [0.02, \, 0.25] \to \mathbb{R},
\]
constructed as follows:
\begin{enumerate}
\item Let \(a_k = 0.02 + (k-1)\frac{0.23}{199}\), \(k=1,\dots,200\), be an equally spaced grid.
\item We generate one set of independent standard normal variables \(\zeta_k \sim \mathcal{N}(0,1)\), for $k=1,\dots, 200$, and keep them fixed.
\item We define $\Upsilon$ as the linear interpolation of the points \((a_k, \zeta_k)\) on $[0.02,0.25]$, which is the range of the true mean values in the data set.
\end{enumerate}

\begin{figure}[htb!]
\begin{center}
\includegraphics[width=0.72\textwidth]{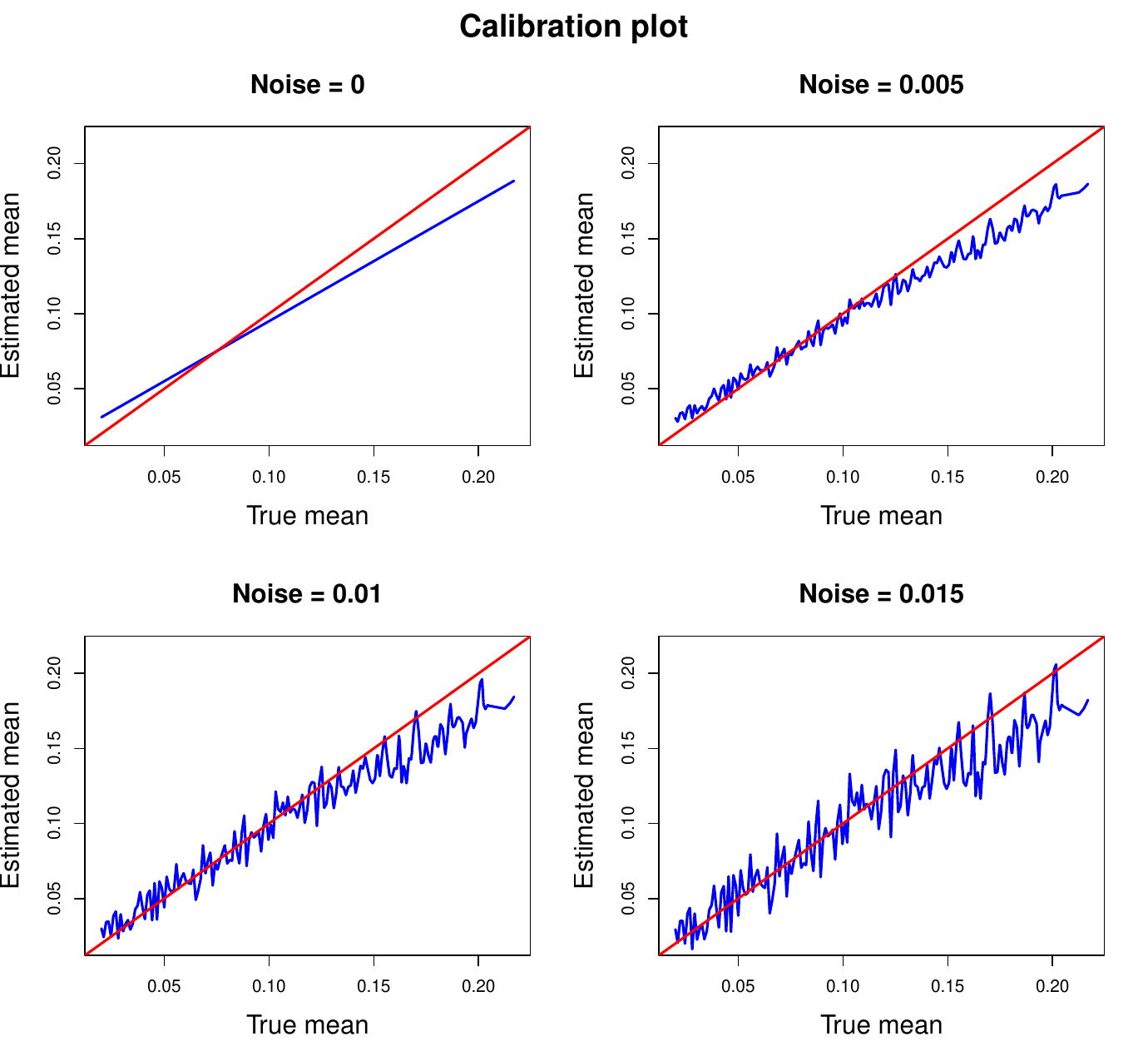}
\end{center}
\vspace{-.7cm}
\caption{Estimated means $\widehat{\mu}_i^{j,\sigma}$ vs.~true means $\mu^{*}_i$ in the miscalibration scenario with the slope equal to 0.8; the red line is the diagonal and the blue points indicate the misspecified relation under different sizes of the noise $\sigma=\{0, 0.5, 1, 1.5\}\times 10^{-2}$ that perturb the original strict monotonic relationship between the mean estimates and the true mean values.}
\label{fig_misspecification_2}
\end{figure}

We consider new four scenarios of miscalibration indexed by $j=1, 2, 3, 4$ and characterized by $\operatorname{slope}_j\in\{1,0.9,0.8,0.7\}$. In the previous examples both assumptions (C) and (D) were satisfied. Now, both (C) and (D) are violated due to the perturbation we introduce. The perturbed mean estimates for the miscalibration scenarios $j=1,2,3,4$ are given by the following formula
\beq\label{perturbated_mean_estimates}
\widehat{\mu}_i^{\,j,\sigma} \;=\; 
\widehat{\mu}_i^j \;+\; \sigma \, \Upsilon\!\bigl(\mu_i^*\bigr), \quad i=1,\dots, n.
\eeq
We consider the noise levels
\[
\sigma \in \bigl\{0,\; 0.005,\; 0.01, \; 0.015\bigr\},
\]
where \(\sigma = 0\) recovers the unperturbed case. The case $\sigma = 0.005$ is interpreted as a moderate perturbation, and $\sigma = 0.01, 0.015$ as strong perturbations. We do not consider mild perturbations, e.g., $\sigma=0.002$, as they do not have any impact on the results. For each miscalibration scenario $j = 1, 2, 3, 4$, we use the same perturbation function $\Upsilon$. The perturbed relationship for the miscalibration scenario $j = 3$ with $\operatorname{slope}_j=0.8$ between the resulting mean estimates and the true mean values is shown in Figure \ref{fig_misspecification_2}. Clearly, the (strict) monotonic relationship between the mean estimates and the true mean values is no longer preserved; in other words, the mean estimates do not preserve the ordering of the true conditional mean values as required by assumption (C) or (D). The larger the noise level $\sigma$ is in \eqref{perturbated_mean_estimates}, the more strongly the ordering of the true conditional mean values is perturbed.

\begin{figure}[htb!]
\begin{center}
\includegraphics[width=0.72\textwidth]{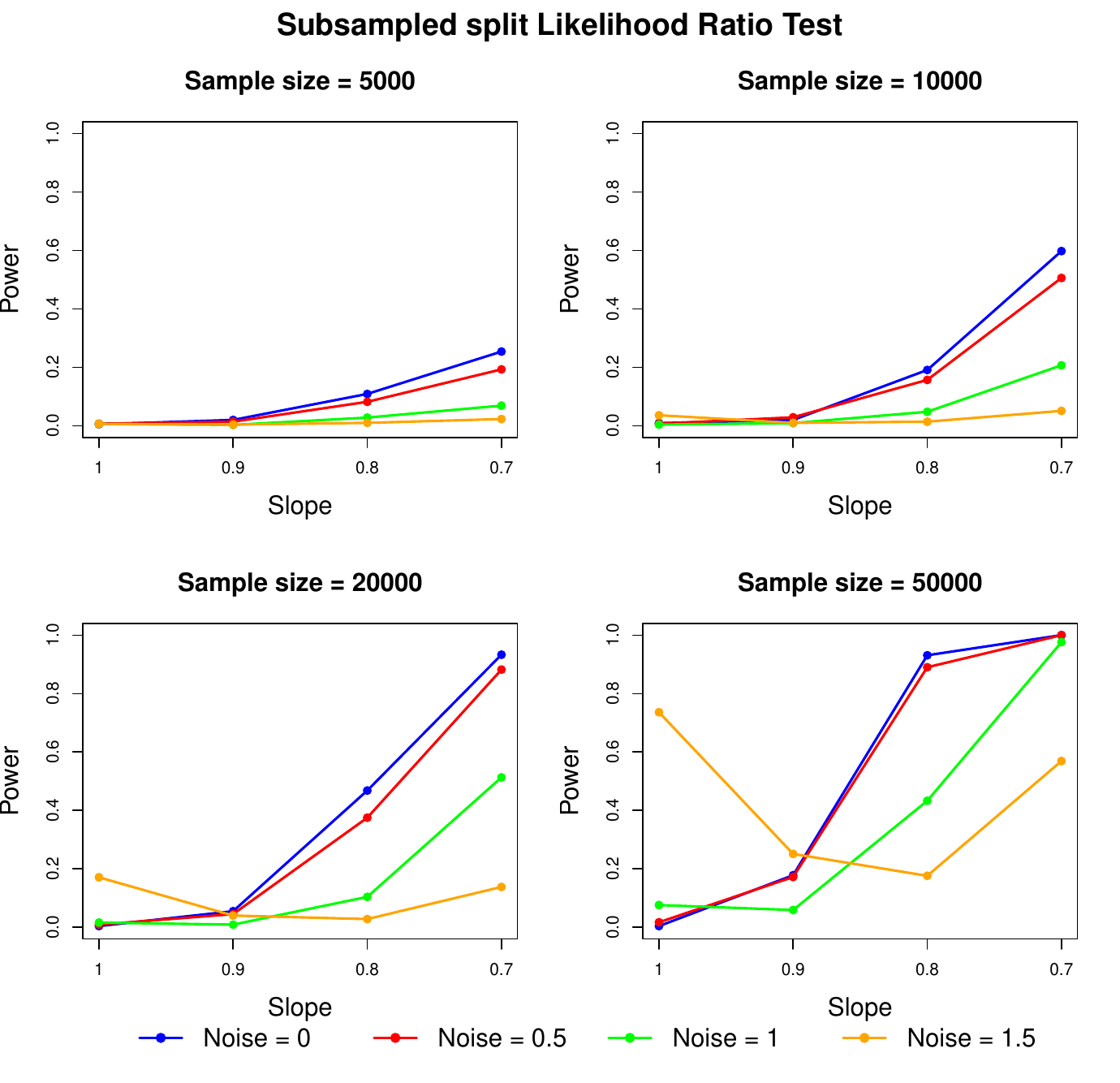}
\end{center}
\vspace{-.7cm}
\caption{Power of the sub-sampled split LRT with the (universal) critical value equal to $1/\alpha=20$ under different sizes of the noise $\sigma=\{0, 0.5, 1, 1.5\}\times 10^{-2}$ that perturb the original strict monotonic relationship between the mean estimates and the true mean values.}
\label{fig_misspecification_3}
\end{figure}

The results of our experiment are presented in Figure \ref{fig_misspecification_3}. The power of the sub-sampled split LRT decreases across all miscalibration scenarios $j=2,3,4$, with $\operatorname{slope}_j\in\{0.9,0.8,0.7\}$, as the noise level $\sigma$ increases. The only exception occurs for $\operatorname{slope}_2=0.9$ and $\sigma=0.015$, where the perturbation is sufficiently strong to increase the power of the test, as it induces greater miscalibration than that resulting from the original strictly monotonic relationship between the estimated and true mean values. Importantly, we conclude that the loss of power of the sub-sampled split LRT in detecting miscalibration remains negligible or small when the perturbation of the strictly monotonic relationship between the mean estimates and the true mean values is mild to moderate. The loss of power becomes more pronounced as the magnitude of the perturbation increases to strong levels, but this loss is mitigated in larger samples and in the presence of larger miscalibration sizes. As discussed in Section \ref{sec_universal_split}, within the framework of universal inference, our sub-sampled split LRT can be used to test mean-specification under (C) or (D) and mean-calibration under (D), with the two being equivalent under (D). If either (C) or (D) is violated, the isotonic estimate of the mean values under the alternative hypothesis of misspecified mean estimates may not be an optimal estimator. Consequently, we expect the power of the test to be low unless the miscalibration is sufficiently large to be detected by the isotonic regression and thereby leads to the rejection of the null hypothesis. Hence, although assumptions (C) and (D) are difficult to verify empirically, we argue that their violation is likely to have only a minor effect on the power of our sub-sampled split LRT in detecting miscalibration, provided that the regression function is estimated accurately enough to approximately satisfy (C) and (D), as is expected in large samples, or that the size of the miscalibration is sufficiently large.

We also note that $\operatorname{slope}_j=1$ previously corresponded to the original scenario $j=1$ of perfect calibration. In the present setting, however, $\operatorname{slope}_j=1$, combined with $\sigma\in\{0.5, 1, 1.5\}\times10^{-2}$, represents scenarios of miscalibration due to the added noise. In this case, we observe that the power of the sub-sampled split LRT increases with the noise level, which is consistent with our intuition. In general, if miscalibration is difficult to detect under assumption (D), substantial perturbations of the rankings that violate (D) may make miscalibration easier to detect. Conversely, if miscalibration is easily detectable under (D), such perturbations may hinder its detection.

\subsection{Universal inference: sub-sampled split L$q$RT}
\label{Modifications of the sub-sampled split LRT}

There are (at least) two known options to increase the power of the sub-sampled split LRT studied in the previous section. A first option is to stop the sub-sampling process once the sub-sampled split LRS exceeds the universal critical value $1/\alpha$ for the first time; see \citet[Section 5.3]{wang_book}. A second option is to introduce a randomization into the decision of rejecting the null hypothesis against the alternative; see  \citet[Section 2.4]{wang_book}. These modifications are feasible in our setting since the sub-sampled split LRS \eqref{def_sub_split_lrt} is an e-variable. We do not investigate the application of these two modifications because our focus is on understanding the performance of our novel test statistics, the split mean and maximal power L$q$RS \eqref{def_split_mean power_lrt} and \eqref{def_split_max power_lrt}, respectively, as well as their sub-sampled versions \eqref{def_sub_split_mean power_lrt} and \eqref{def_sub_split_max power_lrt}.

\subsubsection{Statistical power}

\begin{figure}[htb!]
\begin{center}
\includegraphics[width=0.72\textwidth]{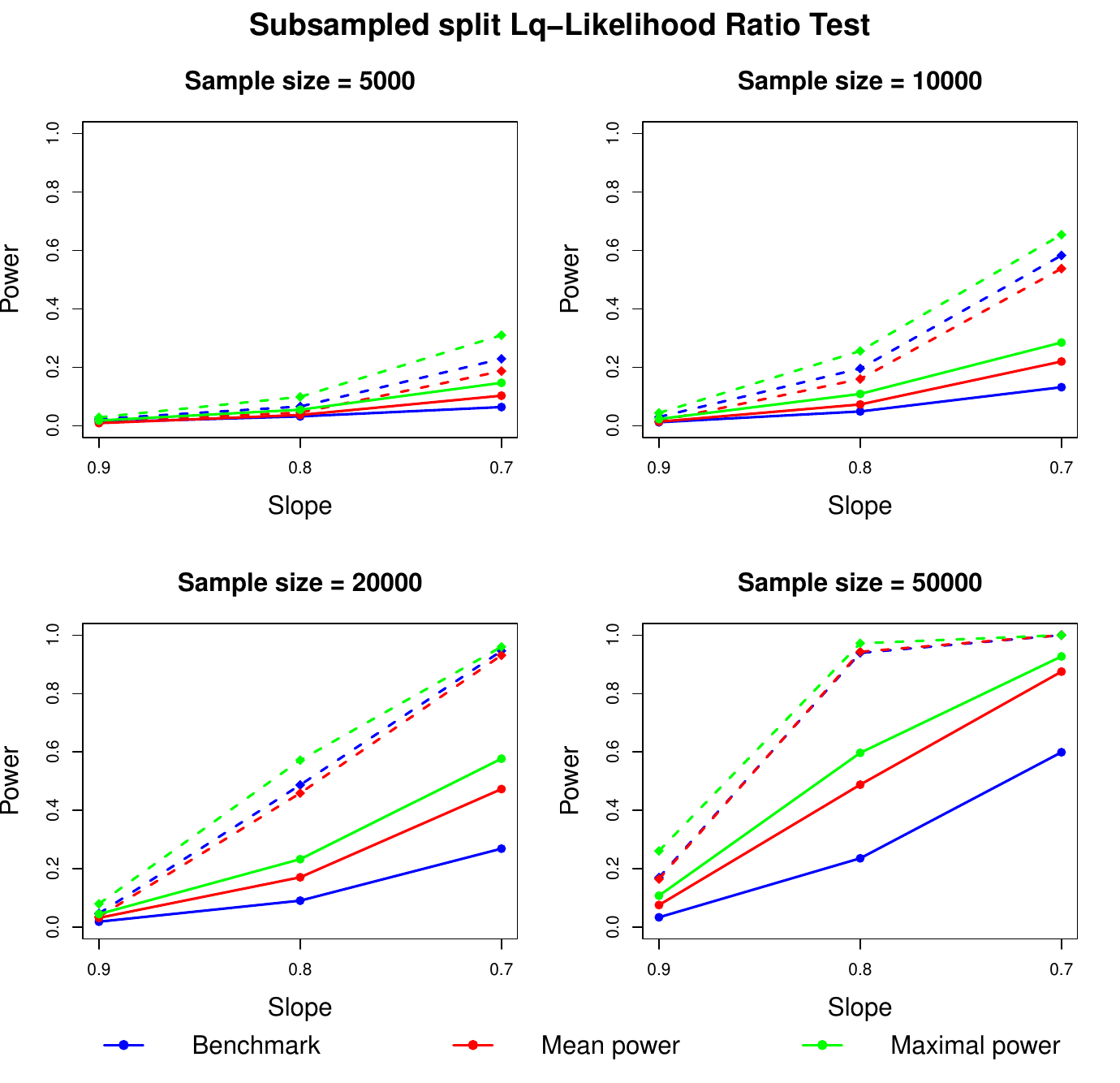}
\end{center}
\vspace{-.7cm}
\caption{Power of the (sub-sampled) split LRT (Benchmark), the (sub-sampled) split mean power L$q$RT and the (sub-sampled) split maximal power L$q$RT with (universal) critical value $1/\alpha=20$: the solid lines show the single sub-sample versions $B=1$, and the dashed lines the sub-sampled versions $B=1000$.}
\label{fig_LRT_best_power}
\end{figure}

The solid lines in Figure \ref{fig_LRT_best_power} show the single sub-sampled versions of the statistical tests with $B=1$: the benchmark is the split LRS $E^{LRT}$, and the split mean and maximal power L$q$RS show the results of the test statistics \eqref{def_split_mean power_lrt} and \eqref{def_split_max power_lrt}. We conclude that in any case these $L_q$-likelihood versions lead to an increased power compared to the benchmark;  ${\cal I}$ was selected by 10 uniformly distributed values $q$ across $(0,1)$. The results suggest that on medium and large sample sizes it is beneficial to work with the split mean power L$q$RT and the split maximal power L$q$RT to identify medium and large miscalibration, as in these cases the improvement in the power is the largest.

\begin{table}[ht]
{\small
  \centering
\begin{tabular}{rrrrr}
  \hline
Sample size & Statistics & Miscalibration=0.9 & Miscalibration=0.8 & Miscalibration=0.7 \\ 
  \hline
  10000 & Benchmark $B=1000$ & 0.03 & 0.22 & 0.61 \\ 
  10000 & Mean power $B=1000$ & 0.02 & 0.17 & 0.55 \\ 
  10000 & Benchmark $B=20$ & 0.02 & 0.17 & 0.54 \\ 
  10000 & Mean power $B=20$ & 0.01 & 0.14 & 0.53 \\ 
  \hline
  20000 & Benchmark $B=1000$ & 0.06 & 0.49 & 0.94 \\ 
  20000 & Mean power $B=1000$ & 0.05 & 0.44 & 0.93 \\ 
  20000 & Benchmark $B=20$ & 0.05 & 0.40 & 0.90 \\ 
  20000 & Mean power $B=20$ & 0.05 & 0.41 & 0.92 \\
  \hline 
  50000 & Benchmark $B=1000$ & 0.21 & 0.96 & 1.00 \\ 
  50000 & Mean power $B=1000$ & 0.20 & 0.96 & 1.00 \\ 
  50000 & Benchmark $B=20$ & 0.14 & 0.89 & 1.00 \\ 
  50000 & Mean power $B=20$ & 0.16 & 0.94 & 1.00 \\ 
   \hline
\end{tabular}
\caption{Power of the sub-sampled split LRTs (Benchmark) based on 1000 and 20 sub-samples and the sub-sampled split mean power L$q$RTs based on 1000 and 20 sub-samples with (universal) critical value $1/\alpha=20$.}
\label{tab_LRT_best_power 2}
}
\end{table}

In Figure \ref{fig_LRT_best_power}, we also look at the sub-sampled versions (dashed lines) of the statistical tests with $B=1000$ sub-samples: the benchmark is the sub-sampled split LRS $\overline{E}_B^{LRT}$, and the sub-sampled split mean and maximal power L$q$RS show the results of the test statistics \eqref{def_sub_split_mean power_lrt} and \eqref{def_sub_split_max power_lrt}. We observe that the sub-sampled split maximal power L$q$RT has the largest power for all sample sizes and all miscalibration sizes. Unfortunately, this test statistics cannot be supported because it lacks the formal type I error bound, although our numerical results verify that the type I error bound holds in our example, and the true type I error for the test with the critical value of $20$ corresponding to the pre-assumed significance level $\alpha=0.05$ fluctuates around $0.5\%$ for all sample sizes considered (similar as for the sub-sampled split LRT, see Figure \ref{fig_LRT_type_1}). Therefore, we shall only focus on the sub-sampled split mean power L$q$RT. The true type I error for this test with the critical value of $20$ fluctuates around $0.2\%$ (lower than for the sub-sampled split LRT). It is interesting to note that if we sub-sample $B=1000$ training and validation sets, the sub-sampled split LRT has now a slightly larger power than the sub-sampled split mean power L$q$RT for all considered sample sizes and miscalibration errors. Hence, the ranking of these two tests based on their power has changed if we switch from $B=1$ to $B=1000$, see also discussion in Appendix \ref{sec_app_3}. Table \ref{tab_LRT_best_power 2} shows that if we sample a lower number of partitions, e.g., if we set $B=20$, the sub-sampled split mean power L$q$RT receives a slight advantage over the sub-sampled split LRT in terms of power. Moreover, their powers, despite using a much lower number of sub-samples, are at a similar level as the power of the sub-sampled split LRT based on $B=1000$ sub-samples. From Figure \ref{fig_splitLRT_trials} we know that $B=1000$ is not needed to have a sound statistical power of a sub-sampled split test, and in practical applications we have recommended to use a lower $B$ with a small effect on the power of the test. We conclude that the sub-sampled split mean power L$q$RT may be preferred over the sub-sampled split LRT in some applications, especially if we choose a low $B$.

\section{Conclusions}

In this paper, we investigated the framework of universal inference to \bl{test mean-calibration and mean-specification} within the exponential dispersion family (EDF). We developed a sub-sampled split Likelihood Ratio Test (LRT) and its modifications based on the $L_q$-Likelihood Ratio Test (L$q$RT). In a case study, we confirmed that the sub-sampled split LRT and its modifications are appealing alternatives to the classical LRT. We identified the beneficial impact of sub-sampling of training and validation sets on the split LRT, and we demonstrated that despite of a low type I error, the sub-sampled split LRT has a high and reasonable power on medium and large samples in the case of medium and large size miscalibration/mispecification errors. \bl{Since large samples are available for many non-life insurance pricing problems, we believe that our universal inference approach has useful applications in actuarial practice, providing reliable results and detecting larger degrees of miscalibration and misspecification in mean estimates.} In addition, our new so-called sub-sampled split mean power L$q$RT may perform better in some scenarios than the sub-sampled split LRT in terms of the power of rejecting the false null hypothesis against the true alternative for mean estimates. \bl{Finally, we demonstrated that the sub-sampled split LRT is robust to small and moderate misspecifications of the assumptions underlying the test in medium and large samples in the case of medium and large miscalibration/mispecification errors. In particular, we considered deviations from the assumptions of a correctly estimated dispersion parameter and of the correct ordering of the true mean values based on the mean estimates.}

\bl{While our case study focused on Poisson responses and claim counts, we also considered gamma responses and claim severities, obtaining similar conclusions. As expected, the power of the tests deteriorates as the dispersion coefficient of the gamma responses increases, making the validation of mean estimates challenging in highly skewed and dispersed settings, i.e., at lower signal-to-noise ratios.}

Based on our findings, we believe that universal inference offers a practical and powerful tool for testing mean-calibration and mean-specification, in particular, for actuarial applications. 
  
\setstretch{1}

\newpage

\setstretch{1.1}

\begin{appendices}

\section{\bl{Verification of assumptions (C) and (D)}}
\label{sec_app_1}

We recall our assumptions (C) and (D):
\begin{itemize}
\item[(C)] There exists a non-decreasing mapping $\mathcal{T}:\br\rightarrow\br$ such that
\beqo
\mathbb{E}\big[Y \mid \widehat{\mu}(\mathbf{X})\big]=\mathcal{T}(\widehat{\mu}(\mathbf{X})), \quad \text{almost surely.}
\eeqo
\item[(D)] There exists a strictly increasing mapping $\mathcal{S}:\br\rightarrow\br$ such that
\beqo
\mathbb{E}\big[Y \mid \mathbf{X}\big]=\mathcal{S}(\widehat{\mu}(\mathbf{X})), \quad \text{almost  surely.}
\eeqo
\end{itemize}

We first discuss (C). Let $\mathcal{V}$ denote an independent validation set. Assumption (C) can be assessed, for example, by fitting a smooth regression to $(y_i,\widehat{\mu}(\bx_i),v_i)_{i\in\mathcal{V}}$ using a reasonably calibrated smoothing parameter. If the fitted curve is not approximately non-decreasing, with clear hills and valleys, the ordering induced by $\widehat{\mu}$ for $\be[Y|\widehat{\mu}]$ may not be correct. As a second approach, we can fit an isotonic regression to $(y_i,\widehat{\mu}(\bx_i),v_i)_{i\in\mathcal{V}}$. If the fitted isotonic regression exhibits substantial flat regions over ranges of $(\widehat{\mu}(\bx_i))_{i\in\mathcal{V}}$, the ordering induced by $\widehat{\mu}$ for $\be[Y|\widehat{\mu}]$ may not be correct. A more statistically oriented analysis would use a test for the resulting Gini score; see \cite{BrauerMenzel}. The assessment may be challenging in cases with a low signal-to-noise ratio. 

We now discuss (D). Since $\mu^*$ is not available, one possible way to assess (D) is to fit an alternative regression function $\widetilde{\mu}$ and compare the rankings induced by $\widehat{\mu}$ and $\widetilde{\mu}$. If the rank correlation between $\widehat{\mu}$ and $\widetilde{\mu}$ is close to one, for a range of potentially challenging alternative models $\widetilde{\mu}$, we may regard this as empirical evidence in support of (D).

We now present a theoretical justification of (D) in large samples. Let us consider a learning sample $\mathcal{L}$ with i.i.d. observations $(Y_i,\mathbf{X}_i,V_i)_{i=1}^m$ sampled from the distribution $F_{(Y,\mathbf{X},V)}$. In the proposition below, in contrast to the results in the main text, the learning sample $\mathcal{L}$ is random. Hence, $\widehat{\mu}_m:\mathcal{X}\times\Omega\mapsto\br$ is treated as a random function since it depends on the observations $(Y_i,\mathbf{X}_i,V_i)_{i=1}^m$ from the random learning sample $\mathcal{L}$. We show that, under suitable conditions, the probability that the estimator $\widehat{\mu}_m$ based on $\mathcal{L}$ reverses the true ordering induced by $\mu^*$ for a pair $(\mathbf{X}_1,\mathbf{X}_2)$ from a test set $\mathcal{D}$, such that $\mathbf{X}_1\neq\mathbf{X}_2$, converges to zero as $m\to\infty$. Clearly, our proposition does not establish (D) directly, but it provides a motivation for this assumption by showing that it is asymptotically justified when the estimation procedure is reasonable (uniformly consistent). 

Formally, we extend the probability space $(\Omega, \mathcal{F},\bp)$ to cover both $(Y_i,\mathbf{X}_i,V_i)_{i=1}^m$ and $(Y_i,\mathbf{X}_i,V_i)_{i=1}^n$ from the learning $\mathcal{L}$ and the test sample $\mathcal{D}$.

\begin{prop}
Let $(Y_i, \mathbf{X}_i,V_i)_{i=1}^m$ denote the i.i.d. observations in the learning set $\mathcal{L}$ and let $\mathbf{X}_1$ and $\mathbf{X}_2$ be independent copies of $\mathbf{X}$, and independent of $\mathcal{L}$. The regression function $\widehat{\mu}_m$ is estimated with $\mathcal{L}$. Assume that:
\begin{itemize}
    \item[(P1)] the estimator $\widehat{\mu}_m$  is uniformly consistent in the sense that
    \beqo
    \Delta_m = \sup_{\bx\in\mathcal{X}}\big|\widehat{\mu}_m(\bx)  - \mu^*(\bx)\big| \xrightarrow{p} 0,
    \eeqo    
    \item[(P2)] the true mean value function $\mu^*:\mathcal{X}\mapsto \br$ is injective in the sense that: let $A=\big\{\omega\in\Omega : \mathbf{X}_1(\omega)\neq \mathbf{X}_2(\omega)\big\}$, then
    \beqo
    \mu^*(\mathbf{X}_1(\omega)) \neq \mu^*(\mathbf{X}_2(\omega)), \quad \text{for} \ \bp\text{-almost all} \ \omega\in A.
    \eeqo
\end{itemize}
Let 
\beqo
R_m=\Big\{\omega\in\Omega : \big(\mu^*(\mathbf{X}_1(\omega)) - \mu^*(\mathbf{X}_2(\omega))\big)\big(\widehat{\mu}_m(\mathbf{X}_1(\omega)) - \widehat{\mu}_m(\mathbf{X}_2(\omega))\big)\leq 0 \Big\}.
\eeqo
We have
\beqo
\lim_{m\rightarrow \infty}\bp\big(R_m \mid \mathbf{X}_1(\omega), \mathbf{X}_2(\omega)\big) =  0 \quad \text{for} \ \bp\text{-almost all} \ \omega\in A.
\eeqo
\end{prop}

\noindent \textbf{Proof:} We fix a pair $\mathbf{X}_1=\bx_1, \mathbf{X}_2=\bx_2$ such $\bx_1\neq \bx_2$ and $\mu^*(\bx_1) \neq \mu^*(\bx_2)$, which holds for almost every such pair by (P2). We have $\mu^*(\bx_1) < \mu^*(\bx_2)$ or $\mu^*(\bx_1) > \mu^*(\bx_2)$, and we assume the former. For a fixed pair $(\bx_1,\bx_2)$ satisfying
$\bx_1\neq\bx_2$ and
$\mu^*(\bx_1)\neq\mu^*(\bx_2)$, consider the event $R_m$ conditional on
$\mathbf{X}_1=\bx_1$ and $\mathbf{X}_2=\bx_2$. In the case
$\mu^*(\bx_1)<\mu^*(\bx_2)$, this event is equivalent to
\beqo
R_m(\bx_1,\bx_2) = \big\{\omega\in\Omega : \widehat{\mu}_m(\bx_1) \geq \widehat{\mu}_m(\bx_2) \big\}.
\eeqo
On the event $R_m(\bx_1,\bx_2)$, we derive
\beqo
0<\mu^*(\bx_2) - \mu^*(\bx_1) &=& \big(\mu^*(\bx_2) - \widehat{\mu}_m(\bx_2)\big) + \big(\widehat{\mu}_m(\bx_2) - \widehat{\mu}_m(\bx_1)\big) + \big(\widehat{\mu}_m(\bx_1) - \mu^*(\bx_1)\big)\\
&\leq& \big(\mu^*(\bx_2) - \widehat{\mu}_m(\bx_2)\big) + \big(\widehat{\mu}_m(\bx_1) - \mu^*(\bx_1)\big)\\
&\leq& \big|\mu^*(\bx_2) - \widehat{\mu}_m(\bx_2)\big| + \big|\widehat{\mu}_m(\bx_1) - \mu^*(\bx_1)\big|\leq 2\Delta_m.
\eeqo
We deduce that
\beqo
R_m(\bx_1, \bx_2) \subseteq \Big\{\omega\in\Omega : \mu^*(\bx_2) - \mu^*(\bx_1) ~\leq~ 2\Delta_m \Big\}.
\eeqo
Consequently, we have the inequality
\beqo
\bp\big(R_m(\bx_1, \bx_2) \mid \mathbf{X}_1=\bx_1,\mathbf{X}_2=\bx_2\big) \leq \bp\Big(\Delta_m \geq \frac{\mu^*(\bx_2) - \mu^*(\bx_1)}{2} \mid \mathbf{X}_1=\bx_1,\mathbf{X}_2=\bx_2\Big).
\eeqo
Since $\Delta_m$ is a function of the learning set $\mathcal{L}$ and the learning sample $\mathcal{L}$ is independent of $\mathbf{X}_1,\mathbf{X}_2$, we have
\beqo
\bp\Big(\Delta_m \geq \frac{\mu^*(\bx_2) - \mu^*(\bx_1)}{2} \mid \mathbf{X}_1=\bx_1,\mathbf{X}_2=\bx_2\Big)=
\bp\Big(\Delta_m \geq \frac{\mu^*(\bx_2) - \mu^*(\bx_1)}{2}\Big).
\eeqo
The probability $\bp(\Delta_m \geq c)$ goes to zero for any $c>0$ as $m\rightarrow\infty$ by (P1). Notice that $c=(\mu^*(\bx_2) - \mu^*(\bx_1))/2>0$ by (P2).\qed

\medskip

We also refer to Appendix \ref{sec_app_4} where we discuss how to verify (C) and (D) in a practical real life example.

\section{\bl{LRT and split LRT with estimated dispersion parameter}}
\label{sec_app_2}

The classical LRT can be easily modified by plugging in an estimate of the dispersion parameter, $\widehat{\phi}$, into the statistics defined in \eqref{def_lrt}. As discussed in Section \ref{sec_classic_lrt}, the distribution of the classical LRS under the null hypothesis is not specified and must therefore be determined by simulation, both when the dispersion parameter is known, $\phi^*$, and when it is estimated by $\widehat{\phi}$.

We now investigate the universal LRT in the case where the dispersion parameter $\phi^*$ is unknown and an estimate $\widehat{\phi}$ of $\phi^*$ is incorporated into the split LRS. Let $\mathcal{D}$ denote the test sample of size $n=|\mathcal{D}|$, and randomly split $\mathcal{D}$ into independent samples $\mathcal{D}_0$ and $\mathcal{D}_1$ of sizes $n_0=[ns]$ and $n_1=n-[ns]$, respectively, where $s$ denotes a split ratio. Recall that the isotonic regression under the alternative hypothesis is estimated using $\mathcal{D}_1$, while the null hypothesis is validated on $\mathcal{D}_0$. We use a consistent estimator $\widehat{\phi}$ of $\phi^*$  within the assumed member of the EDF based on the test sample $\mathcal{D}_1$.

We extend the split LRS \eqref{def_split_lrt} as follows
\beq\label{def_split_lrt_phi}
E^{sLRT}_{\widehat{\phi}_{n}}
&=&\prod_{i\in\mathcal{D}_0}\exp\left(\frac{Y_i\,(\widehat{\Xi}_{i}-\widehat{\Theta}_{i})-(\kappa(\widehat{\Xi}_{i})-\kappa(\widehat{\Theta}_i))}{\widehat{\phi}_{n}/V_i}\right).
\eeq
We can observe that
\beq\label{relation_phi_known_estimated}
\Big(E^{sLRT}_{\widehat{\phi}_{n}}\Big)^{\frac{\widehat{\phi}_n}{\phi^*}}=E^{sLRT},
\eeq
where $E^{sLRT}$ is the split LRS with the known dispersion parameter given by \eqref{def_split_lrt}. We use the notation $\widehat{\phi}_{n}$ to emphasize that our analysis focuses on its limiting behavior as $n\rightarrow\infty$. Other terms also depend on $n$, but we suppress their subscripts for notational simplicity.

\begin{thm}\label{splitLRT_EDF_phi}
Let (A)-(B) and (E) hold. Consider the null hypothesis $\mathcal{H}_0$ against the alternative $\mathcal{H}_1$. The null hypothesis \eqref{h_0_edf} is characterized with measures $\bp\in\mathcal{P}$, and the alternative hypothesis \eqref{h_1_edf} is characterized with measures $\bq\in\mathcal{Q}$. Assume (D) for the statistical test of mean-calibration or (C) for the statistical test of mean-specification. Moreover, assume that
\begin{itemize}
\item[(Q1)] the dispersion parameter $\widehat{\phi}_{n}$ is estimated using $\mathcal{D}_1$, and $\widehat{\phi}_{n}\geq \zeta>0, \ a.s.$ for any $n\geq 1$, 
\item[(Q2)] the sequence $\big(\widehat{\phi}_{n}\big)_{n\geq 1}$ converges in probability to $\phi^*>0$ as $n\rightarrow\infty$ under any $\bp\in\mathcal{P}$,
\item [(Q3)] the set $\mathcal{P}$ of measures is finite.
\end{itemize}
Reject the null $\mathcal{H}_0$ at a pre-assumed significance level $\alpha\in(0,1)$ if the split LRS \eqref{def_split_lrt_phi} exceeds the value $1/\alpha$. The split LRT controls,  asymptotically as $n\rightarrow\infty$, the type I error of rejecting $\mathcal{H}_0$ in favor of $\mathcal{H}_1$ at significance level $\alpha \in (0,1)$, in the sense that
\beqo
\limsup_{n\rightarrow\infty}\sup_{\bp\in\mathcal{P}}\bp\left(E^{sLRT}_{\widehat{\phi}_{n}} \geq 1/\alpha\right) \leq \alpha.
\eeqo 
The assertion also holds for the sub-sampled version of the split LRS \eqref{def_split_lrt_phi}.
\end{thm}
\noindent \textbf{Proof:} Select $\bp\in\mathcal{P}$ from the null hypothesis $\mathcal{H}_0$ set given by \eqref{h_0_edf}. We start as in the proof of Theorem \ref{splitLRT_EDF}. We choose $\alpha\in(0,1)$. For fixed $n$, we compute
\beq\label{proof_split_lrt_phi}
\bp\left(E^{sLRT}_{\widehat{\phi}_{n}}\geq 1/\alpha\right)&=&\be_\bp\left[\bp\left(E^{sLRT}_{\widehat{\phi}_{n}}\geq 1/\alpha\;\middle|\;\mathcal{D}_1,\left(\mathbf{X}_i,\widehat{\Theta}_i, V_i\right)_{i\in\mathcal{D}_0}\right)\right]\nonumber\\
&=&\be_\bp\left[\bp\left(\Big(E^{sLRT}_{\widehat{\phi}_{n}}\Big)^{\frac{\widehat{\phi}_n}{\phi^*}}\geq\Big(1/\alpha\Big)^{\frac{\widehat{\phi}_n}{\phi^*}}\;\middle|\;\mathcal{D}_1,\left(\mathbf{X}_i,\widehat{\Theta}_i, V_i\right)_{i\in\mathcal{D}_0}\right)\right]\nonumber\\
&=&\be_\bp\left[\bp\left(E^{sLRT}\geq\Big(1/\alpha\Big)^{\frac{\widehat{\phi}_n}{\phi^*}}\;\middle|\;\mathcal{D}_1,\left(\mathbf{X}_i,\widehat{\Theta}_i, V_i\right)_{i\in\mathcal{D}_0}\right)\right],
\eeq
where we use property \eqref{relation_phi_known_estimated}. Given $\mathcal{D}_1,(\mathbf{X}_i,\widehat{\Theta}_i, V_i)_{i\in\mathcal{D}_0}$, the variables $\big(\widehat{\Xi}_{i},\widehat{\Theta}_i, V_i, \widehat{\phi}_{n}\big)_{i\in\mathcal{D}_0}$ are fixed, and the responses $(Y_i)_{i\in\mathcal{D}_0}$ are conditionally independent. In particular, we have used (Q1) to conclude that $\widehat{\phi}_{n}$ is fixed conditional on $\mathcal{D}_1,(\mathbf{X}_i,\widehat{\Theta}_i, V_i)_{i\in\mathcal{D}_0}$. The moment generating function of the conditional distribution of $Y_i|\mathcal{D}_1,\mathbf{X}_i,\widehat{\Theta}_i,V_i$ under $\mathcal{H}_0$ is that of the assumed member of the EDF with canonical parameter $\widehat{\Theta}_i$ and dispersion $\phi^*$, 
conditional on fixed $\mathcal{D}_1,\mathbf{X}_i,\widehat{\Theta}_i,V_i$. Using the proof of Theorem \ref{splitLRT_EDF}, we deduce that
\beqo
\be_\bp\left[E^{sLRT}\;\middle|\;\mathcal{D}_1,\left(\mathbf{X}_i,\widehat{\Theta}_i, V_i\right)_{i\in\mathcal{D}_0}\right]=1,
\eeqo
and, by Markov's inequality, it follows that
\beq\label{markov_split_lrt_phi}
\bp\left(E^{sLRT}\geq\Big(1/\alpha\Big)^{\frac{\widehat{\phi}_n}{\phi^*}}\;\middle|\;\mathcal{D}_1,\left(\mathbf{X}_i,\widehat{\Theta}_i, V_i\right)_{i\in\mathcal{D}_0}\right)\leq \alpha^{\frac{\widehat{\phi}_n}{\phi^*}},
\eeq
for $\bp$-a.a. $\mathcal{D}_1,\left(\mathbf{X}_i,\widehat{\Theta}_i, V_i\right)_{i\in\mathcal{D}_0}$. Substituting \eqref{markov_split_lrt_phi} into \eqref{proof_split_lrt_phi}, we show that
\beq\label{proof_split_lrt_phi_2}
\bp\left(E^{sLRT}_{\widehat{\phi}_{n}}\geq 1/\alpha\right)\leq\be_\bp\left[\alpha^{\frac{\widehat{\phi}_n}{\phi^*}}\right].
\eeq
The inequality \eqref{proof_split_lrt_phi_2} holds for any $n\geq 1$. We now take the limit $n\rightarrow\infty$ in \eqref{proof_split_lrt_phi_2}. By (Q2) and the convergence in probability, we deduce that
\beqo
\frac{\widehat{\phi}_n}{\phi^*}\xrightarrow{p} 1, \quad \text{as} \ n\rightarrow\infty,
\eeqo
under $\bp$, and we conclude that
\beqo
\limsup_{n\rightarrow\infty}\bp\left(E^{sLRT}_{\widehat{\phi}_{n}}\geq 1/\alpha\right)\leq \alpha.
\eeqo
Finally, by (Q3) we prove the main result
\beqo
\limsup_{n\rightarrow\infty}\sup_{\bp\in\mathcal{P}}\bp\left(E^{sLRT}_{\widehat{\phi}_{n}} \geq 1/\alpha\right)&=&\limsup_{n\rightarrow\infty}\max_{\bp\in\mathcal{P}}\bp\left(E^{sLRT}_{\widehat{\phi}_{n}} \geq 1/\alpha\right)\\
&\leq&\max_{\bp\in\mathcal{P}}\limsup_{n\rightarrow\infty}\bp\left(E^{sLRT}_{\widehat{\phi}_{n}} \geq 1/\alpha\right)\leq \alpha.
\eeqo
\qed

\medskip

When the dispersion parameter must be estimated, we can also consider an alternative approach. We introduce the split LRS given with
\beq\label{def_split_lrt_phi_2}
\widetilde{E}^{sLRT}_{\widehat{\phi}_{n}}
=\exp\left(\sum_{i\in\mathcal{D}_0}\log\big(Y_i, \widehat{\Xi}_{i}, \widehat{\phi}_n\big)-\sup_{\phi>0}\Big\{\sum_{i\in\mathcal{D}_0}\log\big(Y_i, \widehat{\Theta}_{i}, \phi\big)\Big\}\right),
\eeq
where $\widehat{\phi}_n$ is estimated using $\mathcal{D}_1$. Since \eqref{def_split_lrt_phi_2} is the ratio of two likelihood functions, the results on universal inference presented in Section \ref{sec_universal_intro} imply that, for any fixed sample size $n$, we have
\beqo
\be_\bp\Big[\widetilde{E}^{sLRT}_{\widehat{\phi}_{n}}\Big]\leq 1,
\eeqo
for any $\bp\in\mathcal{P}$. Let us reject the null hypothesis $\mathcal{H}_0$ at a pre-assumed significance level $\alpha\in(0,1)$ if the split LRT \eqref{def_split_lrt_phi_2} exceeds $1/\alpha$. The split LRT \eqref{def_split_lrt_phi_2} controls the type I error of rejecting $\mathcal{H}_0$ in favor of $\mathcal{H}_1$ at significance level $\alpha\in(0,1)$.

\section{Additional analyses of the power of the split L$q$RT}
\label{sec_app_3}

The optimal power parameter $q$ for the split L$q$RT is given by \eqref{optimal_L$q$RT}. In our numerical example, we can calculate $\be_{\bq_\Pi}[\cdot]$ by replacing $\Pi_i$ with $\log(\mu_i^*)$. The optimal $q^*$ is then found by optimizing \eqref{optimal_L$q$RT} over a finite set $\mathcal{I}\subset(0,1]$. In practice, we have to estimate this parameter from data.

\noindent\hrulefill

{\sc Split L$q$RT with estimated power parameter}

\noindent\hrulefill

\begin{itemize}
\item We modify the algorithm for the split LRT:
In addition to the split ratio $s\in(0,1)$, choose a fixed split ratio $r\in(0,1)$.
\item Construct a sub-sample $\mathcal{D}_{0,0}$ by randomly selecting $[[ns]r]$ triples from $(Y_i,\widehat{\Theta}_i, V_i)_{i\in\mathcal{D}_0}$ without replacement, and set $\mathcal{D}_{0,1} = \mathcal{D}_0\setminus \mathcal{D}_{0,0}$.
\item Find $\widehat{q}^*\in\mathcal{I}$ with the rule
\beqo
\widehat{q}^*=\underset{q\in \mathcal{I}}{\arg\max}\sum_{i\in\mathcal{D}_{0,1}}\log\big(E^q_{\widehat{\Xi}_i}\big(Y_i, \widehat{\Theta}_i,V_i\big)\big).
\eeqo
\item Define the split L$q$RS for testing $\mathcal{H}_0$ against $\mathcal{H}_1$
\begin{equation}\label{def_split_lrt_opt}
E^{sLRT,\widehat{q}^*}=\prod_{i\in\mathcal{D}_{0,0}} E^{\widehat{q}^*}_{\widehat{\Xi}_i}\big(Y_i,\widehat{\Theta}_i, V_i\big).
\end{equation}
\end{itemize}
\noindent\hrulefill

\medskip
\noindent
One can show that both $E^{sLRT,q^*}$ and $E^{sLRT,\widehat{q}^*}$ are exact e-variables, this is similar to Theorem \ref{splitLRT_EDF}.

\begin{figure}[htb!]
\begin{center}
\includegraphics[width=.98\textwidth]{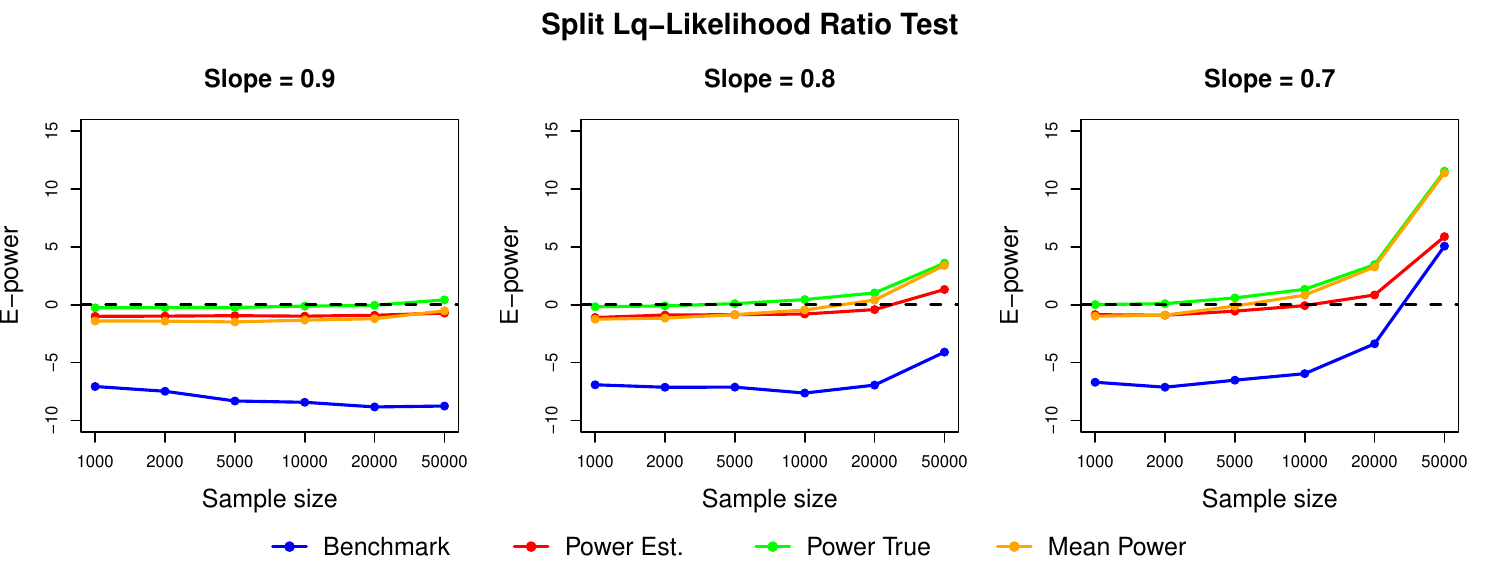}
\end{center}
\vspace{-.72cm}
\caption{E-power of the split LRT (Benchmark), the split L$q$RT with the estimated optimal power parameter (Power Est.), the split L$q$RT with the true optimal power parameter (Power True), and the split mean power L$q$RT with (universal) critical value $1/\alpha=20$ for $B=1$.}
\label{fig_split_LRT_growth_v2}
\end{figure}

\begin{figure}[htb!]
\begin{center}
\includegraphics[width=.98\textwidth]{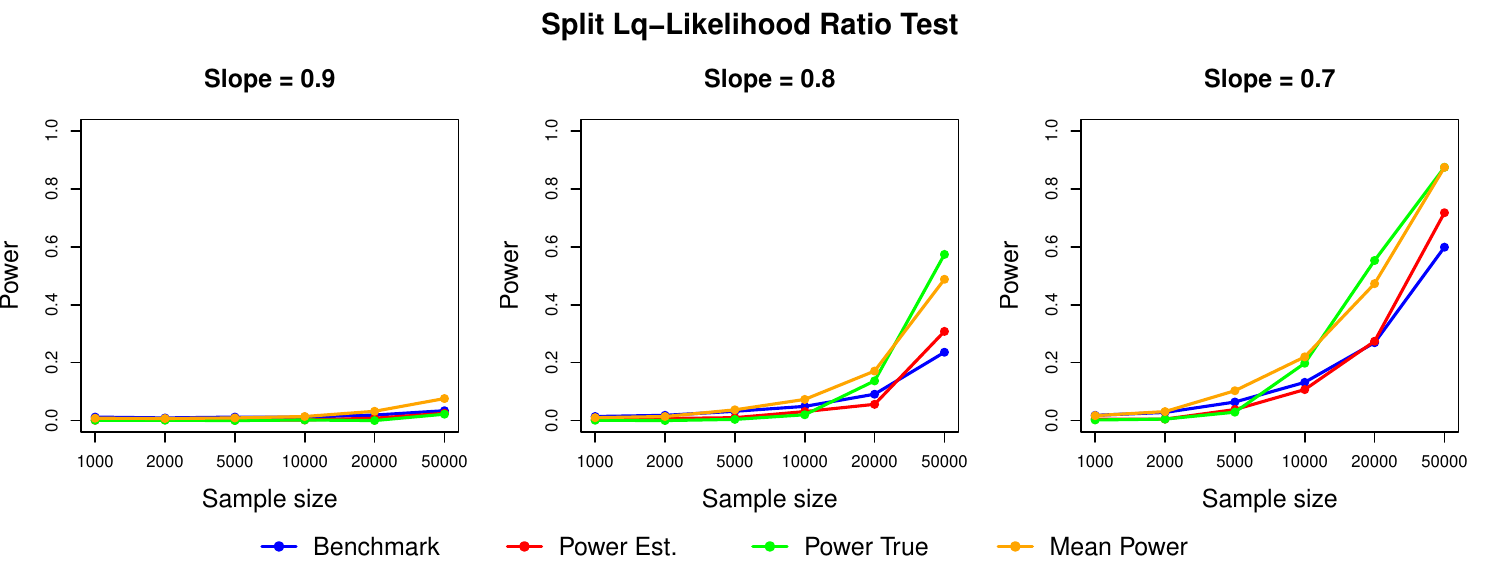}
\end{center}
\vspace{-.7cm}
\caption{Power of the split LRT (Benchmark), the split L$q$RT with the estimated optimal power parameter (Power Est.), the split L$q$RT with the true optimal power parameter (Power True), and the split mean power L$q$RT with (universal) critical value $1/\alpha=20$  for $B=1$.}
\label{fig_split_LRT_power_v2}
\end{figure}

Figure \ref{fig_split_LRT_growth_v2} illustrates the conclusions from Theorems \ref{thm_power_1}-\ref{thm_power_2} and investigates the e-power of the various split test statistics for $B=1$: the benchmark is the split LRS, the split L$q$RS with the estimated optimal power parameter and the true optimal power parameter are given by the split L$q$RS \eqref{def_split_power_lrt} with, respectively, $\widehat{q}^*$ and $q^*$, and the split mean power L$q$RS shows the results of the test statistics \eqref{def_split_mean power_lrt}. The split L$q$RT with $q^*$ achieves the largest e-power in all cases, followed by the split mean power L$q$RT and the split L$q$RT with $\widehat{q}^*$, whereas the split LRT has a significantly lower e-power in all cases. In Figure \ref{fig_split_LRT_power_v2} we look at the statistical power of the test statistics. For a single data split, in the case of medium/large miscalibration sizes in medium/large sample sizes, we see clear advantages of applying the split L$q$RT with $q^*$. A similarly good performance is also observed for the split mean power L$q$RT. In large samples, it is also beneficial to apply the split L$q$RT with $\widehat{q}^*$ since the estimated optimal power parameter becomes close to the true optimal power parameter $q^*$. The split LRT falls behind the other test statistics in all cases. These conclusions have also been confirmed in larger sample sizes of 70000 and 100000; the gap between the red and the blue curve increases and the red curve approaches the green and orange curves as the sample size increases for all miscalibration sizes.

\begin{figure}[htb!]
\begin{center}
\includegraphics[width=.98\textwidth]{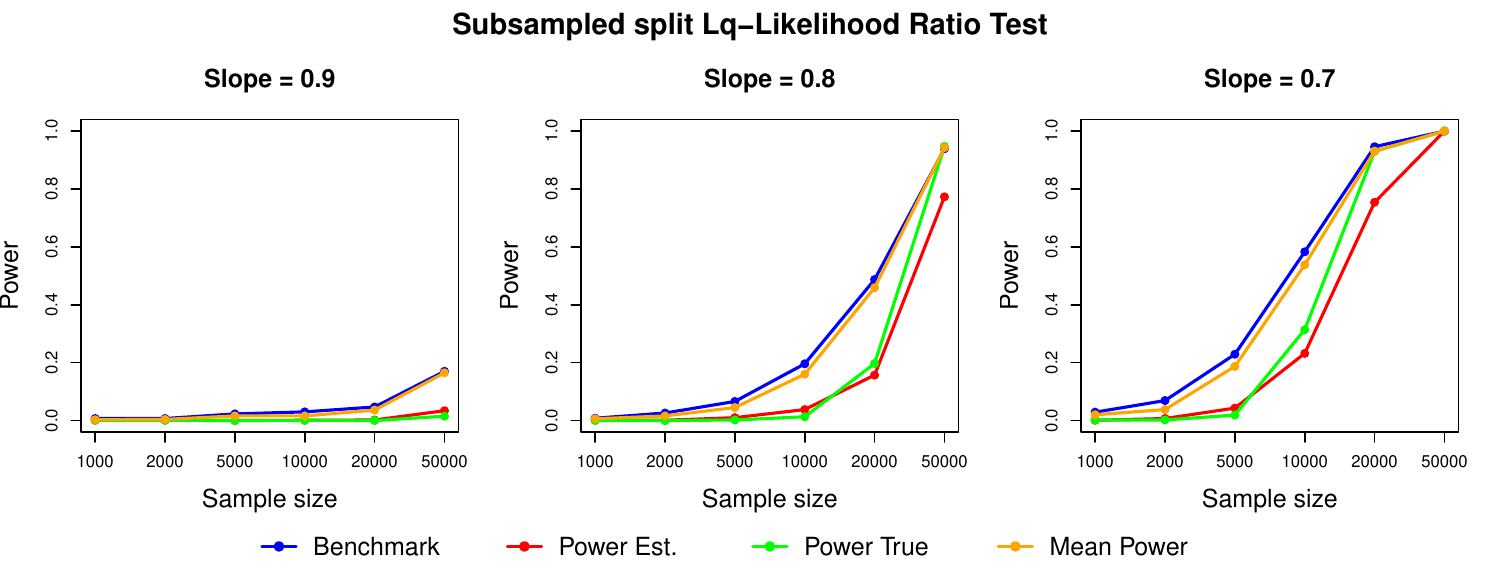}
\end{center}
\vspace{-.72cm}
\caption{Power of the sub-sampled split LRT (Benchmark), the sub-sampled split L$q$RT with the estimated optimal power parameter (Power Est.), the sub-sampled split L$q$RT with the true optimal power parameter (Power True), and the sub-sampled split mean power L$q$RT with (universal) critical value $1/\alpha=20$  for $B=1000$.}
\label{fig_subsampled_split_LRT_power_v2}
\end{figure}

Next, we consider the power of our sub-sampled split test statistics for $B=1000$ in Figure \ref{fig_subsampled_split_LRT_power_v2}. In all cases the power of the sub-sampled split LRT is now the best, followed closely by the sub-sampled split mean power L$q$RT. The power of the sub-sampled split L$q$RTs with $q^*$ and $\widehat{q}^*$ is lower except in large samples. Hence, the ranking of the test statistics based on their power is different for $B=1$ and $B=1000$. The puzzling thing is that the power of the sub-sampled split L$q$RTs with $q^*$ and $\widehat{q}^*$ improves only slightly and moderately in small/medium sample sizes compared to the case with $B=1$, whereas the power of the split LRT and the mean power L$q$RT improves much more due to sub-sampling in all sample sizes (for all miscalibration sizes). 

\begin{figure}[htb!]
\begin{center}
\includegraphics[width=.98\textwidth]{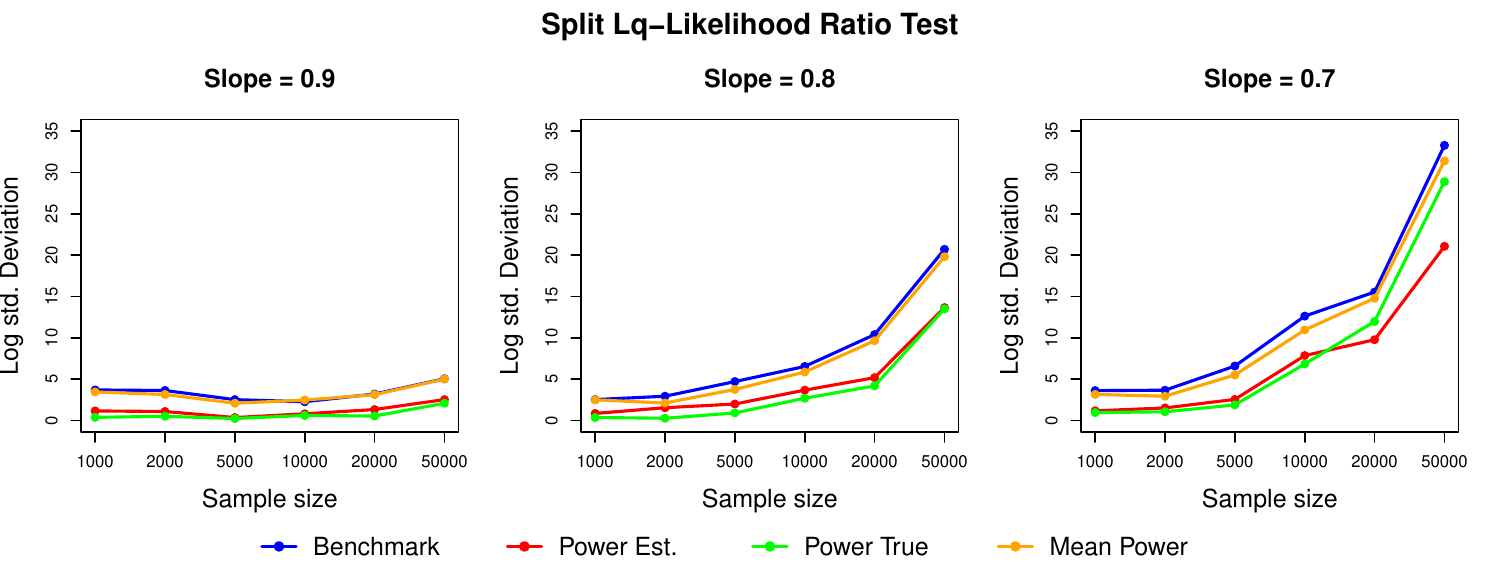}
\end{center}
\vspace{-.72cm}
\caption{Logarithm of standard deviation of the power of the split LRT (Benchmark), the split L$q$RT with the estimated optimal power parameter (Power Est.), the split L$q$RT with the true optimal power parameter (Power True), and the split mean power L$q$RT with (universal) critical value $1/\alpha=20$.}
\label{fig_split_LRT_sd}
\end{figure}

We provide an explanation. Consider the e-power of a sub-sampled split test statistics denoted by $E$. Its i.i.d.~replications are denoted by $(E_b)_{b=1}^B$. It is known that
\beqo
\be\left[\log\Big(\frac{1}{B}\sum_{b=1}^BE_b\Big)\right]\geq \be\Big[\log(E_1)\Big]=\be\left[\log\Big(\Big(\prod_{b=1}^BE_b\Big)^{1/B}\Big)\right],
\eeqo
by concavity of the logarithm. In the inequality above, we compare the arithmetic mean with the geometric mean, that is,
\beqo
\frac{1}{B}\sum_{b=1}^BE_b \quad \text{ against } \quad \Big(\prod_{b=B}^bE_b\Big)^{1/B},
\eeqo
and, intuitively, the larger the variance of $E_b$, the larger the difference in the arithmetic and the geometric mean. In Figure \ref{fig_split_LRT_sd} we study the standard deviation of our split test statistics. We notice that the standard deviation of the split LRT is the largest, followed by the split mean power L$q$RT, the standard deviations of the split L$q$RT with $q^*$ and $\widehat{q}^*$ are much lower. This observation may explain the change in the ranking of the test statistics based on their power. The standard deviation of a split test statistics must reach a certain level of volatility in order to observe significant benefits from sub-sampling.

\section{\bl{Motor insurance data example}}
\label{sec_app_4}

We illustrate the theoretical findings presented in this paper on a popular actuarial data set.
We select the Swiss motor third-party liability (MTPL) insurance data set of \cite{wuthrich_buser}. This data set was synthetically constructed based on a real French MTPL data set of \cite{dutang}. It contains information on insurance policies for 500,000 Swiss car drivers. The response variable is the number of claims, together with the corresponding exposure period and a range of explanatory variables. This data set has been widely used in actuarial applications; see, e.g., \cite{wuthrich_buser}, \cite{denuit_curves2} or \cite{Gatti}. The advantage of this data set is that the true mean values of the response variables are known.

The data set is independently split into a learning set $\mathcal{L}$ and a testing set $\mathcal{D}$ in an 80:20 ratio. Based on the learning set $\mathcal{L}$, we fit four predictive models.
\begin{itemize}
\item \textbf{Model 1:} We fit a generalized linear model (GLM) using all available covariates, which are appropriately pre-processed for the GLM; this model corresponds to Model GLM3 of \cite{wuthrich_buser}.
\item \textbf{Model 2:} We fit a gradient boosting machine (GBM) using all available covariates without pre-processing them, following the procedure described in Listing 7.2 of \cite{wuthrich_buser}. The tree depth is set to 3, the learning rate to 0.5, the shrinkage parameter to 0, and the minimum number of observations per bucket selected by 5,000. The number of iterations is set to 50, which was identified as an optimal number of iterations based on cross-validation performed on the learning set.
\item \textbf{Model 3:} We fit a GBM as in Model 2, but with the number of iterations increased to 100, resulting in a small degree of overfitting.
\item \textbf{Model 4:} We fit a GBM as in Model 2, but with the number of iterations increased to 500, resulting in a larger degree of overfitting.
\end{itemize}

Once the predictive models have been fitted, the goal is to validate the calibration and the specification of the mean estimates on the test set $\mathcal{D}$. In this motor insurance example, the test set contains 100,000 observations, which is substantially larger than the data sets considered in our numerical experiments in Section \ref{sec_examples}, but this is the usual size if we split typical real life insurance portfolios in a 80:20 ratio. We note that all four models satisfy the global balance property on the test set, with the fluctuations between the average mean estimates and the average observed claim counts being approximately of magnitude $0.5\%$.

\begin{figure}[htb!]
\begin{center}
\includegraphics[width=0.72\textwidth]{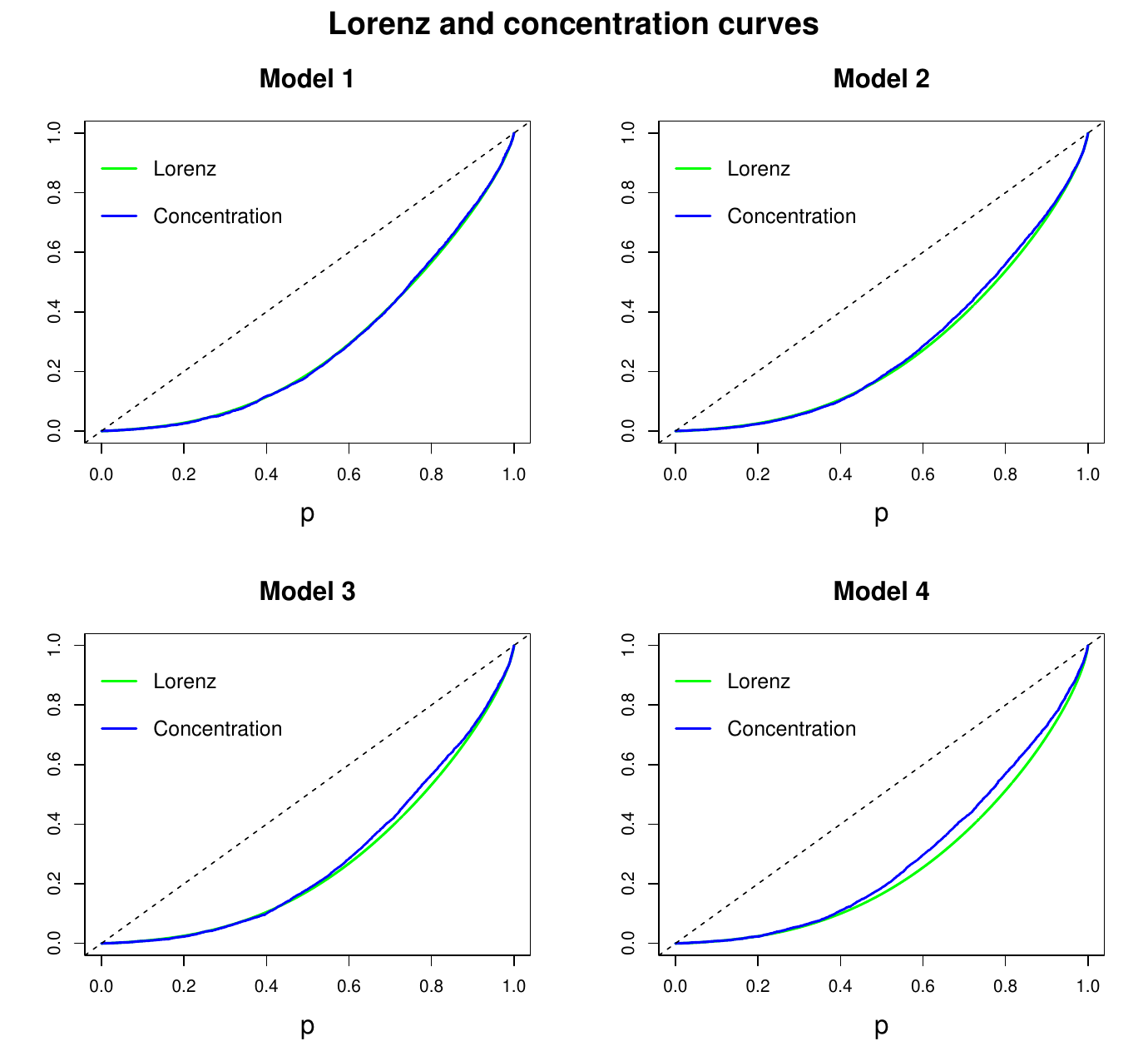}
\end{center}
\vspace{-.7cm}
\caption{Lorenz and concentration curves for the mean estimates on the test set $\mathcal{D}$; the dashed line is the diagonal.}
\label{fig_lorenz_example}
\end{figure}

We begin by applying the auto-calibration test from Section 4 of \cite{denuit_curves} to assess mean-calibration in the sense of \eqref{calibration_def}. The test is based on the Lorenz and concentration curves and, in particular, on the maximum absolute difference between the two curves. In \citet[Proposition 2.1]{denuit_curves}, the authors show that, for globally balanced mean estimates, the property of mean-calibration holds if and only if the Lorenz and concentration curves coincide. Figure \ref{fig_lorenz_example} presents these two curves for the four models; for more information about these curves, we also refer to \cite{Brauer1}. The two most extreme cases among the four models suggest that the mean estimates from Model 1 are likely to be calibrated, whereas the mean estimates from Model 4 are likely miscalibrated. Moreover, the degree of miscalibration appears to increase from Model 1 to Model 4, as the discrepancy between the Lorenz and concentration curves is smallest for Model 1 and largest for Model 4. This intuition is confirmed by the $p$-values of the auto-calibration test of \cite{denuit_curves}, reported in Table \ref{tab_example}. We use a significance level of $\alpha=0.05$. The null hypothesis of mean-calibration is rejected for Models 3 and 4, whereas there is insufficient evidence to reject the null hypothesis for Models 1 and 2. Based on the $p$-values, the conclusions are clear for Models 1, 2, and 4, whereas the evidence is less conclusive for Model 3. This is consistent with the shapes of the Lorenz and concentration curves shown in Figure \ref{fig_lorenz_example}.

\begin{table}[ht]
\centering
\begin{tabular}{rrrrrr}
  \hline
Model & Deviance & Miscalibration & Discrimination & $p$-value AC test\\ 
  \hline
Model 1 &  28.167 & 0.077 & 3.195 & 0.934\\ 
Model 2 &  27.884 & 0.082 & 3.482 & 0.186\\ 
Model 3 &  27.907 & 0.115 & 3.492 & 0.020\\ 
Model 4 &  28.103 & 0.161 & 3.343 & 0.000\\ 
   \hline
\end{tabular}
\caption{Performance metrics on the test set, including the Poisson deviance loss, the miscalibration and discrimination statistics from Murphy's score decomposition of the Poisson deviance loss and the $p$-value of the auto-calibration test from \cite{denuit_curves}.}
\label{tab_example}
\end{table}

The auto-calibration test of \cite{denuit_curves} provides a statistically rigorous method for testing mean-calibration. However, the $p$-values of their test have to be estimated by simulating the distribution of the test statistic under the null hypothesis, which may be less attractive to practitioners.

Next, we analyze the miscalibration statistics obtained from Murphy's score decomposition of the deviance loss, which is the building block of the classical and universal LRTs, see Section \ref{sec_classic_murphy}. We choose the Poisson deviance loss. Since the number of claims follows a Poisson distribution in this data set, we can set $\phi^*=1$ complying with assumption (B). Moreover, the empirical estimates of the dispersion parameter for all four models are close to 1. The results for the miscalibration statistics are presented in Table \ref{tab_example}. They confirm the findings of the auto-calibration test of \cite{denuit_curves} regarding the degree of miscalibration of the mean estimates and the ranking of the models according to their degree of miscalibration. However, based on the standalone values of the miscalibration statistics reported in Table \ref{tab_example}, we cannot make a statistical decision regarding the null hypothesis of mean-calibration.

\begin{figure}[htb!]
\begin{center}
\includegraphics[width=0.72\textwidth]{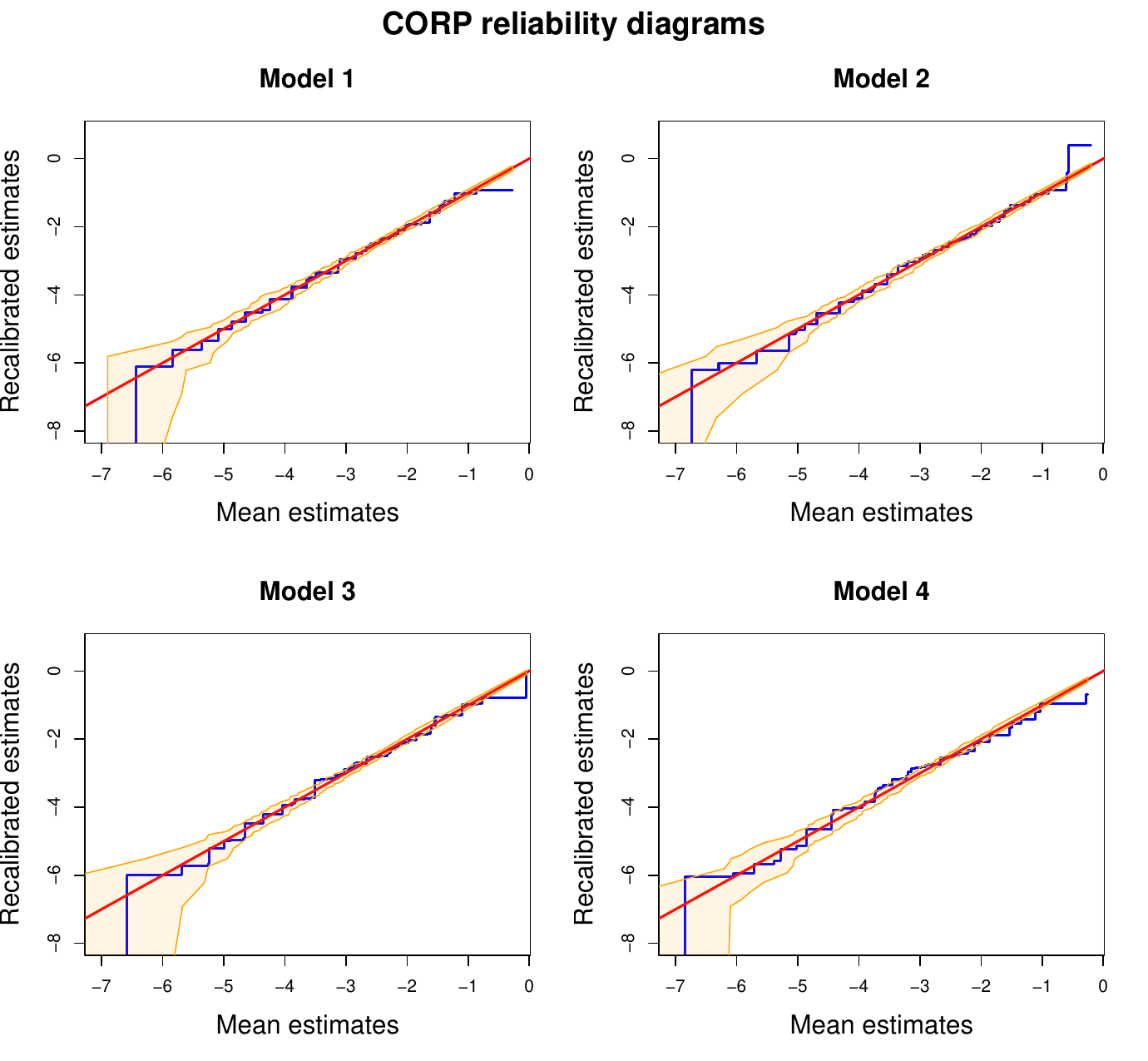}
\end{center}
\vspace{-.7cm}
\caption{CORP reliability diagram (blue) on the log-scale for the mean estimates on the test set together with $95\%$ point-wise consistency band (orange); the red line is the diagonal.}
\label{fig_corp_example}
\end{figure}

To gain further insight into the miscalibration statistics, we present the CORP reliability diagrams for the four models in Figure \ref{fig_corp_example}. We also include the $95\%$ point-wise consistency bands constructed using the parametric method proposed in \cite{delong_iso}. The consistency bands are generated by sampling the number of claims from Poisson distributions with the estimated mean values and re-fitting by an isotonic regression. A graphical inspection of the plots indicates that the deviations of the CORP reliability curve from the diagonal are small for Models 1 and 2, and become larger for Models 3 and 4. The CORP reliability diagrams for Models 1 and 2 are, in general, contained within the consistency bands, whereas those for Models 3 and 4 cross the consistency bands. The largest significant deviations from mean-calibration are observed in the far-right tail of the mean estimates for all models. Overall, the CORP reliability diagrams with the consistency bands provide evidence for mean-calibration of Models 1 and 2 and miscalibration of Models 3 and 4. This diagnostic is useful for assessing calibration of the mean estimates, but, again, it requires simulations of pseudo-responses to construct the consistency bands.

The auto-calibration test of \cite{denuit_curves}, the Lorenz and concentration curves, and the CORP reliability diagram provide information only about the (mis)calibration of the mean estimates. In Table \ref{tab_example}, we also report the Poisson deviance loss and the discrimination statistics, see Section \ref{sec_classic_murphy}. These quantities provide additional information that can be used to assess the specification of the mean estimates. An important observation is that Model 1 has the largest Poisson deviance loss on the test set, despite presenting the best mean-calibration. Model 4 has the second-largest Poisson deviance loss. Models 2 and 3 have smaller Poisson deviance losses than Models 1 and 4,  and they show similar discriminatory power, which is higher than that of Models 1 and 4. This may indicate that Model 1 is misspecified (pre-processing of covariates can be improved) despite being mean-calibrated, whereas Model 4 is both misspecified and miscalibrated.

Finally, we apply our sub-sampled split LRT. To determine whether we shall formally test mean-calibration or mean-specification under assumption (C) or (D), we first assess the plausibility of (C) and (D). We follow the discussion in Appendix \ref{sec_app_1}. To assess assumption (C), we examine the CORP reliability diagrams. Based on Figure \ref{fig_corp_example}, we do not observe any substantial flat regions in the CORP reliability diagrams. Consequently, assumption (C) appears to be satisfied for all four predictive models, at least approximately. Assumption (D) is more challenging to assess. We calculate Spearman's rank correlations between the mean estimates, and also include Model 0 in this analysis. Model 0 corresponds to the true mean values of the claim counts for the observations in the test set, which are provided in this synthetic data set. The correlation matrix is presented in Table \ref{tab:kor}. We observe that the mean estimates from Models 2 and 3 are very strongly correlated with each other, whereas the mean estimates from the remaining models are less strongly correlated. This suggests that the rankings of the mean estimates implied by Models 2 and 3 are similar, whereas the rankings implied by Models 1 and 4 differ more substantially from those implied by Models 2 and 3. Taking also into account the results reported in Table \ref{tab_example}, we conclude that assumption (D) may be more severely violated for Models 1 and 4. Since no further statistical assessment of assumption (D) is possible based on the available data, in practice, we would proceed under the assumption that (D) holds, at least approximately, for Models 2 and 3. For Models 1 and 4 it may be safer to assume (C). Note that at this stage, one could also use the Gini score for assessing the rankings among the models, and there is a statistical test for the Gini score in \cite{BrauerMenzel}.

\begin{table}[ht]
\centering
\begin{tabular}{rrrrrr}
  \hline
 & Model\_0 & Model\_1 & Model\_2 & Model\_3 & Model\_4 \\ 
  \hline
  Model\_0 & 1.00 & 0.94 & 0.98 & 0.98 & 0.96 \\ 
  Model\_1 & 0.94 & 1.00 & 0.95 & 0.94 & 0.91 \\ 
  Model\_2 & 0.98 & 0.95 & 1.00 & 0.99 & 0.97 \\ 
  Model\_3 & 0.98 & 0.94 & 0.99 & 1.00 & 0.98 \\ 
  Model\_4 & 0.96 & 0.91 & 0.97 & 0.98 & 1.00 \\ 
   \hline
\end{tabular}
\caption{Spearman's rank correlations between the mean estimates.} 
\label{tab:kor}
\end{table}

\begin{figure}[htb!]
\begin{center}
\includegraphics[width=0.72\textwidth]{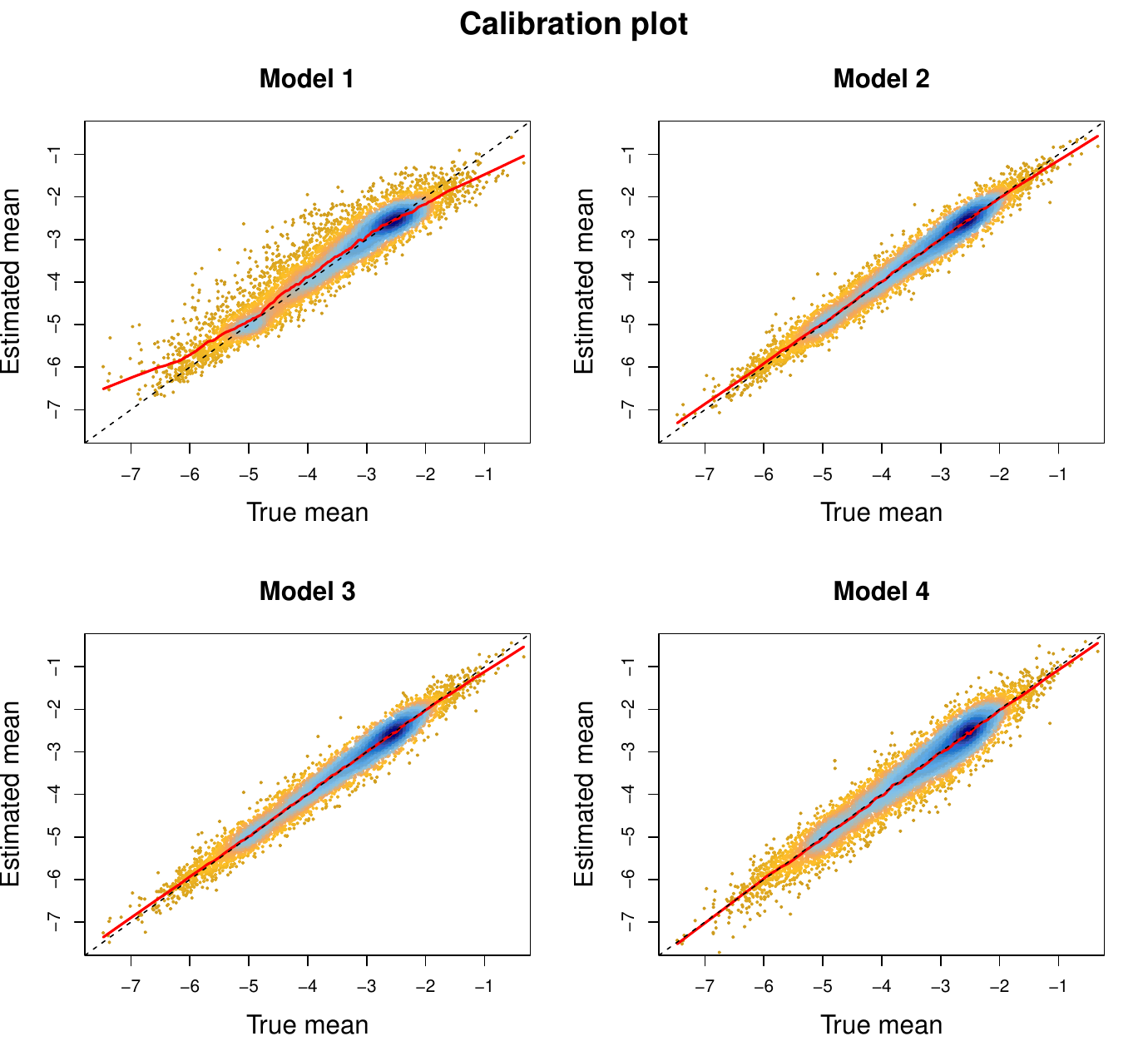}
\end{center}
\vspace{-.7cm}
\caption{Estimated means $\widehat{\mu}_i$ vs.~true means $\mu^*_i$ on the log-scale; the red lines present a local linear regression and the points present the predictions. A color is assigned to each point according to the local density of points: low-density regions appear in warm colors, while high-density regions appear in progressively darker blues.}
\label{fig_ranking_example}
\end{figure}

Since the true mean values are available, Figure \ref{fig_ranking_example} presents the relationships between the mean estimates from Models 1 to 4 and the true mean values from Model 0. The figure confirms that the ranking of the mean estimates is more strongly distorted for Models 1 and 4 than for Models 2 and 3, relative to the ranking of the true mean values from Model 0. This indicates that the violation of assumption (D) is indeed more pronounced for Models 1 and 4 than for Models 2 and 3. For each relationship, we fit a local linear regression using a small bandwidth parameter that nevertheless ensures sufficiently smooth fitted values. The global trend in the relationship between the mean estimates and the true mean values appears to be strictly increasing for all models. Yet, at the individual level, the mean estimates clearly do not preserve the ranking of the true mean values. This is expected as a strictly monotone relationship between the mean estimates and the true mean values is unlikely to hold in real statistical applications. A violation of assumption (D) is not, in itself, a problem for our sub-sampled split LRT. As discussed in Section \ref{sec_examples_c_d}, a violation of (D) may result in a loss of power of the test. The loss of power depends on the severity of the violation of assumption (D), the miscalibration size and the sample size.

\begin{figure}[htb!]
\begin{center}
\includegraphics[width=0.72\textwidth]{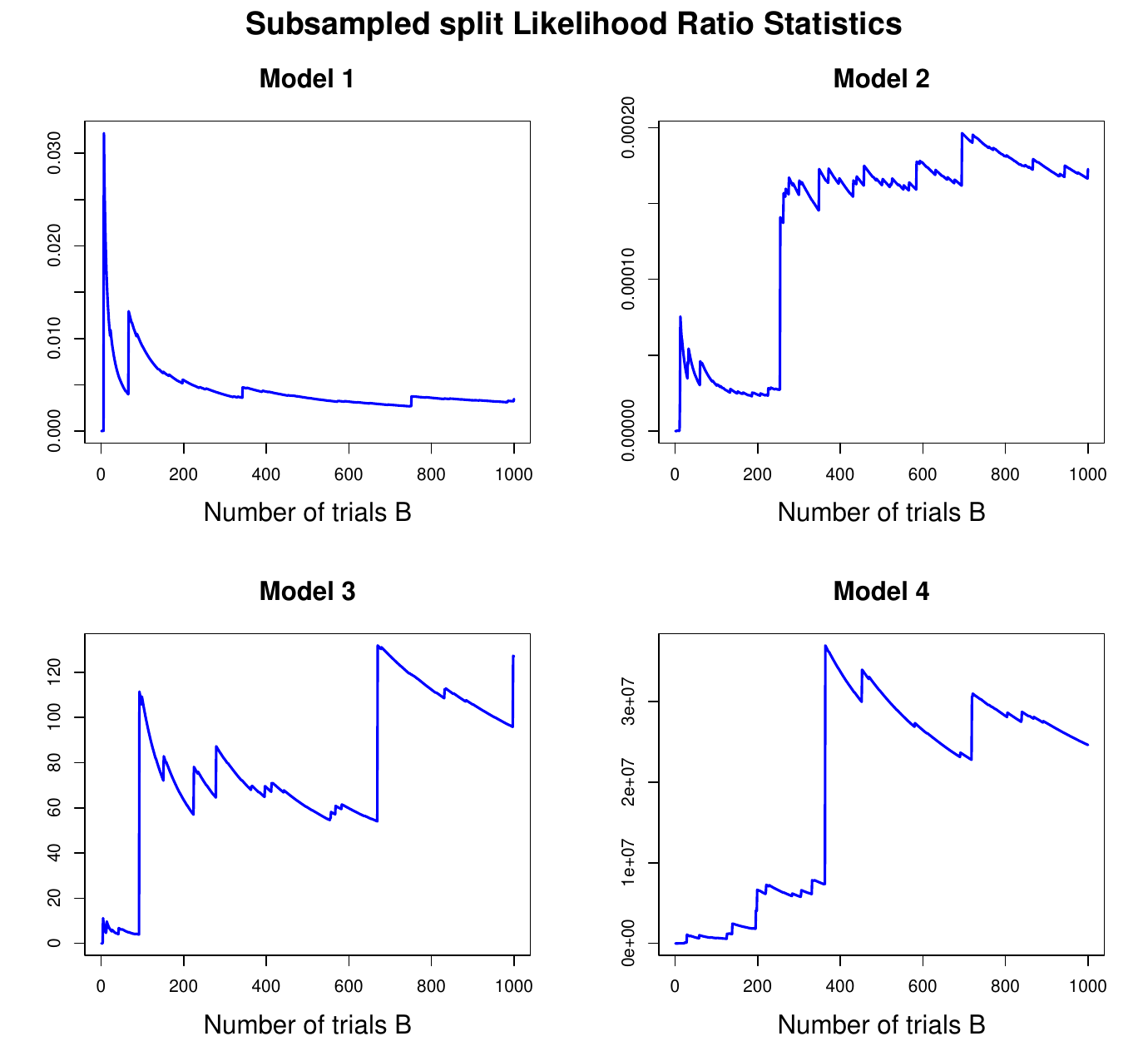}
\end{center}
\vspace{-.7cm}
\caption{Running value of the sub-sampled split LRS in repeated sampling on the test set $\mathcal{D}$, $B\ge 1$.}
\label{fig_universal_example}
\end{figure}

\begin{figure}[htb!]
\begin{center}
\includegraphics[width=0.72\textwidth]{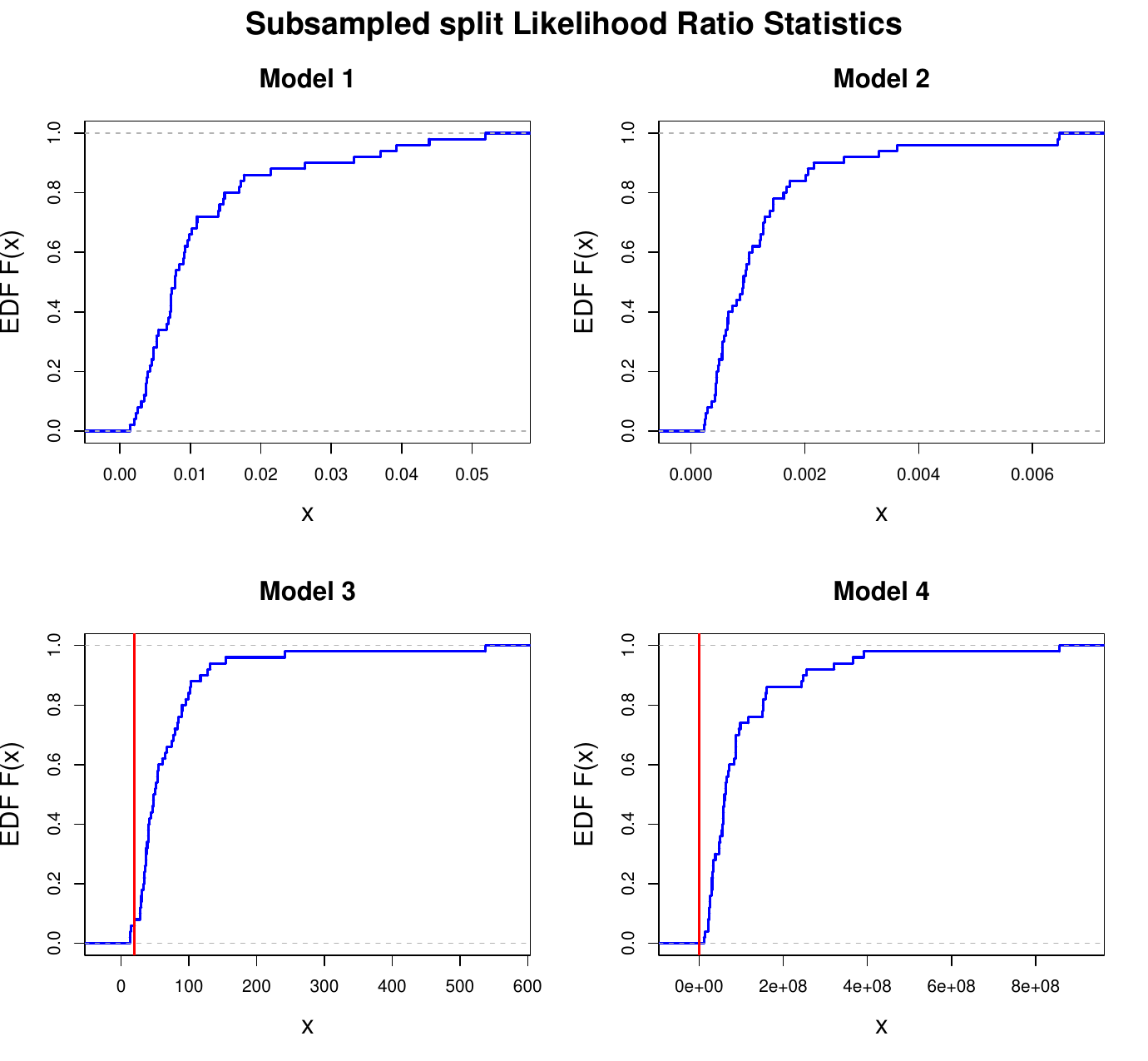}
\end{center}
\vspace{-.7cm}
\caption{Empirical distribution function of the sub-sampled split LRS based on $B=1000$ sub-samples from the test set $\mathcal{D}$; the red line is the (universal) critical value $20$.}
\label{fig_histogram_example}
\end{figure}

We test mean-calibration and mean-specification of the estimates from Models 2 and 3 under assumption (D), and mean-specification of the estimates from Models 1 and 4 under assumption (C). We use $B=1000$ subsamples and a split ratio of $0.5$. The LRS is calculated using the Poisson likelihood. One realization of the running value of the sub-sampled split LRS is presented in Figure \ref{fig_universal_example}, while the empirical distribution of the sub-sampled split LRS based on $B=1000$ splits is presented in Figure \ref{fig_histogram_example}. The empirical distributions are constructed from 50 simulations of the sub-sampled split LRS, with $B=1000$ splits in each simulation. We use a significance level of $\alpha=0.05$ and the universal critical value $1/\alpha=20$.

The sub-sampled split LRT does not reject the null hypothesis for Models 1 and 2, whereas the null hypothesis is rejected for Models 3 and 4. These results are consistent with those obtained using the other metrics presented above. For Model 3, we observe four data-splitting scenarios in which the test statistic is below 20. For Models 1, 2 and 4, the test decision is consistent across all data-splitting scenarios. The previous analyses have also shown that the evidence against Model 3 is less conclusive, its $p$-value for the auto-calibration test is $0.020$, and only a small part of its CORP reliability diagram lies outside the consistency bands. 

The results obtained using the sub-sampled split LRT can be interpreted as follows. For Models 3 and 4, the degree of miscalibration appears sufficiently large for the sub-sampled split LRT to reject the null hypotheses of mean-calibration (Model 3) and mean-specification (Model 4). These results suggest that our statistical test can detect miscalibration under (D) and misspecification arising from miscalibration under (C). Recall that, under assumption (C), the null hypothesis of correct mean-specification can be rejected only when the miscalibration of the mean estimates is sufficiently large. For Model 2, the absence of evidence against the null hypotheses may admit in general two possible explanations. First, assumption (D) may hold approximately, in which case the miscalibration and misspecification of the mean estimates from Model 2 may be small. Under this explanation, the test may have adequate power to detect miscalibration and misspecification, both equivalent under (D), but there may be insufficient evidence of miscalibration and mispecification to warrant the rejection of the null hypothesis. Second, assumption (D) may be substantially violated while assumption (C) remains valid. In this case, the mean estimates from Model 2 may be correctly specified or misspecified, but the resulting miscalibration may be too small for the isotonic regression procedure to detect. The test may have insufficient power to identify misspecification of the calibrated mean estimates. Nevertheless, as argued in Section \ref{sec_universal_split}, the test should retain power to detect miscalibration whenever sufficient evidence of miscalibration is present. Consequently, under either explanation, our statistical test should detect miscalibration under (C) and (D) if it concerns mean estimates. Given the results presented above and our working assumption that (D) holds for Model 2, we consider the first explanation more plausible. We therefore interpret the results for Model 2 as providing evidence in favor of calibrated and correctly specified mean estimates. For Model 1, the sub-sampled split LRT also does not provide sufficient evidence to reject the null hypothesis of mean-specification under assumption (C) or the null hypothesis of mean-calibration under assumption (D). Since assumption (C) is adopted for Model 1, the failure to reject the null hypothesis of correct mean-specification does not necessarily imply that the mean estimates are correctly specified. As already discussed, this potential false negative may reflect the limited power of the sub-sampled split LRT to detect misspecification under assumption (C) when the resulting miscalibration is weak. When testing mean-specification under assumption (C), the isotonic estimate of the mean values under the alternative hypothesis is not the MLE. Consequently, when the mean estimates are misspecified but only weakly miscalibrated, a more efficient estimator of the mean values under the alternative hypothesis may increase the power to detect misspecification. At the same time, although assumption (C) serves as our working assumption and we formally test mean specification for Model 1, the construction of the test statistic allows us to interpret a failure to reject the null hypothesis as evidence consistent with the calibration of the mean estimates, even if assumption (D) is violated for Model 1. This interpretation also agrees with the other results presented above. We conclude that our statistical test based on universal inference yields valid decisions in all four cases.

We believe that the sub-sampled split LRT provides an attractive approach for verifying calibration and specification of mean estimates. It can complement other well-established measures of miscalibration and misspecification. Its main advantage lies in its simplicity of application, it requires only repeated splitting of the test sample and avoids the need for sophisticated simulation procedures.

We point out that since \cite{denuit_curves} do not consider weights associated with responses in their approach, the analysis above is performed for the number of claims as responses, not the claim frequency. The prediction of the number of claims is given by a regression function $(V,\mathbf{X})\mapsto V\widehat{\mu}(\mathbf{X})$. Consequently, mean-calibration and mean-specification of a regression function for predicting the number of claims are understood as
\beqo
\mathbb{E}\left[Y \mid V\widehat{\mu}\left(\mathbf{X}\right)\right] &=& V\widehat{\mu}\left(\mathbf{X}\right), \qquad \text{almost surely},\\
\mathbb{E}\left[Y \mid V,\mathbf{X}\right] &=& V\widehat{\mu}\left(\mathbf{X}\right), \qquad \text{almost surely}.
\eeqo
The former referes to $PZ$-calibration in \cite{W2026}, and this reference also provides further versions of mean-calibration under exposures that can be treated in a similar way.

\end{appendices}

\end{document}